\documentclass{aa}  

\usepackage{graphicx}
\usepackage{txfonts}
\usepackage{longtable}
\usepackage{supertabular}  
\usepackage{color}
\usepackage[dvipsnames]{xcolor}
\usepackage{float}
\usepackage[normalem]{ulem}
\newcommand{\ind}[1]{_{\mathrm{#1}}}

\begin{document} 

   \title{Towards the true number of flaring giant stars in the \emph{Kepler} field}
   \subtitle {Are there flaring specialities associated with the giant nature?}

   \author{K. Ol\'ah
          \inst{1}
          \and
          Zs. K\H{o}v\'ari
          \inst{1}
          \and
          M.~N.~G\"unther \inst{2,}\thanks{Juan Carlos Torres Fellow}
          \and
          K. Vida
          \inst{1}
          \and   
          P. Gaulme
          \inst{4}
          \and
          B. Seli
          \inst{1,5}
          \and
          A. P\'al
          \inst{1}
          }

   \institute{Konkoly Observatory, Research Centre for Astronomy and Earth Sciences, Budapest, Hungary\\
\email{olah@konkoly.hu}
         \and
         Department of Physics, and Kavli Institute for Astrophysics and Space Research, Massachusetts Institute of Technology, Cambridge, MA 02139, USA
         \and
        Max-Planck-Institut f\"ur Sonnensystemforschung, G\"ottingen, Germany 
         \and
        E\"otv\"os Lor\'and University, Department of Astronomy, Budapest, Hungary 
             }

   \date{Received mm dd, yyyy; accepted mm dd, yyyy}

 
  \abstract
   {} 
    {We aim to give a reliable estimate of the number of flaring giant stars in the {\it Kepler} field. By analyzing the flaring activity of these stars we explore their flare statistics and the released flare energies. The role of oscillation in suppressing magnetic activity is also investigated. On a sample of flaring giant stars we search for flaring specialities which may be associated with the giant nature.}
   {We search for flares using the $\approx$4\,yr long \emph{Kepler} data on a sample of 706 stars compiled from two lists of flaring giants ($\log g\leq3.5$) found in the literature. To lessen the probability of false positives two different pipelines are used independently for flare detection.
   Tests are carried out to correct the detection bias at low flare energies for a subsample of 19 further studied, frequently flaring stars. For these 19 stars flare energy distributions and flare frequency diagrams (FFDs) are constructed. For comparison purposes KIC~2852961 \citep{2020arXiv200505397K} is re-analysed with our present approach. 
   }
   {From the 706 \emph{Kepler} flaring giant candidates, we ruled out those where oscillations or pulsations were misclassified and those that turned out to be dwarf stars. In the end, we confirm only 61 stars as flaring giants.
   Among these 61 flaring giants we found only six which also show oscillations;  we suggest that a large fraction of the 61 flaring giants are members of spectroscopic binaries which are proven already for 11. The number of detected flares on giant stars correlate only weakly with the rotational periods.
   The FFDs for the most flaring 19 stars were fitted by power-law functions. On log-log representation the slopes of the individual fits lead to an average $\alpha = 2.01\pm 0.16$ power law index, but the ranges of flare energies scatter within almost two orders showing the inherent heterogeneity of the sample of flaring giants. Broken power-law fits are applied for two giant stars which have similar flare energy ranges, however, the energy at the breakpoints of the power laws are different showing possible differences in the magnetic field strengths and atmospheric structures of these stars. The average power law index of $\alpha\approx$\,2 is the same for the flaring giants, for the (super)flaring G-dwarfs and for dwarf stars between spectral types M6--L0. 
   }
   {The 61 confirmed flaring giant stars make up only $\approx$0.3\% of the entire giant star population in the {\it Kepler} database, in contrast with previous estimates of about an order higher percentage. We found that most of the false positives are in fact oscillating red giants. No strong correlation was found between the stellar properties and the flaring characteristics. The wide scale of the flaring specialities are hardly related to the giant nature, if at all. This, together with the finding that the observed flare durations correlate with flare energies, regardless of the flare energy level and stellar luminosity class, suggest common background physics in flaring stars, or in other words, a general scaling effect behind.}


   \maketitle
%

\section{Introduction}\label{intro}

Magnetism accompanies stars from cradle to grave, magnetic fields play a crucial role from the formation of stars to the end of their existence. Most stars at a certain stage of stellar evolution exhibit enhanced magnetic activity, showing phenomena like the most easily observable flares. Stellar flares are sudden eruptions of energy through magnetic reconnection which occur mostly (but not exclusively) on late-type stars. The incidence of flaring stars among different types of stars including solar-type stars, cool dwarfs,  active giants, and even early-type stars marks the presence and strength of magnetic field, and this way has serious impact in studying the evolution of such stars in general \citep{2016ApJ...824...14C}, therefore knowing their true number is essential. Magnetically active red giant stars with strong flares gather on the red giant branch (RGB) of the Hertzsprung$-$Russell-diagram (HRD). This is an exciting episode of stellar evolution. Depending on their mass, the stars spend only a short period of their life on the RGB, with continuously expanding atmosphere and basic changes in their inner structure and energy production. 

Flares on stars (including the Sun) originate through magnetic reconnection in emerging flux tubes. Theoretical description of the magnetohydrodynamic processes based on data of the only star where flares are directly observable, the Sun, is reviewed by \citet{2011LRSP....8....6S}. For a flare eruption a working magnetic dynamo originating from the counteraction of the differential rotation and convective motions in the stellar envelope is needed. A comprehensive review of solar-stellar magnetism was recently presented by \citet{2017LRSP...14....4B}. These works however, deal mostly with late-type dwarf stars. The role of the magnetic field in the activity of giant stars is only little studied and not well understood. To our knowledge no later study about emerging (and trapped) magnetic flux tubes, causing the observable magnetic features like spots or flares, exists, than the one by \citet{2001A&A...377..251H}. 

The existence of the magnetic field is the base of the flare activity on giant stars as well as on dwarfs, however, measuring it is not an easy task. The interested reader finds a summary of such measurements on stars in \citet{2012LRSP....9....1R}. Indirect evidence of magnetic activity on giant stars is the rotational modulation of brightness caused by starspots which is monitored using photometry for decades on a few active giants \citep[][for example]{2009A&A...501..703O} and more recently are studied by Doppler imaging, see e.g., \citet[][see their Table 1.]{2017AN....338..903K}, while a summary of the observables of cool star dynamos, including giant stars, is presented by \citet{2014SSRv..186..457K}. Recently, spots on giant stars' surfaces were even directly imaged \citep{2016Natur.533..217R}.

\begin{figure}
   \centering
   \includegraphics[width=\columnwidth]{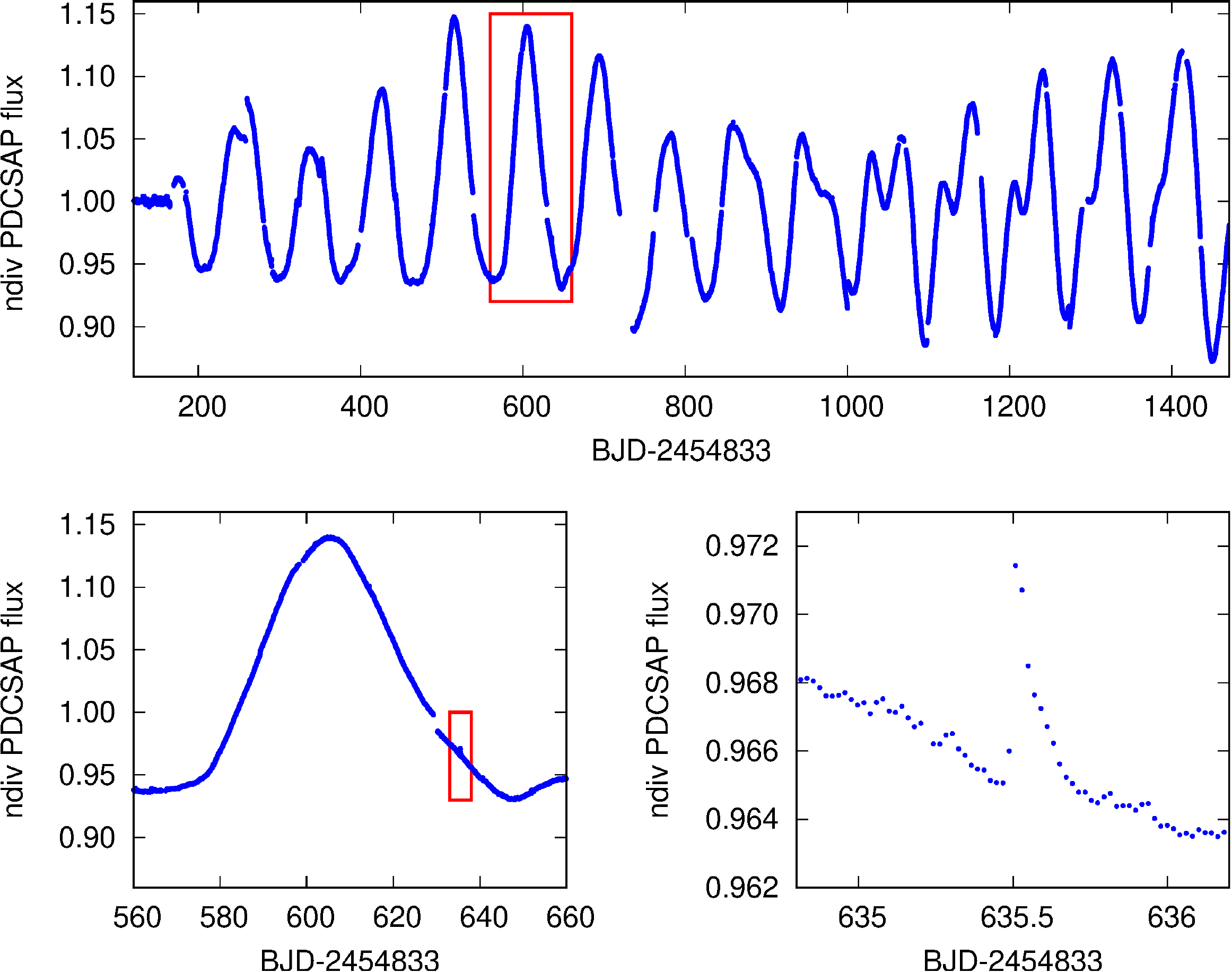}
      \caption{One flare on KIC~6861498 (from the 16 observed) lasting about five hours with negligible amplitude compared to the rotational modulation of $\approx89$ days long. Top: the whole dataset, bottom left: a flare on the descending branch of the rotational modulation, bottom right: the flare enlarged.}
         \label{flare_example}
   \end{figure}

Before the era of space-borne photometry, observations of flares on giant stars were restricted to very few studies. The reason behind this is the fact that flares on giants are hardly observable from the ground: {\it (i)} the day-night period on the Earth generally restricts continuous monitoring of stars. With rotational periods over about a day to weeks, even one single rotation cannot be fully covered - except the polar regions of the Earth with only a few instruments. As an example, see the results of a continuous 10.15 days $BVR$ photometry with 155 sec cadence for the RS~CVn active binary V841~Cen (and the non-radially pulsating $\delta$\,Scuti star V1034~Cen) from the Antarctica by \citet{2008A&A...490..287S}. {\it(ii)} the signal-to-noise (S/N) of the ground-based photometry is limited by atmospheric effects and weather conditions, {\it(iii)} due to the high luminosity of giant stars the flares appear on a luminous background resulting in only a small amplitude brightness increase relative to the star. Fig.~\ref{flare_example} shows a flare lasting a few hours on an active giant star with a long period and high amplitude rotational modulation; this flare would be hardly (or even not) observable from the ground.

Among the very few flares observed on giant stars from the ground there is a remarkable six days long event. This huge eruption, observed in H$\alpha$ by \citet{1994A&A...287..575C}, occurred on HK~Lac, a close binary with an active giant component. The flare released $1.28\times10^{37}$\,erg total in H$\alpha$, and it was correlated with the appearance of a new active region on the star \citep{1991A&A...251..531O}. Another interesting example is an almost 10 days log flare event lasting for about one stellar rotation on YY~Men, an FK~Com-Type K1III giant star, observed photometrically in $UBV(RI)_\mathrm{C}$ broadband colors by \citet{1992A&A...263L...3C} obtaining one datapoint/night. The total released energy in the optical passbands was about $1.8\times10^{39}$\,erg \citep{1992A&A...263L...3C}. 

Continuous monitoring of stars by space instruments in optical wavelengths (\emph{Kepler}, \emph{TESS}) both in 2-min and 30-min cadence immediately initiated studying the incidence of flaring objects among stars with different stellar parameters such like their spectral type, mass, effective temperature etc. A number of different algorithms have been developed to find flares in the enormous photometric datasets  provided by the satellites (in the order of $10^5-10^6$ stars from each instrument). In the same time the question also arose, namely, how trustful are the algorithms, considering that flares can be mixed up with several real and erroneous features like stellar oscillations or photometric errors. Of course, false positive detections are hardly inevitable but the goal is to minimize their number.

\citet{2016ApJ...829...23D} investigated the available \emph{Kepler} data from Quarters 0-17 and catalogized 4041 flare stars but did not list their luminosity classes. 
\citet[][hereafter D17]{2017ApJS..232...26V} analysed the long-cadence \emph{Kepler} data from Quarter 15 and tabulated the flaring stars by spectral types. They gave a list of 695 flaring giant stars out of the 21875 analysed giants yielding a 3.18\% flare incidence, slightly higher than the rate for the G-stars. These cited results used fully automatic pipelines for flare detection which resulted in a high number of false positives. The possible origin of the detected false positives in \citet{2016ApJ...829...23D} was explained by \citet{2019ApJ...871..241D} in their own results, namely, that the earlier version of the {\tt appaloosa} code insufficiently treated the periodic signals in the data leaving in sharp or peaked structures in the residuals which then were recognized as flares.

Finally, applying all \emph{Kepler} long-cadence, DR~25 data  \citet[][hereafter YL19]{2019ApJS..241...29Y} discussed in detail the origin of the high number of false positive flare detections in the earlier studies and summarised the characteristics of all previous flare-finding techniques together with their own approach which included the possibility of visually checking the results. In their Table~3 YL19 compared the flare incidences published by different authors with their own results, and specifically, they found 0.33\% flare incidence for the giant stars. From the final dataset of 3420 flaring stars of YL19 the published stellar parameters made us possible to select the giant flaring stars of the paper for our present study. 

In this paper we revise the earlier detections of D17 and YL19 on the incidence of flaring stars among the giants in the \emph{Kepler} field and give details about the flare characteristics of the confirmed flaring giants. 

Recently we studied KIC~2852961, a giant star in the {\it Kepler} field discussing its flaring properties in \citet[][hereafter Paper~I]{2020arXiv200505397K}, with a title "Superflares on the late-type giant KIC~2852961". Studying now dozens of flaring giants a perhaps naive question arises: what does "superflare" indeed mean? Since superflares are commonly defined by energies $\log E_\mathrm{bol} > 32.0$\,[erg] all detectable flares on giant stars would automatically be classified as superflares. But this definition is only an energy scale: the environments where the flares originate are different on ultracool dwarfs (full convenction, high gravity), on main-sequence G, K and M dwarfs (convective shell, high gravity) and on giants (convective shell, low gravity). This question is discussed in Sect.~\ref{ffd:19stars}.

The outline of the paper is as follows: the applied data are summarized in Sect.\,\ref{sect_data}, the description of the applied methods are found in Sect.\,\ref{sect_methods}. In Sect.\,\ref{sect_results_disc} and Sect.\,\ref{sect_results_flares} the results are given and discussed, and are summarized in Sect.\,\ref{sect_summary}.


\section{Data}\label{sect_data}

For checking the flaring events in giant stars we used a set of 706 stars compiled from the lists of D17 with radii less than 30 solar radius (666 stars), and from YL19 (80 stars in this paper, 40 stars are common in the two lists). The stars in the aforementioned lists were selected taking $\log g\leq3.5$ as well. Both these studies originally used the stellar parameters from the Kepler Input Catalog (KIC) database; D17 used KIC~DR10 while YL19 used KIC~DR25. However, recently, \citet{2018ApJ...866...99B} revised the radii of the \emph{Kepler} stars, and used the $T_{\rm eff}$ values from \citet{2017ApJS..229...30M} supplemented with information from other sources. The Revised \emph{TESS} Input Catalog \citep{2019AJ....158..138S} (hereafter TICv8) contains most of the stars we planned to study. Thus, it is timely to revise the flaring giant statistics in the light of the revised stellar parameters.

\begin{figure}
   \centering
   \includegraphics[width=\columnwidth]{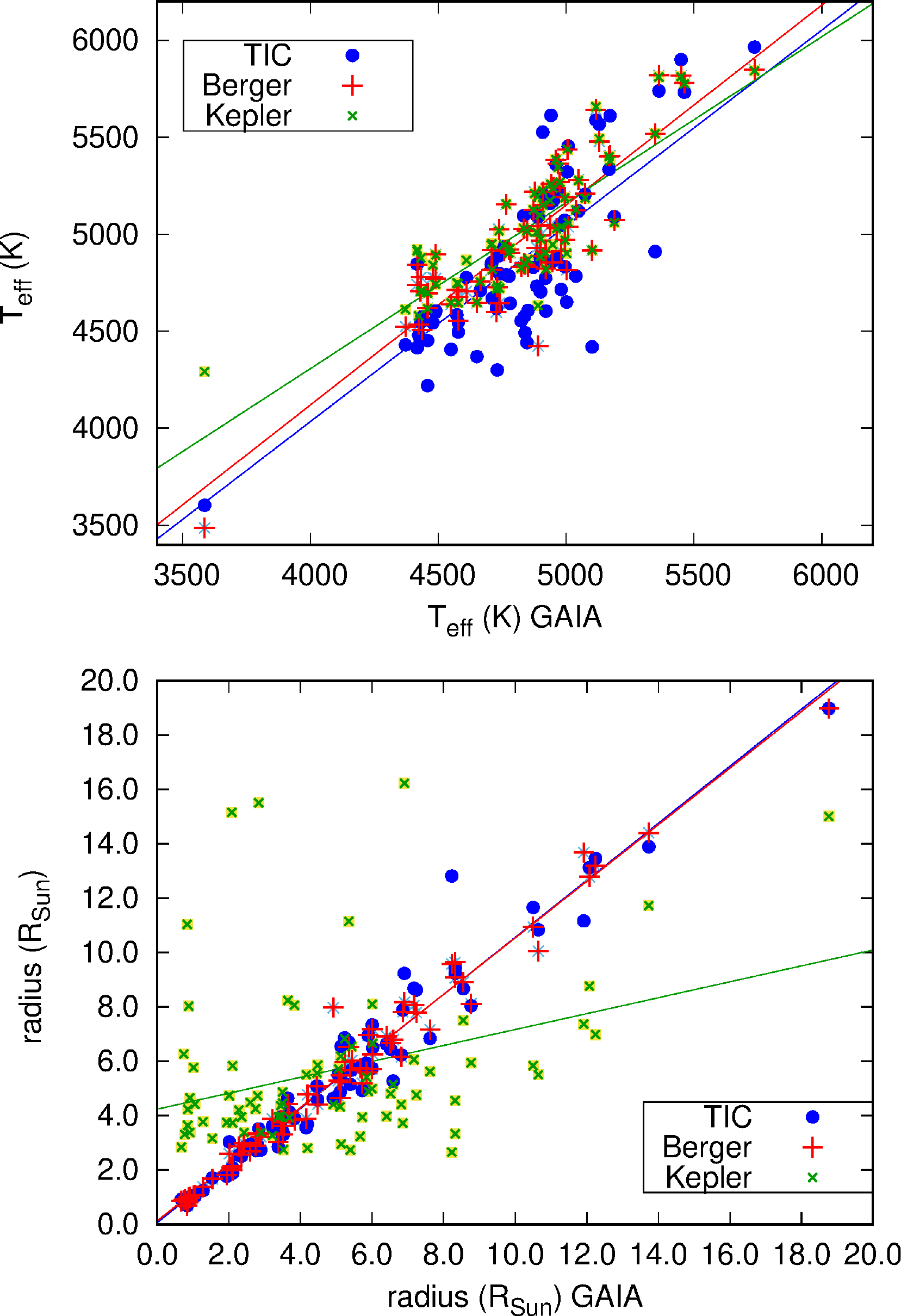} 
      \caption{Comparison of effective temperature and radius values of 77 flaring stars from the resulting 86 using TICv8, Berger \citep{2018ApJ...866...99B}, \emph{Kepler} and \emph{Gaia}~DR2 databases. See the text for more.}
         \label{compare_params}
   \end{figure}

To decide the most useful catalog to represent stellar parameters for our study, in Fig.~\ref{compare_params} we plotted $T_{\rm eff}$ and radii of those 77 flaring stars from the resulting 86 (see Sec.\ref{sect_results_disc}, Table~\ref{tab:long}) which have values in all of these catalogs: KIC, \emph{Gaia} DR2, TICv8 and that of \citet{2018ApJ...866...99B}. The result shows that concerning the KIC data, the radii of the stars strongly deviate from the values comparing to the other three sources; data from
\citet{2018ApJ...866...99B}, \emph{Gaia}~DR2 and TICv8 agree well with each other. The temperature values of \citet{2018ApJ...866...99B} and KIC have a small, on average $\approx$100\,K systematic difference comparing to the \emph{Gaia}~DR2 and TICv8 data. The mean effective temperature of the 77 stars from the \emph{Gaia}~DR2 and TIC 
catalogs are $4831\pm305$ and $4872 \pm428$, whereas from \citet{2018ApJ...866...99B} and KIC are $4977\pm366$ and $5018\pm312$, respectively. We note, that the temperature values in \emph{Gaia}~DR2 and TIC catalogs, and that of in 
KIC and Berger catalogs are mostly of the same origin, as shown by these averages as well. For the full D17+YL19 sample of giants used in this paper comparisons of the temperatures, radii and distances (which are important for the radius determinations) are compared in Appendix~\ref{A1} on Figs.~\ref{dist}--\ref{temp}.

In this paper we used the long cadence \emph{Kepler} Pre-search Data Conditioning (PDC) light curves from DR25 with the help of various facilities of the NASA Exoplanet Archive.
Finally,  $T_{\rm eff}$ and radii of the stars in the tables of results of the present paper are taken from TICv8 \citep{2019AJ....158..138S}, if not available, than from \emph{Gaia}~DR2 and KIC~DR25, in this order. 

\section{Methods}\label{sect_methods} 
 
To identify the flaring giant stars we used conventional (visual inspection) and automated pipeline search in combination. The first approach to check if the stars were indeed flaring in D17 and YL19 was simply to inspect their light curves one by one visually. First we kept the seemingly truly flaring and the questionable cases and got a list of stars that were good candidates as indeed flaring giants.

Next, a flare search made use of an automated pipeline accompanied by visual inspection.
To this end, we ran an updated version of the flare detection algorithm from \citet{2020AJ....159...60G} on the detrended \emph{Kepler} light curves. First, we computed a Lomb-Scargle periodogram to identify (semi-)periodic modulation caused by stellar variability or rotation. Then, we removed the dominating periodic signal using a cubic spline with knots spaced at one tenth of this period. During this step, we also identified and masked all outlier datapoints at 3-$\sigma$ by sigma clipping the data residuals. We repeated this entire process two more times or until no new periods were found. Finally, from the collected list of all outliers, we considered those to be flare-like events which contained a series of at least three subsequent 3-$\sigma$ outliers. 

These candidates were then visually inspected to decide if they should be declared a flare or a false positive. For this, we used the prior knowledge of typical M-dwarf flare profiles in white light, which typically show a rapid rise followed by exponential decay. However, we also identified and recorded various shapes that were clearly distinct from the stellar variability, but did not exactly match the classical rapid rise and exponential decay profile. These might have been altered by quasi-period oscillations, scatter, instrumental effect, multiple flare events — or a different physical process leading to the flare origin. 

We cross-matched the lists from the two approaches (purely visual, and automated pipeline combined with visual inspection) and checked again the questionable cases. Those stars which show brightenings that we considered unlikely to be flares were deleted from the list. This way we got a final list of 86 flaring stars. 

Next, we run the pipeline FLATW’RM (FLAre deTection With Ransac Method) of \citet{2018A&A...616A.163V}\footnote{https://github.com/vidakris/flatwrm/} for each target of the final list of the 86 flaring stars for independent automated flare detections. Finally, the individual flare detections carried out by the two independent pipelines from \citet{2020AJ....159...60G} and FLATW’RM of \citet{2018A&A...616A.163V} were cross-matched, and a flare was considered real when it appeared as a result from both pipelines. 

The observed flare duration is the time difference between the last and first datapoints of the flares which are above the 3-$\sigma$ detection limit, and, due to the long (30\,min) cadence, this underestimates the true flare duration especially for shorter flares.

For calculating the $\varepsilon_f$ relative flare energies (that is the flare energies relative to the quiescent star) we made use of the program FLATW’RM \citep{2018A&A...616A.163V} as well. To run FLATW’RM we set the minimum number of flare points to two, and the detection level to 3-$\sigma$. This way we could minimize the number of false positives at the cost of missing some smaller events. As template a simple "classical" flare shape was used \citep{2014ApJ...797..122D}; we discuss the effect of this simplification in the flare energy calculations of complex flares in Section~\ref{flares_number_energy}. 
From the $\varepsilon_f$ relative flare energies the $E_f$ absolute flare energies were derived by multiplying $\varepsilon_f$ values by the $L_{\star{\rm Kep}}$ base luminosity of the quiescent flare hosting star. Base luminosities in most cases are obtained using stellar effective temperatures and radii from TICv8 (if not available, then from \emph{Gaia}~DR2 and KIC~DR25, see Table~\ref{tab:long}) and assuming blackbody radiation through the \emph{Kepler} filter. For more details on flare energy calculations see Paper\,I (their Sect.\,6).

Rotational periods of the flaring stars were calculated using the Lomb-Scargle algorithm provided by the NASA Exoplanet Archive Kepler Stellar Table.

   \begin{figure}
   \centering
   \includegraphics[width=\columnwidth]{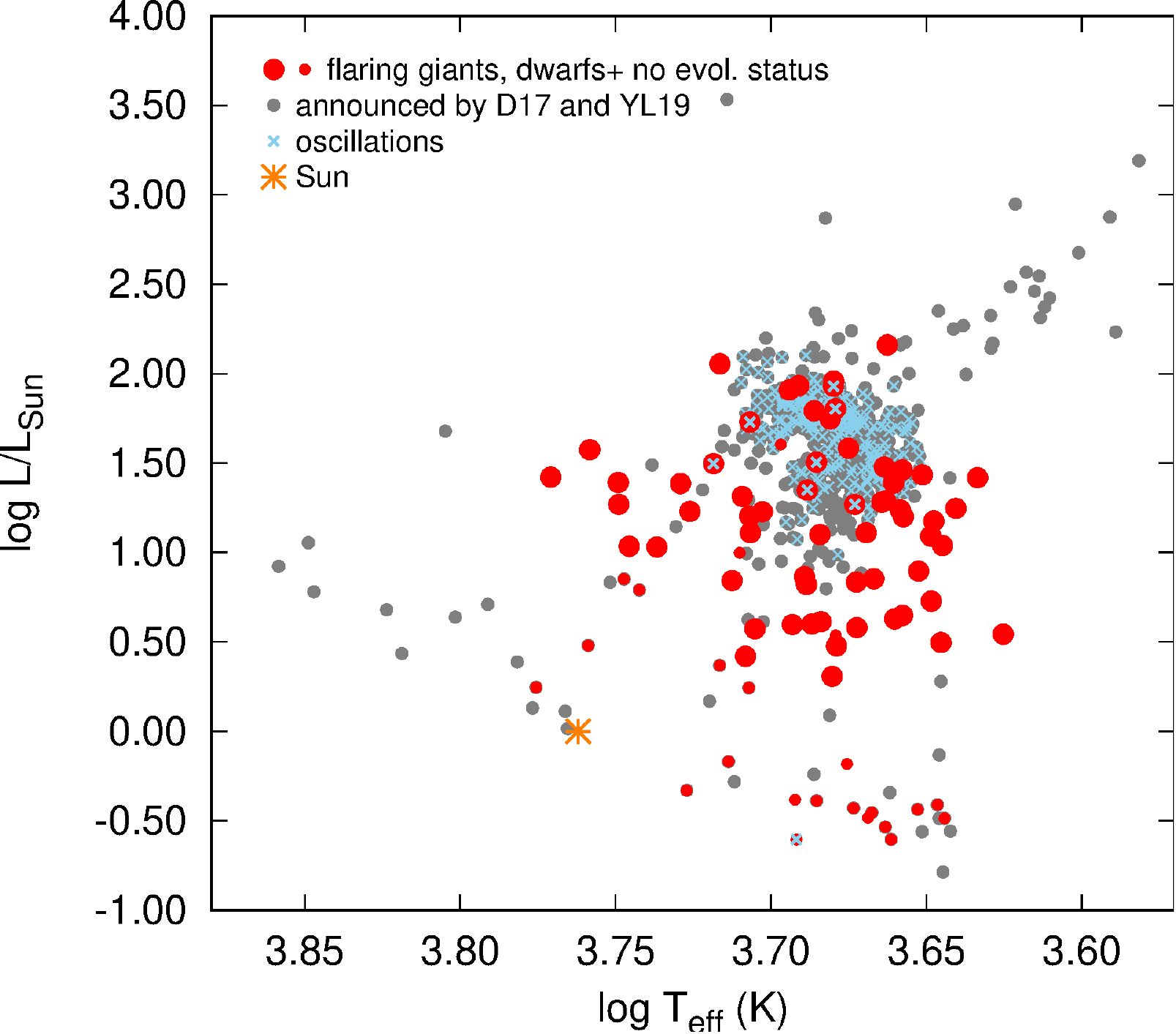} 
      \caption{Locations of the D17+YL19 giant sample in the HRD. Symbols: grey dots: D17+YL19 giant sample, big red dots: flaring giants verified in the present paper, small red dots: contaminating dwarfs and stars with unknown evolutionary status. Light blue crosses mark the oscillating giant stars from \citet{Gaulme_2020} for stars brighter than Kpmag=12.5 which coincide with the stars in our D17+YL19 sample} overlapping them on the figure. Orange star marks the location of the Sun. See the text and Fig.\ref{HRD_enlarge} for more.
         \label{flares_hrd}
   \end{figure}

   \begin{figure}[h!]
   \centering
   \includegraphics[width=8cm]{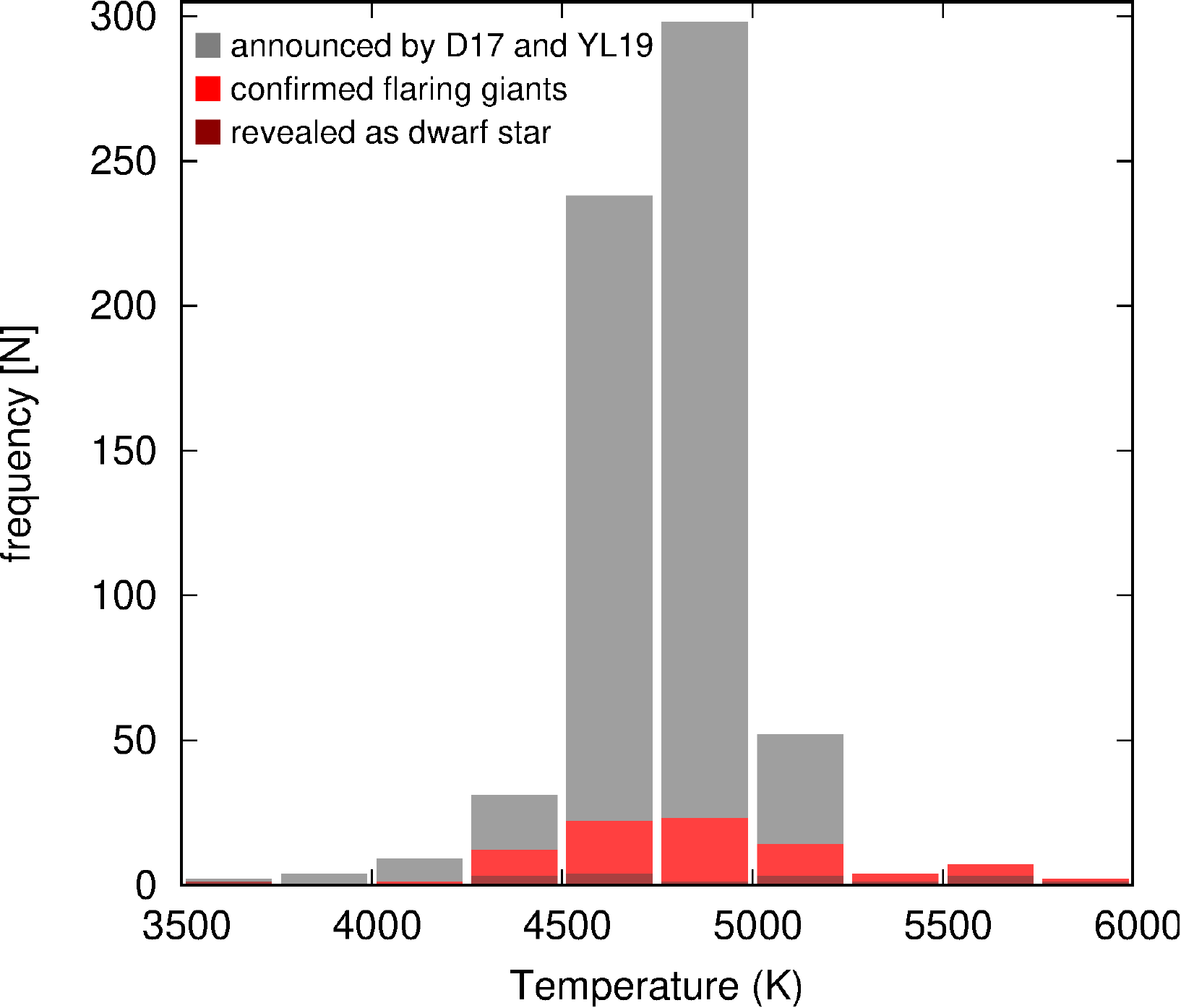}
   \vspace{2mm}
   \includegraphics[width=8cm]{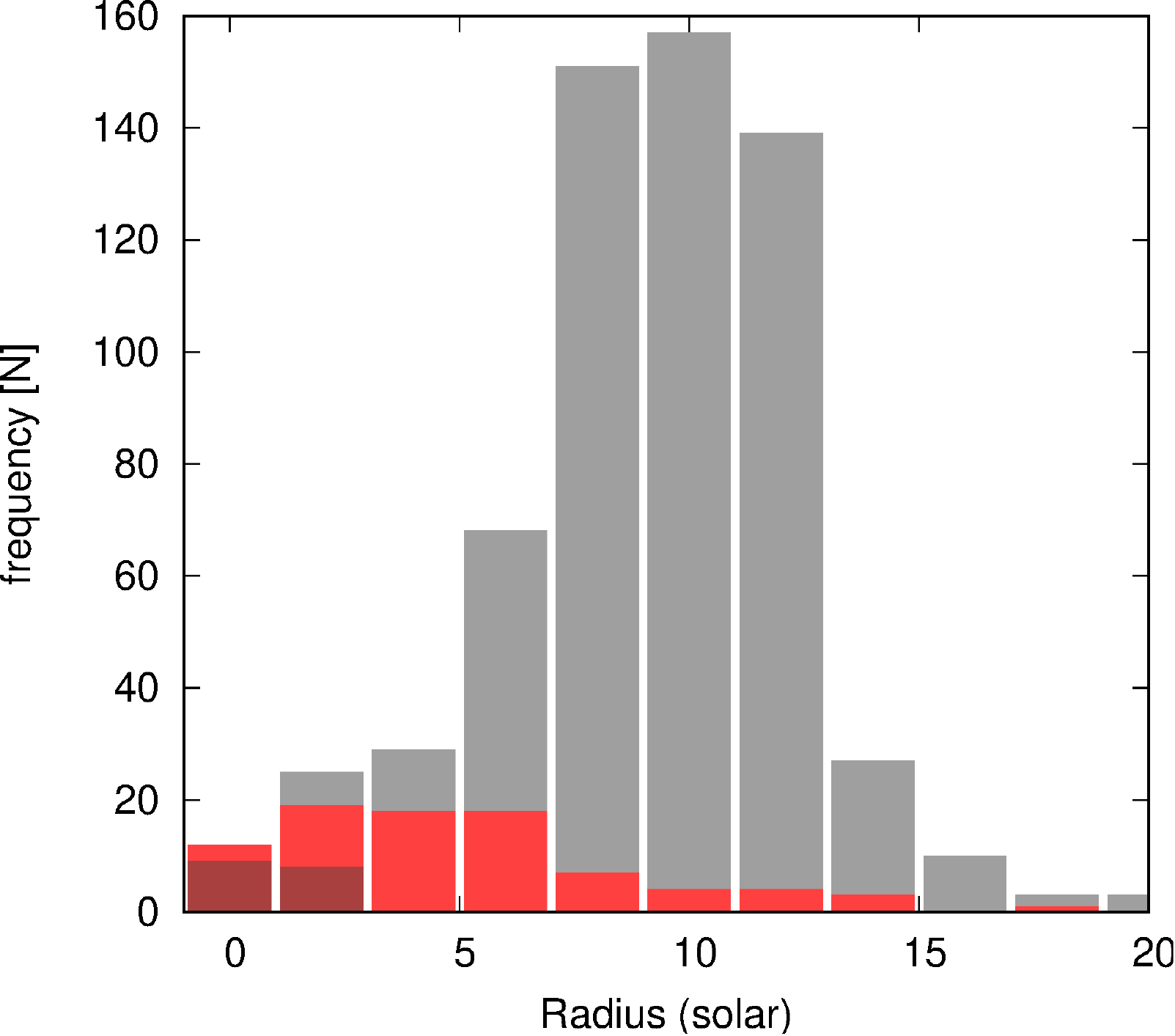}
      \caption{Distribution of the effective temperatures ({\it top}) and radii ({\it bottom}) of the full sample (originally identified as flaring giants by D17 and YL19) and the number of stars we found as flaring giants plotted in grey and red, respectively. Dark red marks the stars classified as dwarfs by TICv8. We note that the histogram of radii does not exceed 20 solar radius (18 stars) for better visibility, since no flaring star is found above this limit.}
         \label{flares_params}
   \end{figure}


   \begin{figure}
   \centering
   \includegraphics[width=\columnwidth]{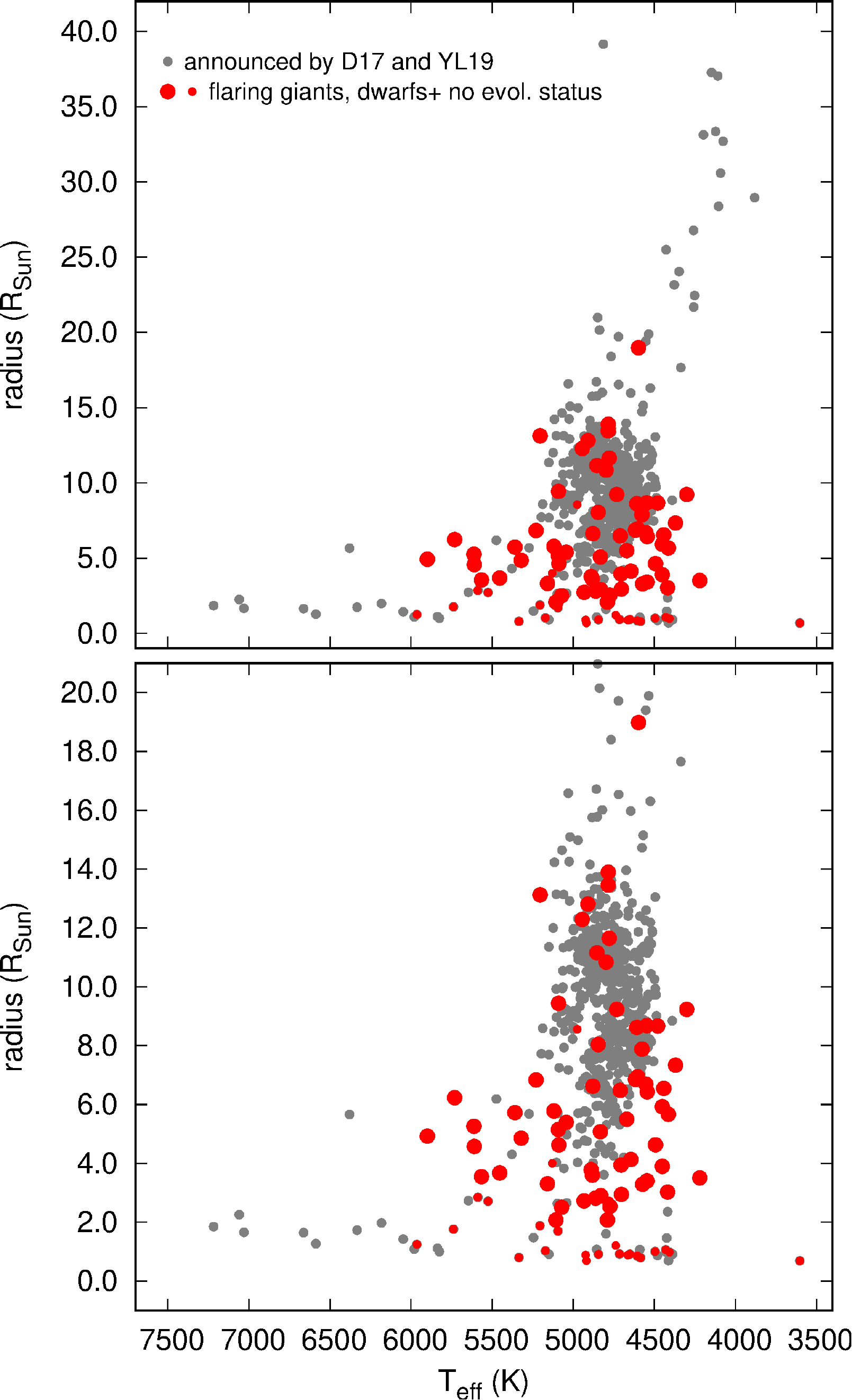} 
      \caption{Radii of the flaring giants. Top: the whole investigated sample, bottom: enlarged region of radii. Grey dots represent the D17+YL19 giant sample, big red dots mark those flaring giants which are verified in the present paper, while small red dots represent the contaminating dwarfs and stars with unknown evolutionary status.} 
         \label{flares_radii}
   \end{figure}

\section{Results}\label{sect_results_disc}

\subsection{The confirmed flaring giant stars}\label{true_flaring_giants}

Among the 706 giant stars that we compiled from the lists of D17 and YL19, we confirm the detection of flares in 86 cases, and do not measure any significant flare signal in the rest of the sample. We also confirm the detection of flares for 53 of the 80 flaring red giants that were found by YL19.

 As already mentioned in the Introduction, according to \citet{2019ApJ...871..241D} the false positive flare detections in \citet{2016ApJ...829...23D} originated from the earlier version of the {\tt appaloosa} code which left sharp or peaked pseudo-structures in the residuals from the periodic signals. Such a failure might happen when using automated search for flares in other, earlier codes as well. When discussing what kind of variability was behind the false positives in D17, YL19 mentioned pulsating stars, but without any further detail (see their Sections 2.2 and 3.2). The origin of the high number of false positive flare detections has finally been unveiled with the help of the recent papers by \citet{2019MNRAS.485.5616H} and \citet{Gaulme_2020}. 
 
 Cross-matching the 666 stars from the D17 flaring giant sample with the results of \cite{2019MNRAS.485.5616H} we found 577 common stars meaning that these could be in fact oscillating giants. These 577 oscillating stars by \cite{2019MNRAS.485.5616H} include those 310 brighter ones (Kpmag $< 12.5$) from \citet{Gaulme_2020} detected independently. Visualizing these results, the HRD of the studied sample and our results are shown in Fig.\ref{flares_hrd}, where the brighter subsample of the oscillating giants derived by \cite{Gaulme_2020} are plotted with light blue crosses. In Fig.\ref{HRD_enlarge} an enlarged version of the dense part of Fig.\ref{flares_hrd} is found including also the oscillating stars from \cite{2019MNRAS.485.5616H} cross-matched with our sample. Taking these evidences into account we can conclude that the vast majority of the false positive flaring giants we found are in fact oscillating stars. Of course during co-adding and cross-matching samples of many stars from different sources may result in a few misclassifications, but we are confident that this number is very small. It turns out that flaring stars show oscillations as well (e.g., the Sun), but among our verified flaring giants we found only a few. We discuss the incidence of oscillations in flaring giants in Sect.\,\ref{osc}.

The stellar parameters from the TICv8 (if not available, then from \emph{Gaia}~DR2 and KIC~DR25), the rotational periods calculated using the Lomb-Scargle algorithm provided by the NASA Exoplanet Archive Kepler Stellar Table, and the number of detected flares are given in Table~\ref{tab:long}. We note that the luminosity classes reported in the TICv8 indicate that 17 of the 86 flaring stars are actually dwarfs, and another 8 stars are not found in the TICv8, so for these no evolutionary status is given. Therefore, the number of flaring giants we confirm is at best 69 -- only confirming a tenth of the originally claimed discoveries of D17. Fig.\ref{flares_hrd} shows a clear main sequence with the position of the Sun marked, drawn by the polluting flaring dwarfs.

There are 17 stars in the combined sample of 706 flaring giant candidates from D17+YL19 which we find uncertain cases to classify as flaring giants for various reasons, like too short or no rotational periods, blends and possible oscillating or pulsating nature of the variability. They are listed Table~\ref{table:uncertain} with explanations; six stars in the table are considered as dwarfs in TICv8.

The distributions of effective temperatures and stellar radii of both the full sample and the confirmed flaring giants are displayed in Fig.~\ref{flares_params}. The two panels show that only 10-12\% of the giants of the D17+YL19 candidates do exhibit flares. The histogram of effective temperatures shows that the distribution of the confirmed flaring stars is very similar to that of the whole sample. To the contrary, the distribution of radii is significantly different: the whole sample peaks at about 10\,$R_\odot$ whereas the distribution of confirmed flaring stars peaks between 3 and 7\,$R_\odot$. This is well in line with the finding of \citet{Gaulme_2020} (their Fig. 9., upper part of the bottom panel), showing that highest number of the radii of active giants on the RGB is around 5-6\,$R_\odot$. The part of the sample with the smaller radii mostly consists of dwarf stars (dark red color in Fig.~\ref{flares_params}), which contaminate the sample according to TICv8. 

Stellar radii vs. temperature is shown in Fig.~\ref{flares_radii}, in the upper panel the whole investigated sample while in the lower the most dense part, containing all the resulted flaring stars, are plotted. The dense group of false positive flaring giant stars (grey dots) with radii between 6-14\,$R_{\odot}$ in the temperature range between 4500-5200\,K are mostly oscillating red giants as it is suggested by our cross-matching results detailed above.

Magnetic fields are measured in 29 single G-K giant stars via the Zeeman-effect by \citet[][Fig. 5]{2015A&A...574A..90A} depicting a well-defined "magnetic strip" on the HRD indicating the maximum convective turnover time ($\tau_\mathrm{c}$) on the tracks and presenting Rossby-numbers (ratio of the rotational period and the convective turnover time, $R_o= P_\mathrm{rot}/\tau_c$) of the studied stars. The origin of these fields is thought to be working $\alpha-\Omega$ dynamos, whose existence is predicted by \citet[][their Fig. 6., upper right panel]{2017A&A...605A.102C} showing that low Rossby-numbers, which is essential for the generation of the magnetic field, are characteristic on the segments of the theoretical evolutionary tracks where the giant stars with the measured magnetic fields are found \citep{2015A&A...574A..90A}. The location of the "magnetic strip" drawn both by observations and theory agrees well with the position of the flaring giant stars shown in Fig.~\ref{flares_hrd}, confirmed in this paper.

\subsection{Activity and rotation}\label{act_rot}

\begin{figure}
   \centering
   \includegraphics[width=\columnwidth]{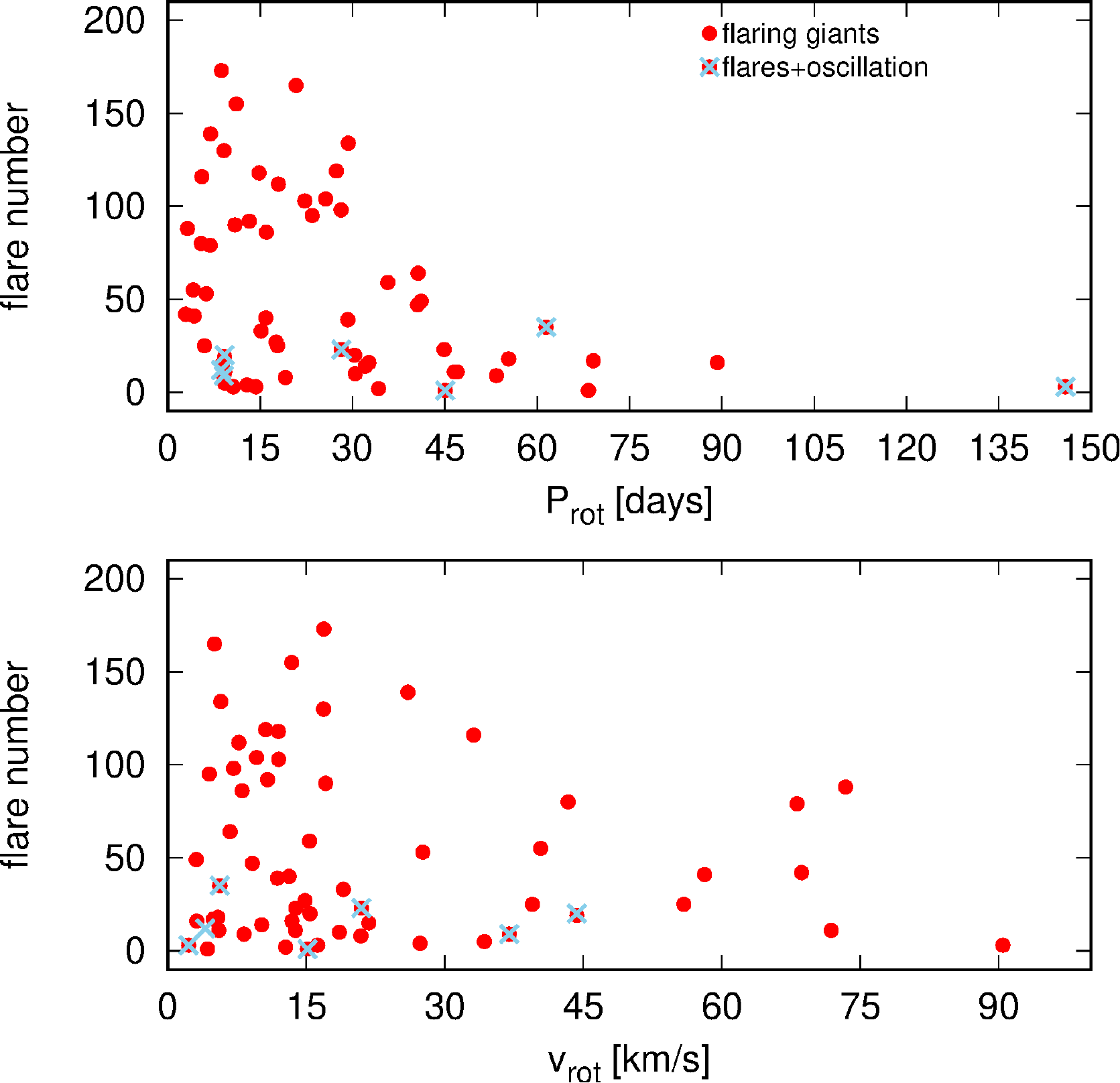} 
      \caption{Number of flares at different rotational periods (top) and rotational velocities (bottom). Light blue crosses mark those flaring red giants which also show oscillations (see Sect.\,\ref{osc}).}
         \label{flares_numbers}
   \end{figure}

We plotted the number of flares as a function of the rotational periods and rotational velocities in Fig.~\ref{flares_numbers}. Looking at the upper panel of the figure it is seen that flaring giant stars with rotational periods from the shortest ones to about 30-40 days can produce both higher and lower numbers of flares. There are 15 stars on Fig.~\ref{flares_numbers} with rotational periods longer than 35 days, with flare numbers below 65, all of them are giants, and constitute about one quarter of the confirmed flaring giant population (minimum 61 stars). From the lower panel of Fig.~\ref{flares_numbers} no real disctinction of flare numbers between stars with lower and higher rotational velocities is seen, though a slight decrease of showing less flares with higher rotational velocity is maybe present. However, due to the low number of stars with higher rotational velocities, this change is not significant. The investigation of the connection (if any) between the rotational properties and flare numbers of giant stars needs detailed knowledge of the physical parameters of these giants which extends the scope of the present paper. 

\subsection{Binarity and oscillations among the flaring red giants}

\subsubsection{Binarity}\label{bin}

\begin{figure}[t!]
\includegraphics[width=\columnwidth]{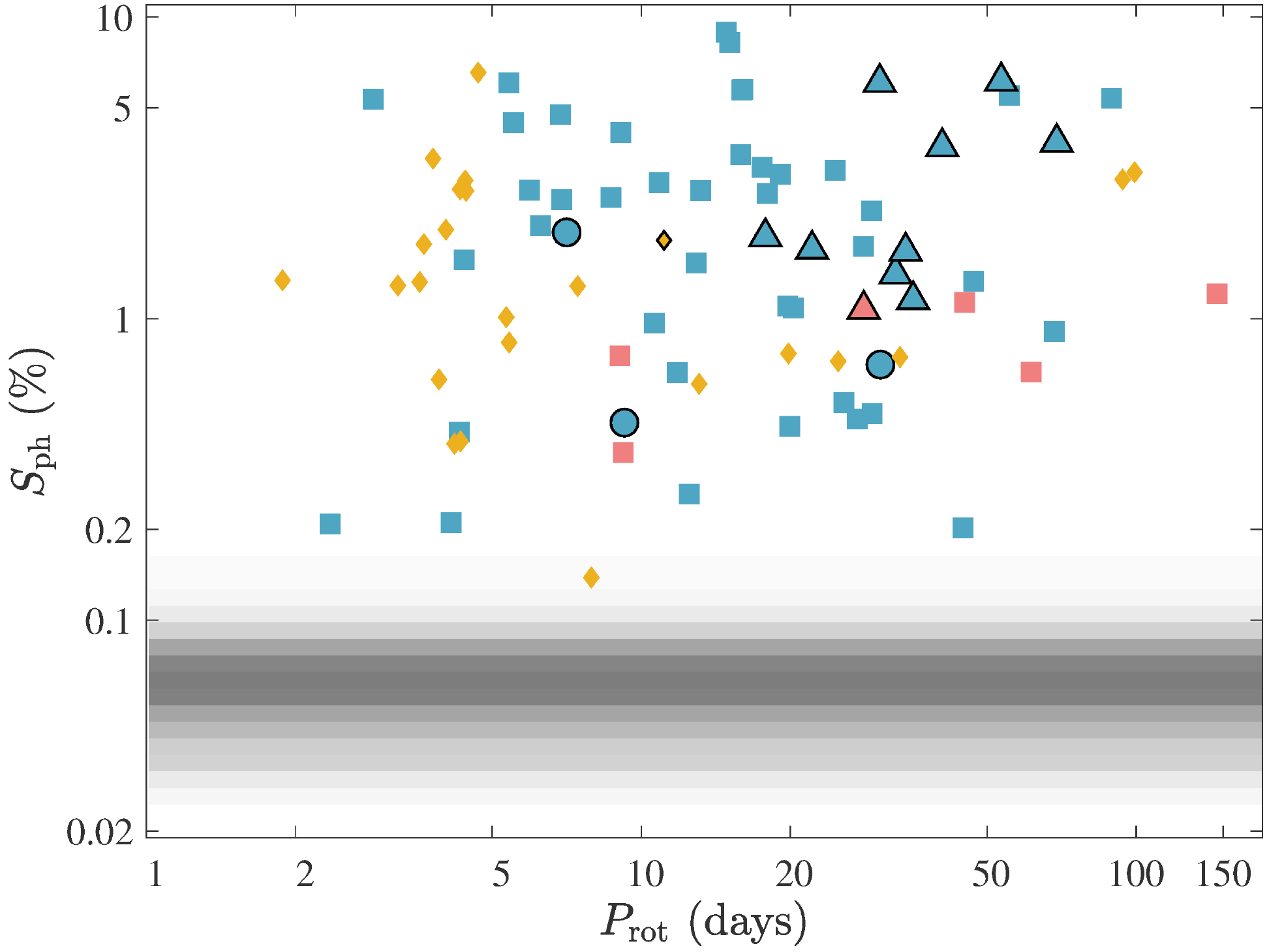} 
\caption{Photometric index $S\ind{ph}$ (percent) as a function of rotation period $P\ind{rot}$ (days) for the 86 flaring stars from Table~\ref{tab:long}. Both axes are in logarithmic scale. Blue and red symbols indicate the non-oscillating and oscillating giant stars, respectively. Orange markers indistinctly indicate dwarf and subgiant stars. The upward-pointing triangles are confirmed close spectroscopic binaries including at least one red giant, and circles RV-stable red giant stars \citep{Gaulme_2020}. The orange diamond with black edges is a confirmed subgiant belonging to a close binary system (KIC-11551404, \citealt{Gaulme_2020}). 
The gray background reflects the distribution of $S\ind{ph}$ of the regular inactive red giant stars (the darker, the higher) according to \citet{Gaulme_2020}. }
\label{fig_Sph_Prot}
\end{figure}

\citet{Gaulme_2020} searched for the incidence of binary systems among the red giants that show photometric rotational modulation. They obtained multi-epoch spectroscopic observations for a subsample of 85 stars.  Among these 85 stars, 14 belong to the 61 flaring giants studied in the present paper (see  Table~\ref{tab:long}). Of these 14 stars, one is a subgiant (or a small red giant) spectroscopic binary (KIC 11551404), nine are clear red giants in spectroscopic binaries with no oscillations, one is an oscillating (weak oscillations) red giant in a spectroscopic trinary (KIC 8515227), one is a single fast rotating red giant (KIC 11087027), and two are fast rotators (KICs 5821762 and 8749284) for which the spectroscopic signal is too noisy to assess any radial-velocity variability (Fig. \ref{fig_Sph_Prot} and Table \ref{tab:long}). Even though only a complete spectroscopic survey would be needed, this information confirms that close binary systems are likely to compose a large fraction of the sample: among the 61 giants, 11 SBs represent already 18\,\%.

We also computed the photometric index $S\ind{ph}$ (percent) for each target. This parameter is the mean standard deviation of the time series computed over five rotational periods \citep{2014JSWSC...4A..15M}. It is commonly used as a proxy of magnetic activity in photometric time series: the larger $S\ind{ph}$, the larger the surface magnetic field.
Fig.~\ref{fig_Sph_Prot} shows this parameter of all 86 stars from Table~\ref{tab:long} as a function of the rotational period. No particular trend is visible but it is worth noticing that its value is always higher than 0.2\% for all the flaring giants, whereas regular inactive red giants have $S\ind{ph} < 0.1$\% \citep{Gaulme_2020}. In addition, a large fraction of the targets display values of $S\ind{ph}$ larger than 1\%. According to the spectroscopic study of 85 red giants with activity in \citet{Gaulme_2020}, red giants with $0.1\leq S\ind{ph} \leq 1$\% are fast single rotators, whereas the vast majority of red giants with $S\ind{ph} \geq 1$\% belong to close binary systems (see their Fig. 12). We can therefore rightly suspect that a large fraction of our flaring giants do belong to close binary systems. 

\subsubsection{Oscillations}\label{osc}

Red giants are known to display oscillations in the vast majority of cases \citep[$\geq$99\%][]{Gaulme_2020}. The only situation where oscillations are not detectable corresponds to red giants in close binary systems: more than 85\% of the red giants with no oscillations are in close binaries \citep{Gaulme_2020}. The presence of oscillations is thus a parameter that is complementary to large values of $S\ind{ph}$ to indicate the presence of close binary systems.
We therefore looked for oscillations in all the 86 targets of the sample. We note that we could not search for oscillations of main-sequence or subgiant stars of the sample because the sampling of the Kepler light curves (30-minutes cadence) is not fast enough to access stars with frequency at maximum amplitude larger than about 300\,$\mu$Hz, that is, not smaller than red giants.

Despite 20 targets were already listed in \citet{Gaulme_2020} and 24 in \citet{2019MNRAS.485.5616H} (the same plus four), we thoroughly restudied them one by one. Indeed, light curves with flares are tedious to deal with as a single uncorrected flare event is able to ruin the power spectrum of the light curve, in which we search for oscillations. \citet{2019MNRAS.485.5616H} report the detection of oscillations for the 24 stars that are common to our list. To the contrary, \citet{Gaulme_2020} reports the detection of oscillations in only four red giants of the same list. We reprocessed one by one the light curves thanks to the flare date and duration produced by the flare automatic pipeline. We systematically cut the flares and bridged them with a smooth of the light curves during the gaps. We also manually removed a few remaining flares that were not picked by the algorithms. Except for the removal of the flares, we processed the data in the same way as performed in \citet{Gaulme_2020} and we refer to this paper for more details.   

Of the 86 flaring stars, we detect red-giant oscillations in seven of them (KIC~4680688, 5286780, 5808398, 6445442, 7363468, 8515227, and 8951096) and we rule out all of the others. The discrepancy with the results of \citet{2019MNRAS.485.5616H} are explainable by the fact that they make use of an automatic pipeline based on machine learning that deals with almost all of the Kepler red giants observed by Kepler. Individual errors can happen here and there, and it is not surprising that it did so on light curves that have large flares, which are a terrible source of systematic noise in the Fourier domain. The oscillating stars are indicated in Fig. \ref{fig_Sph_Prot} in the form of red symbols. We also detect a $\gamma$\,Doradus--$\delta$\,Scuti type oscillator in the sample.

It has been observed that red giants in close eclipsing binary (EB) systems have peculiar photometric properties \citep{2014ApJ...785....5G}. Among the 35 \emph{Kepler} red giants that are confirmed to belong to EBs, 18 display regular solar-like oscillations -- with their amplitudes matching the empirical expectations \citep[e.g.,][]{Kallinger_2014} -- and no rotational modulation. All of the remaining systems display rotational modulation, seven with partially suppressed oscillations, and ten with no detectable oscillations \citep[][Benbakoura et al. submitted]{2014ApJ...785....5G, 2016ApJ...832..121G}. \citet{2016ApJ...832..121G} showed that the non-detection of oscillations is not an observational bias. This is observed in the closest systems, where most orbits are circularized, and rotation periods are either synchronized or in a spin-orbit resonance. Such a configuration is observed for systems whose orbital periods are shorter than about $P\ind{orb}\sim150$ days.
\citet{2014ApJ...785....5G} suggested that the surface activity and its concomitant mode suppression originate from tidal interactions. During the red giant branch, close binary systems reach a tidal equilibrium where stars are synchronized and orbits circularized \citep[e.g.,][]{Verbunt_Phinney_1995, Beck_2018}. Red giant stars are spun up during synchronization, which leads to the development of  a dynamo mechanism inside the convective envelope, leading to surface spots. The magnetic field in the envelope likely reduces the turbulent excitation of pressure waves by partially inhibiting convection. Since spots can also absorb acoustic energy, these two effects lead to the suppression of oscillations.


\begin{table*}
\small
\caption{Flaring stars from the 86 of Table~\ref{tab:long} with oscillations derived by \citet{Gaulme_2020} (G+20) and in the present paper (pp).}             
\label{table:oscill}      
\centering          
\begin{tabular}{c r r | c|  c | c | l  }     
\hline\hline 
\multicolumn{1}{c}{KIC No.}  &  \multicolumn{1}{c}{Kpmag}  &   \multicolumn{1}{c}{$P_{\rm rot}$}  &   \multicolumn{1}{c}{lum. class (TICv8)} & \multicolumn{1}{c}{source} &   \multicolumn{1}{c}{flare no.} & \multicolumn{1}{c}{remarks} \\
\hline
4680688   &   13.476  &     9.190   &   GIANT  &   pp   &  20 &  single? \\
5080290   &   9.507   &     ---      &   GIANT  &  ---    &  6 &  $\gamma$~Dor~/~$\delta$~Sct hybrid  \\
5286780   &   13.963  &     8.625   &   ---    &   G+20     &  12 & \\
5808398   &   11.455  &     145.81   &   GIANT  &   G+20   &  3 & \\     
6445442   &   13.194  &     45.032   &   GIANT  &   G+20   &  1 & \\ 
7363468   &   12.386  &     61.445   &   GIANT  &   G+20   &  41 &\\ 
8515227   &   11.176  &     28.157   &   GIANT  &   G+20   &  23 &  triple-lined sp. binary \\
8951096   &   13.336  &     9.074    &   GIANT  &   G+20   &  9 & \\
\hline
\end{tabular}
\end{table*}


   \begin{figure}[htbp]
   \centering
   \includegraphics[width=\columnwidth]{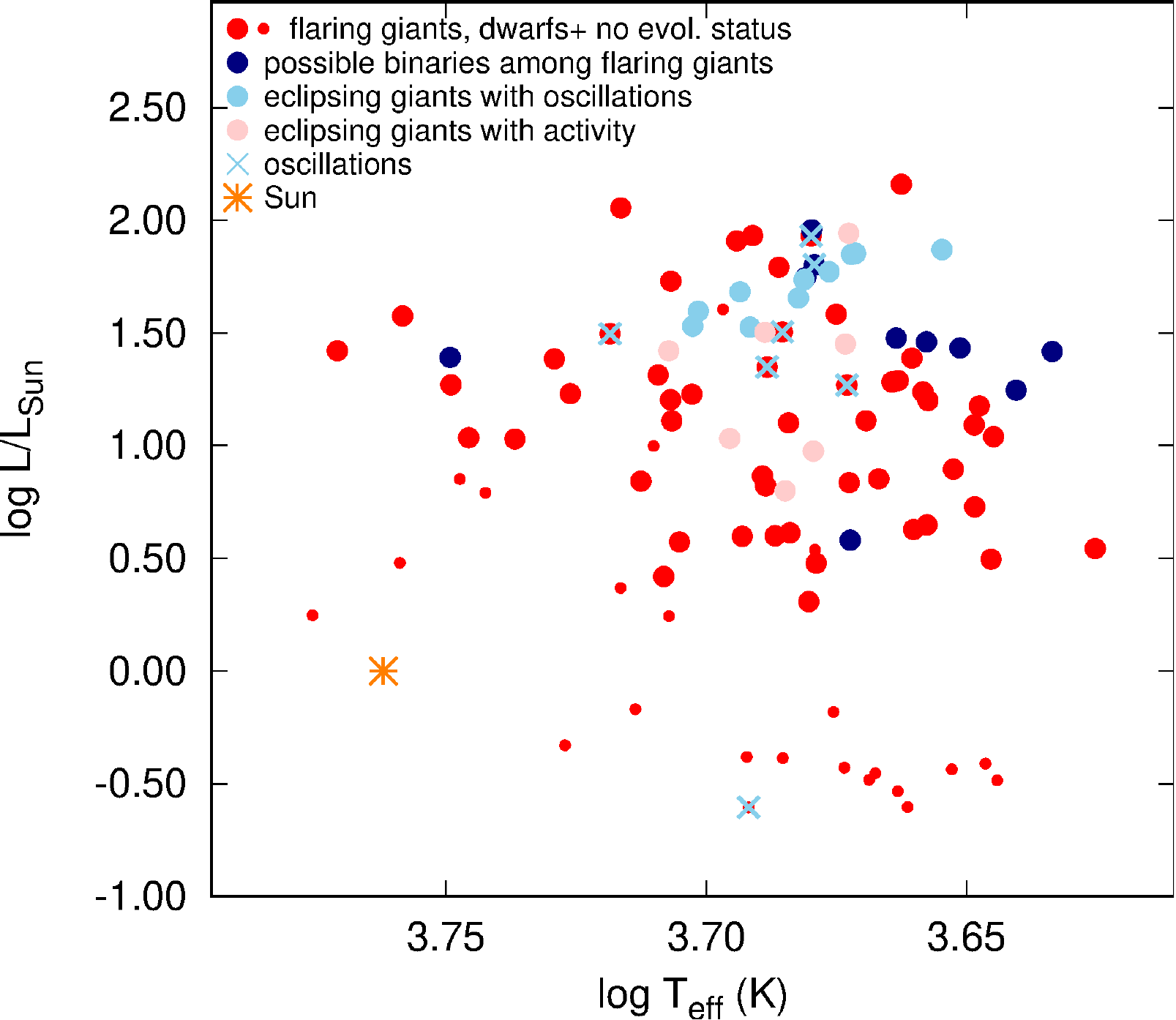} 
      \caption{Locations of the flaring giants (present paper, red dots, small symbols mean dwarfs and with unknown evolutionary status), possible binaries from Table~\ref{tab:long} by \citet{Gaulme_2020} (dark blue dots) and eclipsing giants with oscillations (light blue dots) or activity (pink dots) from  \citet{2016ApJ...832..121G} and Benbakoura et al. (submitted] in the Hertzsprung$-$Russell-diagram. Light blue cross over red dots mark the flaring {\it and} oscillating stars from the present paper. Orange star marks the location of the Sun.}
         \label{hrd_binaries}
   \end{figure}

Summary of the results on binarity and oscillations of the flaring red giants are plotted in
Fig.~\ref{hrd_binaries}. The list of six giant stars plus one of unknown evolutionary status that are both flaring and show oscillations, and one $\gamma$\,Dor~/~$\delta$\,Sct type star is given in Table~\ref{table:oscill}. The star without given evolutionary status (KIC~5286781) seems to be a dwarf star concerning its position in the HRD, see Fig.~\ref{hrd_binaries}. The flaring rate of the six giant stars with oscillations in Table~\ref{table:oscill} (16 flares on average) is much lower than the average flare number of the 61 giants (52 flares on average), which is well seen in Fig.~\ref{flares_numbers} where the oscillating red giants are marked with light blue crosses. This fact is in line with the result of \citet{2016ApJ...832..121G} showing that the presence of oscillations suggests weaker activity. At last, we added the four non-oscillating red giants (KIC~3955867, 4569590, 7943602, 9291629) that belong to eclipsing binary systems studied by \citet{2016ApJ...832..121G}, which are observed to display flares too. Their parameters are known from dynamical analysis, that is, a combination of high-resolution spectroscopy and eclipse photometry. 

\subsubsection{The case of KIC~5080290}

We list KIC~5080290 in Table ~\ref{table:oscill} as $\gamma$~Dor~/~$\delta$~Sct hybrid without derived rotational period. This star was identified as a $\delta$~Sct star earliest by \citet{2011A&A...534A.125U} listing four independent $\delta$~Sct-type frequencies. However, this star had at least six unambiguous flares concerning our cross-matched results with logarithmic energies between $35.7\leq\log E_{Kp}\leq37.2$\,[erg]. KIC~5080290 is a bright star, it is 9.51~mag. in G-band, the neighbour stars, not even close by, are several magnitudes fainter. 

Just recently however, \citet{2020arXiv200904784Z} showed that $\beta$\,Cas, an F2III $\delta$\,Sct type pulsator with only three p-mode frequencies of low amplitude, has a complex magnetic field very possibly of dynamo origin. $\beta$\,Cas is a terminal-age main-sequence (TAMS) star with $\log g = 3.53$$\pm$0.58 and $T_{\rm eff} = 6920$$\pm$140\,K, rotating very fast with $P_{\rm rot}=0.898$\,d which probably helps to maintain the dynamo even at such high temperature \citep{2020arXiv200904784Z}. Thus, KIC~5080290 could be another example of a pulsating star with magnetic activity.

The mass of $\beta$\,Cas is about 2.1\,$M_\odot$ (or 1.91$\pm$0.02 as given in \citet{2011ApJ...732...68C}) and is already moving away from the TAMS. Its age is 1.18$\pm$0.05\,Gyr derived by \citet{2011ApJ...732...68C}. We may suppose that such a star as $\beta$\,Cas could be a progenitor of flaring giant stars investigated in this paper. Specifically, KIC~2852961, having the highest energy flares among the flaring giants confirmed in this paper, has a mass of 1.7$\pm$0.3\,$M_\odot$ and its age is about 1.7\,Gyr. As discussed in Paper~I it is imaginable that the progenitor of KIC~2852961 was an A5-F0 type star.

\section{Discussion of flares}\label{sect_results_flares}

\subsection{Reliability of the derived flare numbers and their energies}\label{flares_number_energy}

For testing the performance of the cross-matched results of flare recoveries from the D17+YL19 sample by the pipeline from \citet{2020AJ....159...60G} and FLATW’RM of \citet{2018A&A...616A.163V}, we compared the flare list of KIC~2852961 from the present paper with the visually found and checked flares published in Paper~I, Table 3. From the 59 flares in Paper~I our automated detection found 51. From the missing eight flares seven belong to the low energy regime $\log E_{Kp} < 36.6$\,[erg] (equivalent with $\varepsilon_f\approx$1-min relative energy) or less, where the probability of detecting a flare starts to diverge noticeably from 100\%, according to the test results presented in Paper~I (see Sect.~7 in that paper). One non-detection belongs to a complex flare. However, the total number of automatically detected flares of KIC~2852961 in this paper is, by coincidence, also 59, due to detecting complex events with multiple peaks as series of single events by the pipelines. This comparison shows that we successfully avoid false positives on the cost of missing some real, mostly small energy flares, and the overall picture of the detected flare numbers and energies are not misleading.

We compared the derived flare energies of 51 flares of KIC~2852961 which are available both from Paper~I and from the present paper.  From these flares 24 are complex and 27 are simple events. Since we use a simple flare template for automatically derive the flare energies by  FLATW’RM  \citep{2018A&A...616A.163V} we also checked how this simplification affect the resulting flare energies.
The simple mean of the difference between the two approaches (in the sense of energies from the present paper minus from Paper~I) is $(-0.326\pm1.481)\times10^{37}$ showing that there is a small shift between the results due to the different $L_{\star\rm Kep}$ used in the two sources. Separately from the complex and simple flares the results are  $(-0.293\pm1.941)\times10^{37}$ and  $(-0.355\pm0.940)\times10^{37}$, respectively, which is only a small difference. The higher standard deviation of the mean of the complex flares points towards a slightly higher average scatter of the resulting flare energies when using the simple template for calculating the energies of complex events, but this does not modify the FFDs noticeably as it is shown in Fig.~\ref{test:ffd} later, see Sect.~\ref{completeness}.

\subsection{Distribution of flare energies}\label{energy_distribution}

\begin{figure}
   \centering
   \includegraphics[width=\columnwidth]{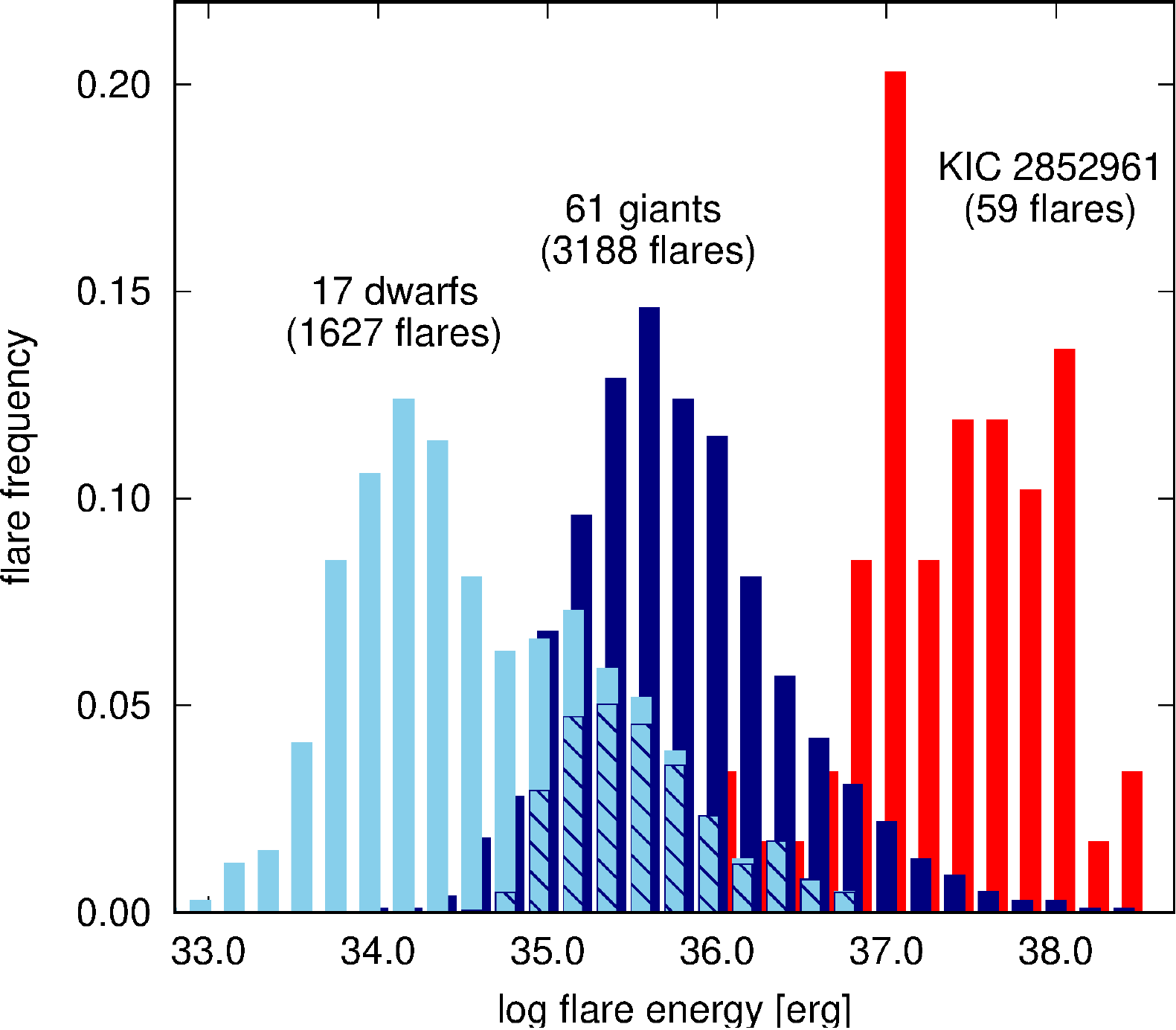}
      \caption{Comparison of the distribution of flare energies in the \emph{Kepler} bandpass observed on 17 dwarfs (light blue), 61 giants (dark blue) and KIC~2852961 (red). A part of the dwarf star's light blue histogram striped with dark blue lines are occupied by the flares of two, possibly subgiant stars. See the text for more.}
         \label{histo_all}
   \end{figure}

In Fig.~\ref{histo_all} flare energy distributions of the 17 dwarfs (light blue), 61 giants (dark blue) and KIC~2852961 (red) from Table~\ref{tab:long} are compared. Missing are those eight stars for which no evolutionary status is given. The difference in the mean flare energies between dwarfs and giants are clear. The extended tail of the flare energy distributions of giants is mostly made up by the excess high energy flares of KIC~2852961.

On the histogram of the dwarfs (light blue) a small secondary maximum is found towards higher energy. This hump is mostly due to two stars listed as dwarfs in TICv8, KIC~9116222 (260 flares) and KIC~11515713 (186 flares) which have approximately one order higher energy flares than rest of the dwarfs. The common energy distribution of these two stars are marked with slanted blue lines in Fig.~\ref{histo_all}. These dwarfs could be G-type superflaring stars, both of them have effective temperatures about 5500\,K, see also \citet[][their Fig.~8]{2013ApJS..209....5S}. On the other hand, radii are 2.7 and 2.8\,$R_\odot$ (Table~\ref{tab:long}) and $\log g$ values from LAMOST\footnote{http://dr5.lamost.org/} are 3.69 and 3.88 for KIC~9116222 and KIC~11515713, respectively, suggesting slightly evolved, subgiant evolutionary status of these stars. (In the LAMOST catalog 45 stars out of the 86 from Table~\ref{tab:long} have spectra, the LAMOST parameters ($T_{\rm eff}$, $\log g$ and [Fe/H]) of these stars are given in Table~\ref{table:LAMOST} as supplementary information.) 

   \begin{figure}[]
   \centering
   \includegraphics[width=\columnwidth]{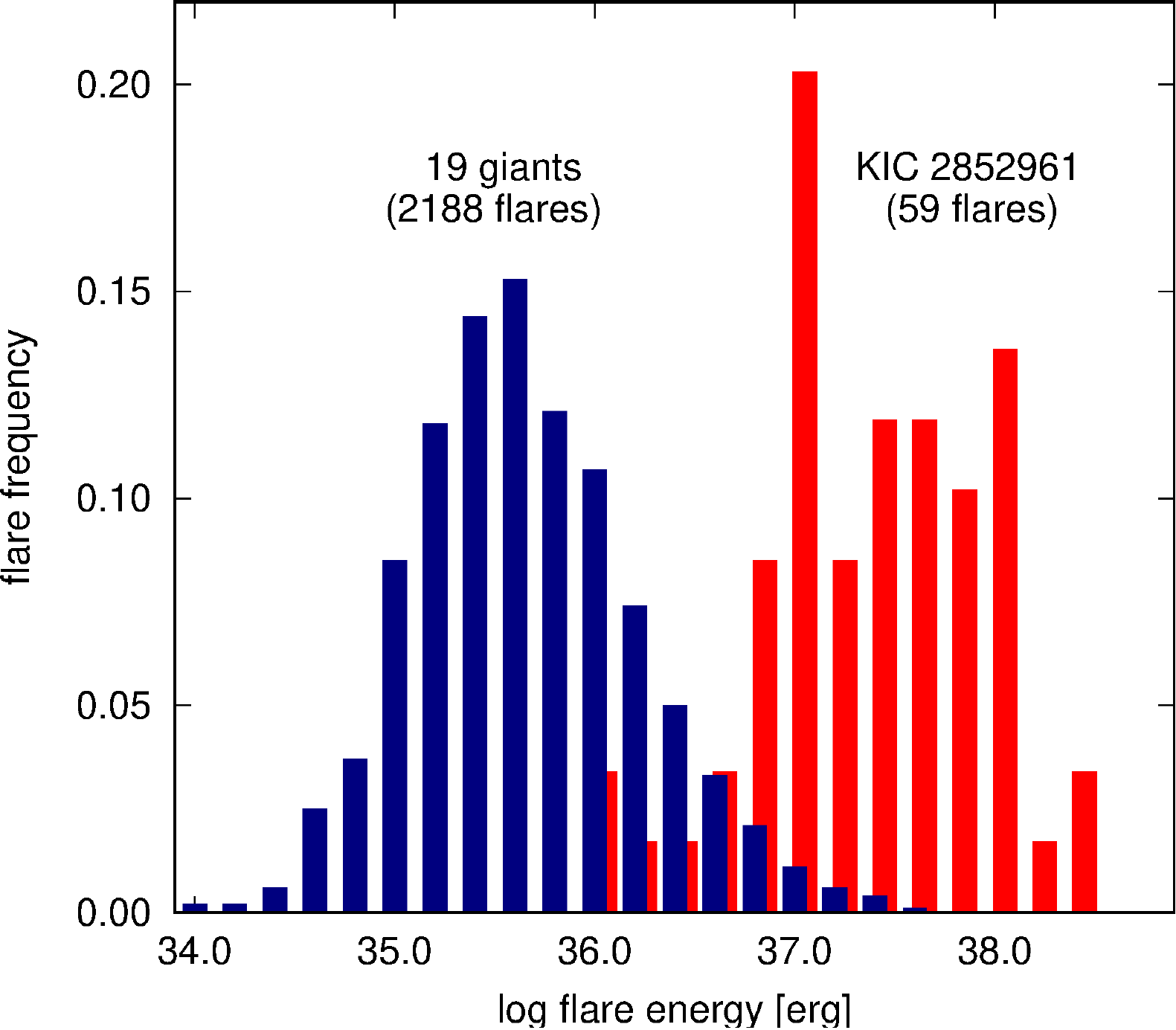}
      \caption{Comparing the combined distribution of flare energies in the \emph{Kepler} bandpass of the most flaring 19 giants (dark blue columns) and that of KIC~2852961 (red columns), see also Fig.~\ref{histo:19stars}.}
         \label{histo:19stars_combined}
   \end{figure}

From the final results of the 86 stars in Table~\ref{tab:long} we selected 19 giant stars with the highest number of observed flares in order to compare their flare energy distributions. The basic stellar parameters ($T_{\rm eff}$, $R_{\odot}$, $P_{\rm rot}$), quiescent stellar luminosity values in the \emph{Kepler} bandpass, average flare energies, equivalent times (=relative flare energies), histogram skews, and remarks if the star is a member of a binary system, are given in Table~\ref{table:flares}.  Comparing these stars with that of KIC~2852961 (Table~\ref{table:flares}, bottom line) we find that the latter has on average much higher energy flares than the selected 19 flaring giants. The flare energy distribution of KIC~2852961 reveals mostly high energy flares (Paper~I, Fig. 15) opposite to the ones observed for the selected 19 stars which all show declining number of flares towards higher energies. The low energy parts of the histograms are affected by the recovery bias.

The individual flare energy distribution histograms of the 19 stars are plotted in Appendix~\ref{C}, Fig.\ref{histo:19stars} together with that of KIC~2852961. Most of the histograms of Fig.\ref{histo:19stars} are approximately symmetric or moderately positively skewed (skewness is less than 0.5, or between 0.5-1.0, respectively), except two cases with highly positive skew. Only the histogram for KIC~2852961 shows a small negative skew meaning that the mean flare energy is smaller than the peak of the distribution, i.e., the high energy flares dominate. The comparison of the distribution between 2188 individual flare energies of 19 stars together, and that of KIC~2852961 in Fig.~\ref{histo:19stars_combined} depicts the striking difference in the energy distributions in one figure.

From Table~\ref{table:flares} it is seen that the flare characteristics of the two confirmed spectroscopic binaries (both SB2) are not different from the rest of the stars, therefore the binarity itself cannot be the ultimate cause of the difference between the flaring nature of the 19 stars and KIC~2852961.
It has already been shown by \citet{1990ApJS...72..191S, 1994A&A...281..855S} from observations of chromospheric activity at Ca{\sc II}\,H\&K and H$\alpha$ that, to the contrary of MS active stars in binaries, in case of evolved stars binarity does not play a crucial role in the activity level.

\begin{table*}
\small
\caption{Basic stellar parameters, average flare energies ($E_f$, in the \emph{Kepler} bandpass) and equivalent times ($t_{\rm eq}$) of the 19 most flaring stars, and characteristics of the flare distributions.}             
\label{table:flares}      
\centering          
\begin{tabular}{r r r r r r r r r r r}     
\hline\hline 
\multicolumn{1}{r}{KIC No.} & \multicolumn{1}{c}{$T_{\rm eff}$}  & \multicolumn{1}{c}{$R/R_{\odot}$} & \multicolumn{1}{c}{$P_{\rm rot}$} &  \multicolumn{1}{c}{number} & \multicolumn{1}{r}{$L_{\star\rm Kep}$} & \multicolumn{1}{r}{mean $E_f$}  & \multicolumn{1}{c}{mean $t_{\rm eq}$}  & \multicolumn{1}{c}{histogram}  &  \multicolumn{1}{c}{remark\tablefootmark{a}} \\
& \multicolumn{1}{c}{[K]}  &   &  \multicolumn{1}{c}{[d]} &  \multicolumn{1}{c}{of flares} & \multicolumn{1}{c}{[$10^{33}$\,erg/s]} & \multicolumn{1}{c}{[$10^{35}$\,erg]} & \multicolumn{1}{c}{ [s]} & \multicolumn{1}{c}{skew} & \\ 
\hline 
1872340  &  5454 &  3.674 &  10.884  &   90  &  11.478  &   6.257  &   54.5  &  1.12   & \\              
2968811  &  4220 &  3.504 &  14.814  &  118  &   2.966  &   4.988  &  168.2  &  0.76   & \\              
4068539  &  4862 &  2.818 &  13.226  &   92  &   3.967  &   3.280  &   82.7  &  0.73   & \\              
4157933  &  5567 &  3.547 &   6.914  &  139  &  11.703  &   9.777  &   83.5  &  0.66   & \\              
4273689  &  5159 &  3.309 &  29.301  &  134  &   7.242  &   3.417  &   47.2  &  0.82   & \\              
5181824  &  4732 &  9.231 &   6.861  &   79  &  37.279  &  24.350  &   65.3  &  0.08   & \\              
5480528  &  4882 &  3.603 &   5.506  &  116  &   6.616  &   3.620  &   54.7  &  0.89   & \\              
6707805  &  5613 &  5.261 &  22.216  &  103  &  26.680  &  38.036  &  142.6  &  0.20   & SB2  \\         
7676676  &  4774 &  2.539 &  15.979  &   86  &   2.946  &   4.687  &  159.1  &  0.36   & \\              
7740188  &  4419 &  3.028 &   9.094  &  130  &   2.831  &   3.274  &  115.7  &  0.35   & \\              
7838958  &  5360 &  5.727 &  27.369  &  119  &  25.813  &   8.202  &   31.8  &  1.01   & \\              
9237305  &  5107 &  2.074 &  23.458  &   95  &   2.715  &   1.003  &   37.0  &  0.09   & \\              
9419002  &  5322 &  4.857 &  25.640  &  104  &  17.982  &   6.477  &   36.0  &  0.04   & \\              
10603977 &  4933 &  2.728 &  17.962  &  112  &   3.987  &   2.573  &   64.5  &  0.89   & \\              
10646009 &  4493 &  4.634 &   5.412  &   80  &   7.228  &   9.840  &  136.1  &  0.71   & \\              
10666510 &  4790 &  2.074 &  20.845  &  165  &   1.998  &   0.882  &   44.2  &  0.53   & \\              
11135986 &  4830 &  2.899 &   8.692  &  173  &   4.067  &   2.803  &   68.9  &  0.55   & \\              
11551404 &  4703 &  2.944 &  11.120  &  155  &   3.678  &   3.077  &   83.7  &  0.22   & SB2 \\          
11970692 &  4705 &  3.944 &  28.116  &   98  &   6.615  &  10.629  &  160.7  &  0.41   &  \\ 
\hline 
\rule{0pt}{4ex}  2852961  &  4797 & 10.836 &  35.715  &   59  &  54.940  & 215.427  &  392.1  & -0.27  & SB1 \\  
\hline
\end{tabular}
\tablefoot{
\tablefoottext{a}{SB1=single-lined spectroscopic binary; SB2=double-lined spectroscopic binary.}
}
\end{table*}

\subsection{Flare recovery completeness}\label{completeness}

We made flare recovery tests for the 19 most flaring giant stars and for KIC~2852961 to compare the results of the present paper with that of Paper~I, using the code \texttt{allesfitter} \citep{allesfitter-code,allesfitter-paper}. We note that the tests were done and used only for long-cadence, 30-min time resolution data of \emph{Kepler}  with the method described in detail in Paper~I, Sect.~7. The photometric precision of the 20 stars with their Kp magnitudes between 10.2-14.0 are similar\footnote{https://keplergo.arc.nasa.gov/CalibrationSN.shtml}, so the effect of S/N has no, or very minor, influence of the test results. 

   \begin{figure}
   \centering
   \includegraphics[width=\columnwidth]{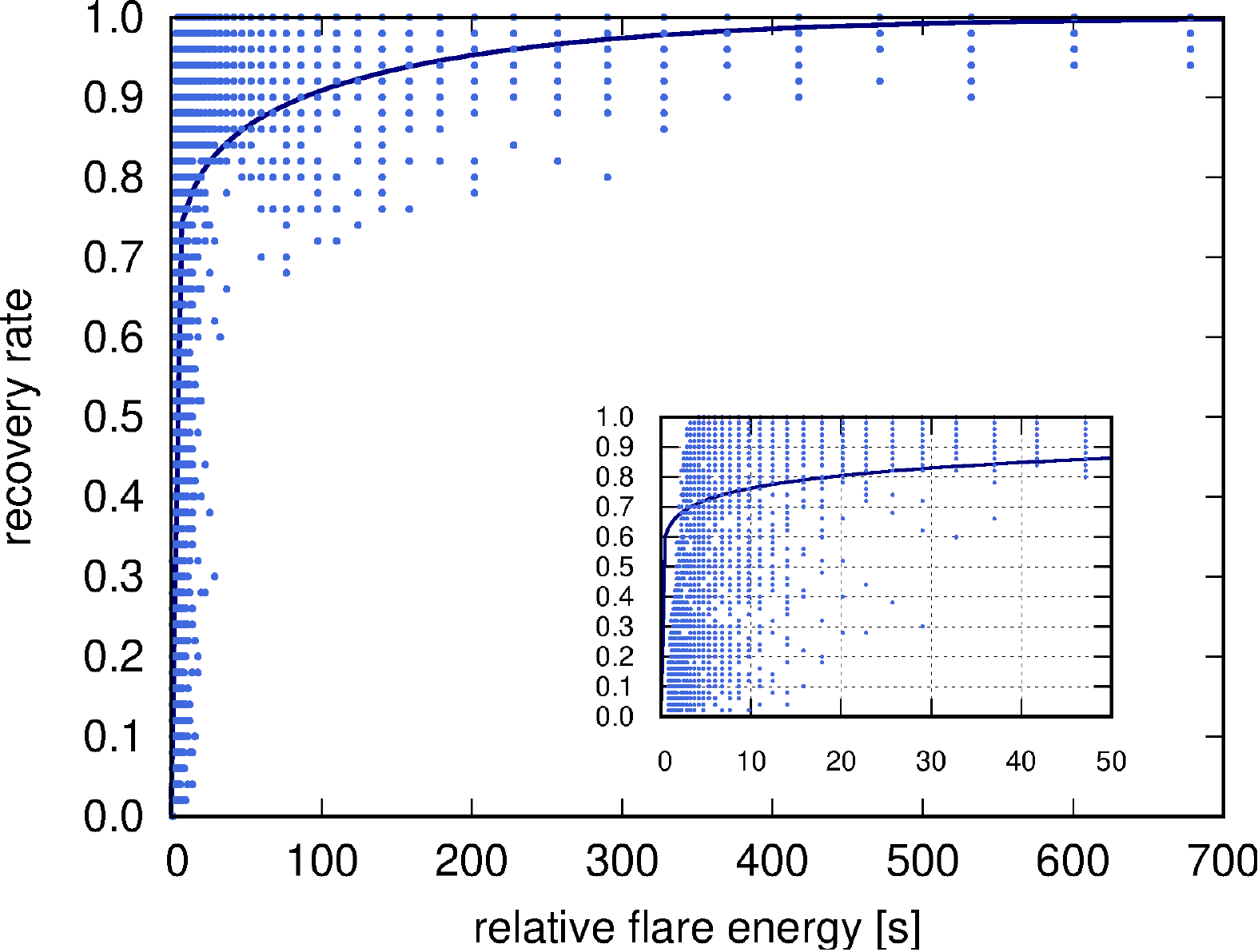}
      \caption{Flare recovery rate in the function of relative flare energy (equivalent time in sec). The recovery rates (dots) were derived from a recovery test using artificial flares injected to the original 20 datafiles. The test results are generalized by using a parameterized hiperbolic tangent function (blue line). The inset shows the details below 50 sec equivalent time. See the text for more.}
         \label{flare_recovery}
   \end{figure}

\begin{figure}
   \centering
   \includegraphics[width=\columnwidth]{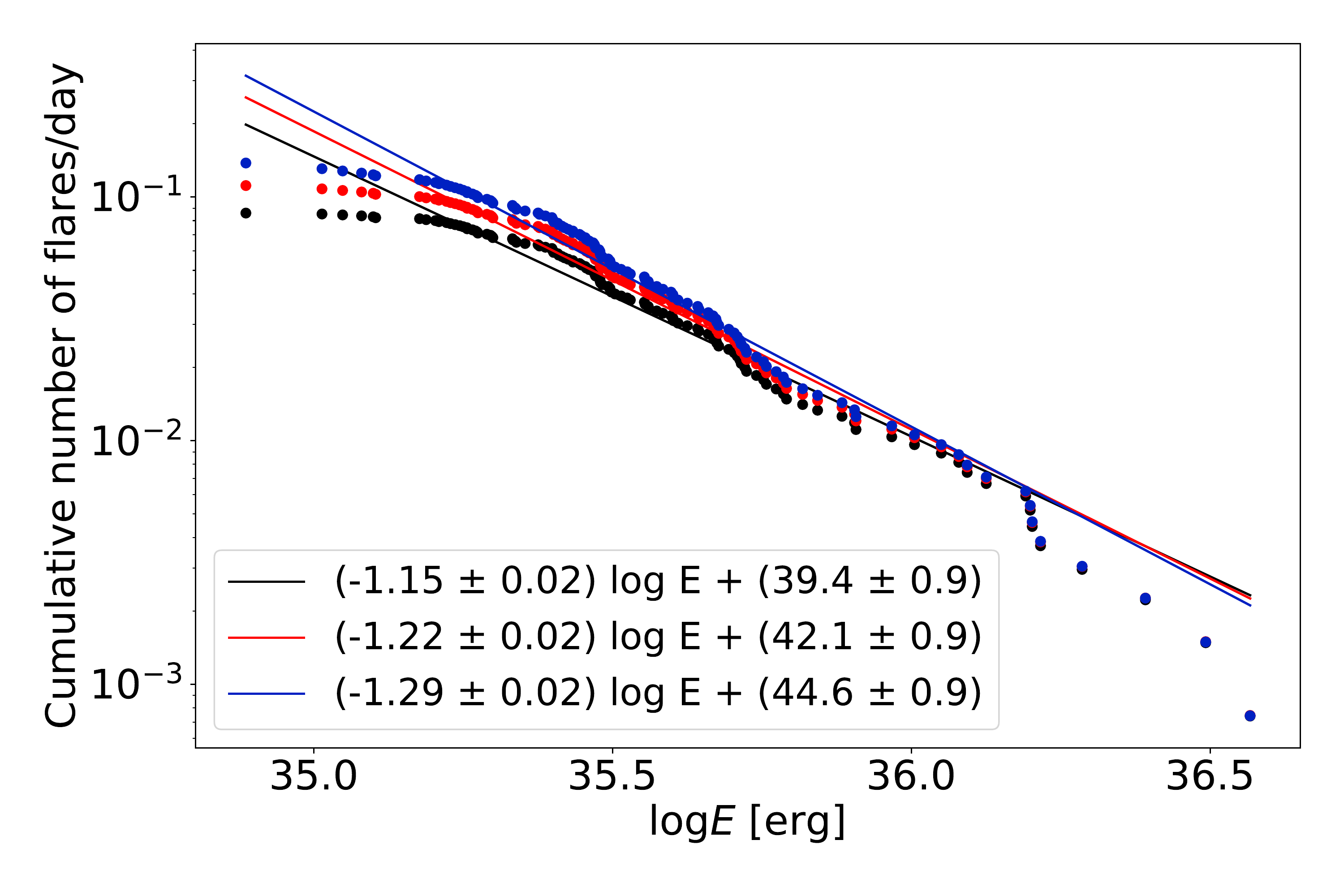}
      \caption{The largest deviation between applying the common recovery correction and the one realised for the star only (own correction). The object is KIC~5480528, black, red and blue colors mark the data and fits to the original data, own corrections, and common corrections, respectively. The difference between the slopes of the fits for the two different corrections (red and blue) is within two sigma from each other. Flare energies are derived for the \emph{Kepler} bandpass. (The five lowest energy datapoints were excluded from the fits.)}
         \label{FFD_test}
   \end{figure}

The recovered flare rates for 20 stars (19 most flaring giants plus KIC~2852961) were plotted together in the function of the relative flare energies of the injected flares in Fig.~\ref{flare_recovery}. The shapes of the individual test results proved to be very similar, so we tried to use the test as a single recovery application by fitting the test results with an arbitrarily chosen parameterized hiperbolic tangent ($\tanh$) function tracking the envelope of the test results. This generalized function allows to automatically correct the detection bias of the cumulative number of flare frequencies towards lower energies for all stars in the same time. However, there are
deviating points in Fig.~\ref{flare_recovery} between $50\leq\varepsilon_f\leq300$\,s relative flare energies ($\equiv t_{\rm eq}$ equivalent times), but most of these come from the recovery tests of three stars (out of 20) only. Fig.~\ref{FFD_test} shows the difference between applying the common recovery correction and the one realised for the star only in the most deviating case of KIC~5480528. The very small difference between the slopes of the fits to the cumulative number of flare frequencies proves that the use of a common correction function for all stars is appropriate.

The inset of Fig.~\ref{flare_recovery} shows the lowest part of the relative flare energies of the test.  The total number of flares to be corrected for recovery bias is 2247, and from these only 30 flares fall below 10\,s equivalent time (1.3\%) where the applied function does not follow well the envelope of the test results. We note that at the lowest relative flare energies (say, below $\varepsilon_f\approx$5\,s) the test gave quite uncertain results, because at this range the data noise exceeds the lowest flare amplitudes. The other reason could be the limited (30-min) time resolution of the observing data.

\begin{figure}[h!]
   \centering
   \includegraphics[width=\columnwidth]{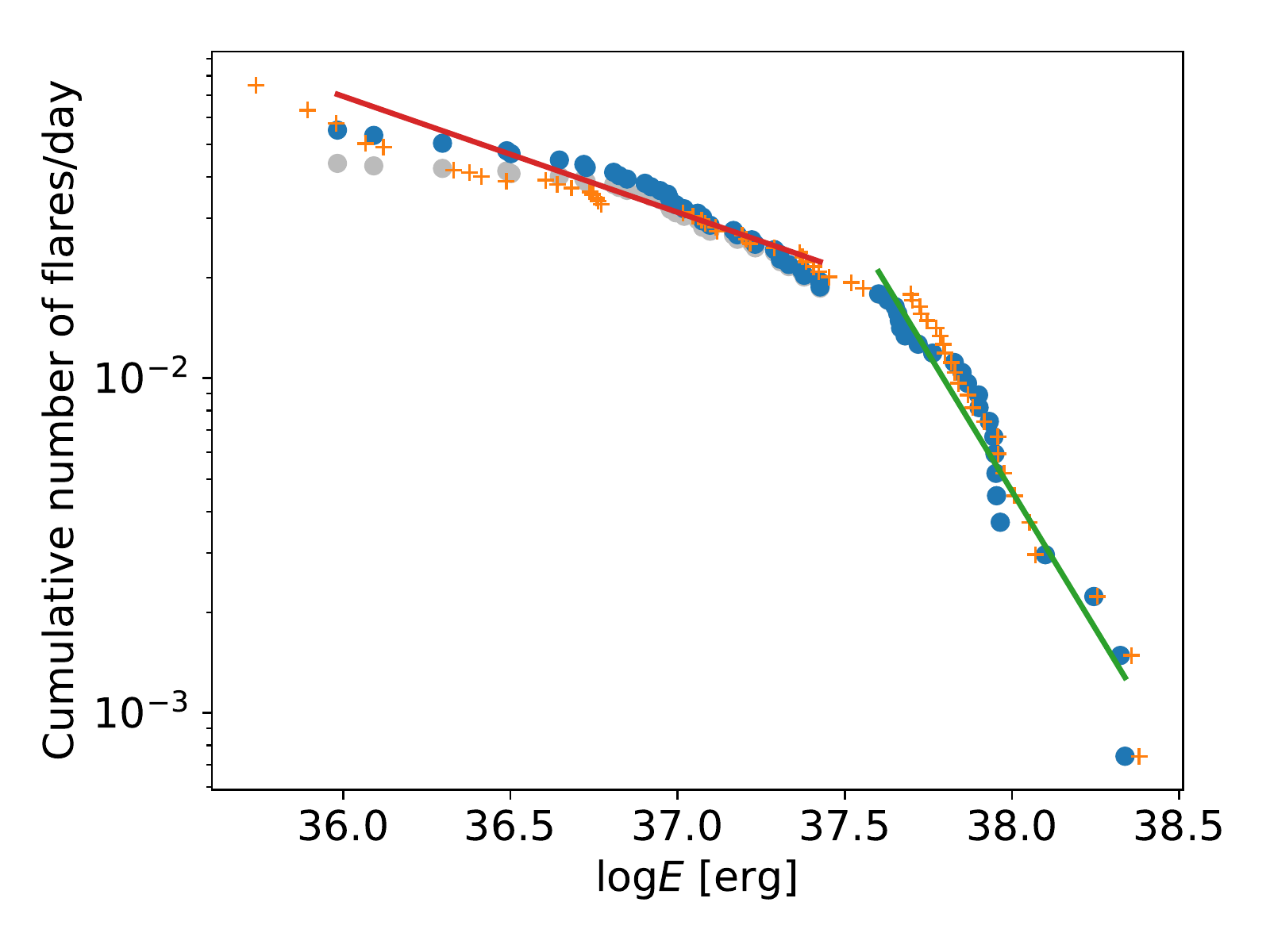} 
      \caption{Detection bias-corrected cumulative FFD for KIC~2852961 using the two different, cross-matched, automatic flare detection results introduced in the present paper. The original datapoints are plotted in grey, the blue points show the detection bias corrected values. Orange crosses show the detection bias corrected results from Paper~I. Flare energies are derived for the \emph{Kepler} bandpass.}
         \label{test:ffd}
   \end{figure}

\begin{figure*} [th!]
   \centering
   \includegraphics[width=\textwidth]{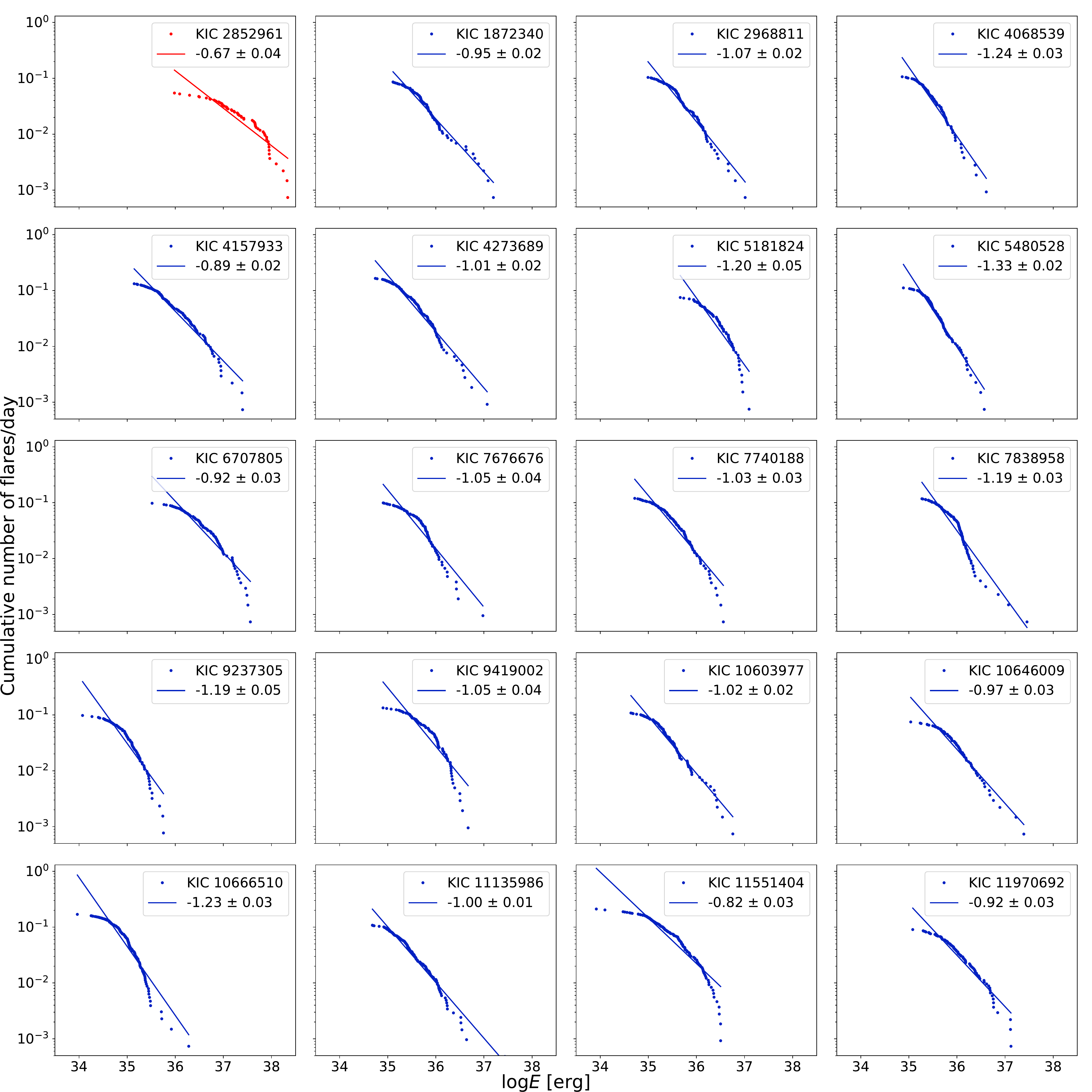} 
      \caption{FFDs for KIC~2852961 (red) and the 19 most flaring giant stars (blue) of the \emph{Kepler} field. Flare energies are derived for the \emph{Kepler} bandpass. A simple linear fit is drawn to the plots, in the upper right corner the slopes of the fits and their errors are indicated.}
         \label{all:ffd}
   \end{figure*}

At the end, from the generalized tanh recovery function we derived correction factors (multiplicative inverses) to estimate the real flare numbers which is significant only at the low energy ranges. These correction factors are used to provide the detection bias-corrected cumulative flare-frequency diagrams (FFDs) in Figs.~\ref{FFD_test}--\ref{test:ffd}--\ref{common:ffd}.

\subsection{Reliability of the flare frequency diagrams}


A flare frequency diagram (FFD) was drawn for KIC~2852961, with the method described in detail in Paper~I (Sect.~6). This time we used the stellar parameters as given in TICv8, i.e., everything was done in the same way for KIC~2852961 as for the selected most flaring 19 stars. The result is plotted in Fig.~\ref{test:ffd} using grey and blue points from the original data and from the corrected flare recovery, respectively. The bias-corrected recovery values from Paper~I (its Fig.~11) are plotted with orange crosses. The comparison with the blue points and  orange crosses shows that mostly the low energy flares were not detected by our method as discussed above in Sect.~\ref{energy_distribution}. A very small shift in the cumulative energy derived in the present paper towards the lower values is also seen due to the different base stellar flux calculated from the temperature and radius given in TICv8 and in Paper~I. However, the two distributions are very similar.

The result for the fit to the lower energy range (red line) yields a power law index of $\alpha$=1.31$\pm$0.02, while the result in Paper~I is $\alpha$=1.29$\pm$0.02. Similarly, the fit for the high energy range (green line) yields $\alpha$=2.65$\pm$0.09, while the result in Paper~I is $\alpha$=2.84$\pm$0.06. From these very similar results we conclude that the method we apply for the 19 frequent flaring giants give reliable frequency diagrams and fits.

\subsection{Flare frequency diagrams of 19 giant stars}\label{ffd:19stars}


   \begin{figure}
   \centering
   \includegraphics[width=\columnwidth]{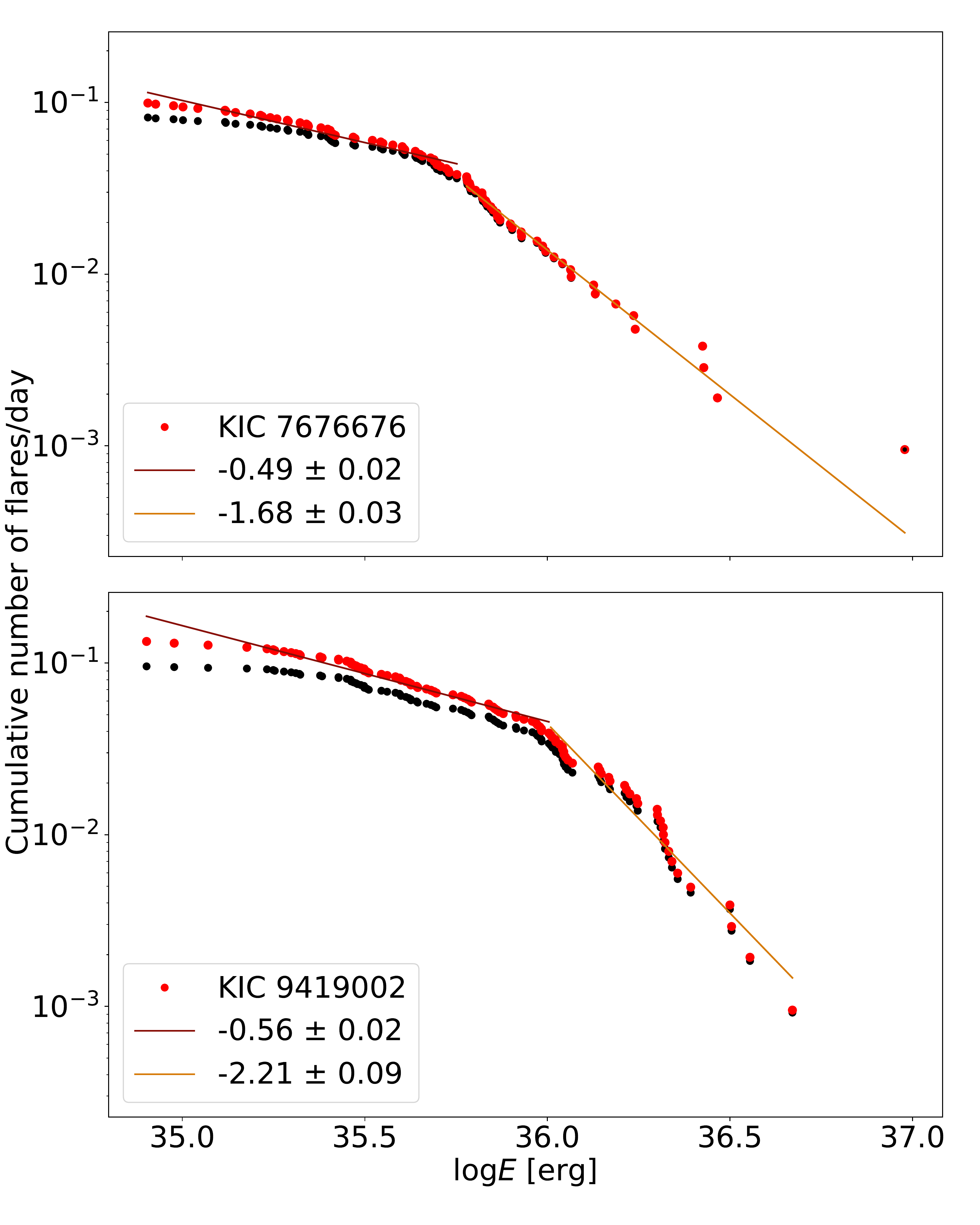} 
      \caption{Two-component fits to the FFDs for KIC~7676676 and KIC~9419002. Black dots correspond to the original values while red dots are bias corrected. Note the similar flare energy range and the different breakpoints of the fits for the two targets. Flare energies are derived for the \emph{Kepler} bandpass. See the text for more.}
         \label{ffd:2broken}
   \end{figure}


   \begin{figure}
   \centering
   \includegraphics[width=9cm]{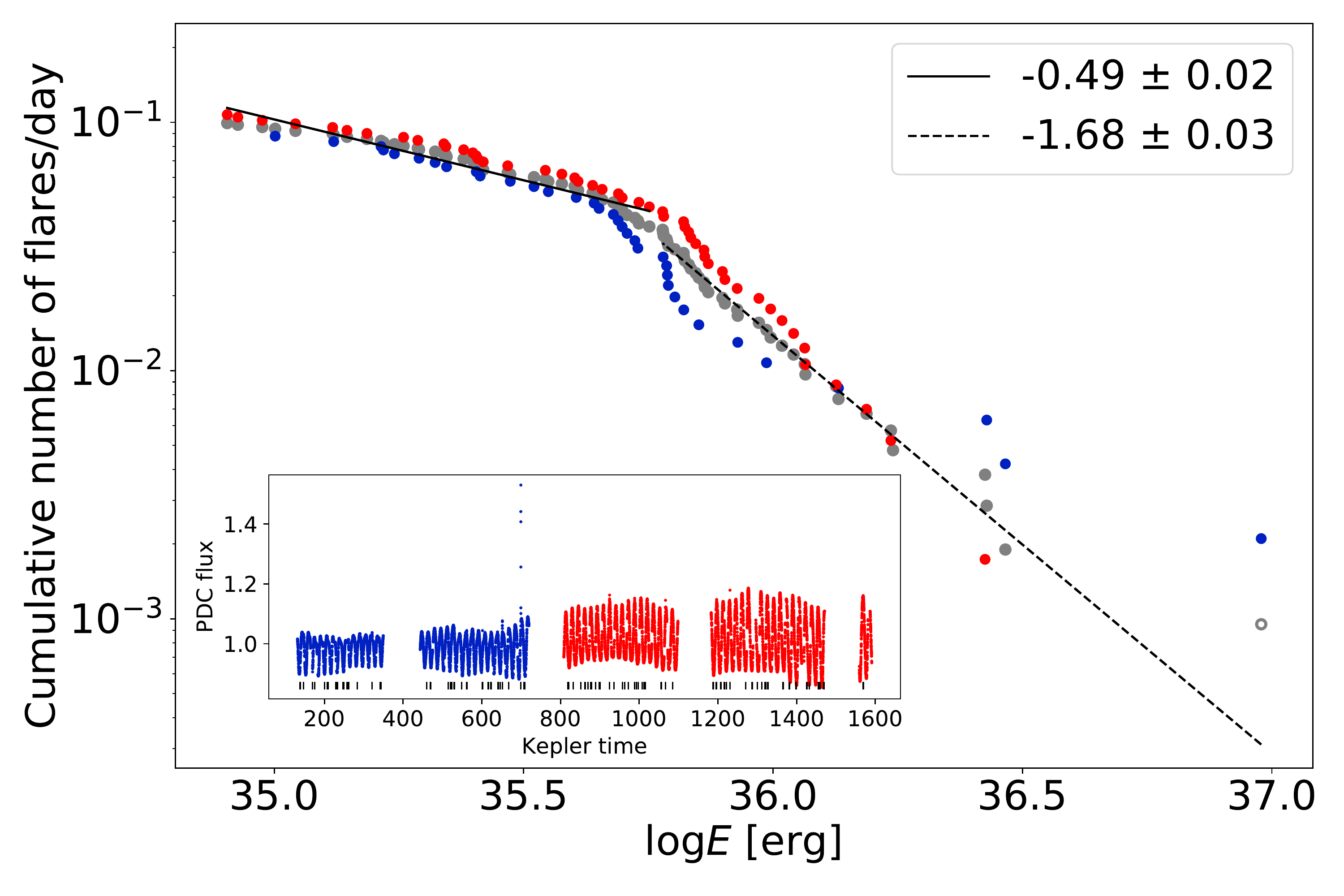} 
      \caption{FFD for KIC~7676676 in two activity levels. The datapoints showing lower and higher amplitude rotational modulation are marked with blue and red dots in the inset, and the same colors mark the corresponding FFDs. The broken power law fit (grey line) corresponds to all the data (grey dots), shown in Fig.~\ref{ffd:2broken}, upper panel. The highest amplitude flare occurred near the end of the lower amplitude rotational modulation. Flare energy values are derived for the {\it Kepler} bandpass.}
         \label{ffd:time_var}
   \end{figure}

After correcting for the detection bias FFDs were constructed for the 19 most flaring giants stars from Table~\ref{tab:long}, plotted in Fig.~\ref{all:ffd} applying simple linear fits to the data. Most of the FFDs can well be fitted with a single line, except KIC~11551404 whose FFD shows a continuous curvature. The slopes of the linear fits (without KIC~2852961) spread between $-$0.82 and $-$1.33 with errors between 0.02-0.05. The mean of the slopes translates to $\alpha = 2.01\pm 0.16$ power law index for the 19 giants (cf. Fig.~\ref{all:ffd}). 

For KIC~7676676, and KIC~9419002 two-component linear fits shown in Fig.~\ref{ffd:2broken} look more appropriate. 
The breakpoints between the high- and low energy parts of the  FFDs are at $\log~E_{Kp} = 35.75$, 36.0, and 37.6\,[erg] for KIC~7676676, KIC~9419002 (Fig.~\ref{ffd:2broken}), and KIC~2852961 (Fig.~\ref{test:ffd}), respectively. 

Studying flare energy spectra of dwarf stars and that of the Sun with two distinct power-law regions \citet{2018ApJ...854...14M} showed, that the breakpoint in an FFD appears at a certain critical energy $E(c)$ which depends on the local magnetic field strength and the local density, thus it should not be the same for all stars. The energy range within which the flares are detectable on a given star depends on observational selection, for example low activity stars exhibit smaller number of flares and low energy flares cannot be detected at high S/N. Thus, it is not necessary that the critical energy $E(c)$ lies inside the observable energy range of flares. If it is outside than no break appears in the FFD, as was suggested by \citet{2018ApJ...854...14M}. We may extend this important finding to FFDs of more evolved stars discussed in the present paper. The two FFDs plotted in Fig.~\ref{ffd:2broken} span about the same energy range, but the breakpoints of the power laws are different, suggesting different magnetic field strengths and densities of the atmospheres, which probably originate from different masses and evolutionary stages on the RGB. 

An interesting question is to explore the changes of the flare activity during different strengths of the overall activity of stars (activity cycles). We addressed this problem already in Paper~I (see their Fig.~13). From the 19 selected stars KIC~7676676 shows continuous increase of the light curve amplitude suggesting increasing activity level of the star. In Fig.~\ref{ffd:time_var} FFDs are plotted for the time intervals of the lower (blue, 35 flares) and higher (red, 51 flares) amplitude rotational modulation. While at low energies the two distributions are nearly similar, from about $\log E_{Kp} = 35.7$\,[erg] the energies of the flares observed after day $\approx$700 (\emph{Kepler} time, plotted in red) are significantly higher, together with the increased overall activity manifested in higher amplitude rotational modulation. At the highest energies the number of flares are small, but it is interesting that the highest energy flare occurred near the end of the lower activity phase of the star.

\begin{figure}
   \centering
   \includegraphics[width=\columnwidth]{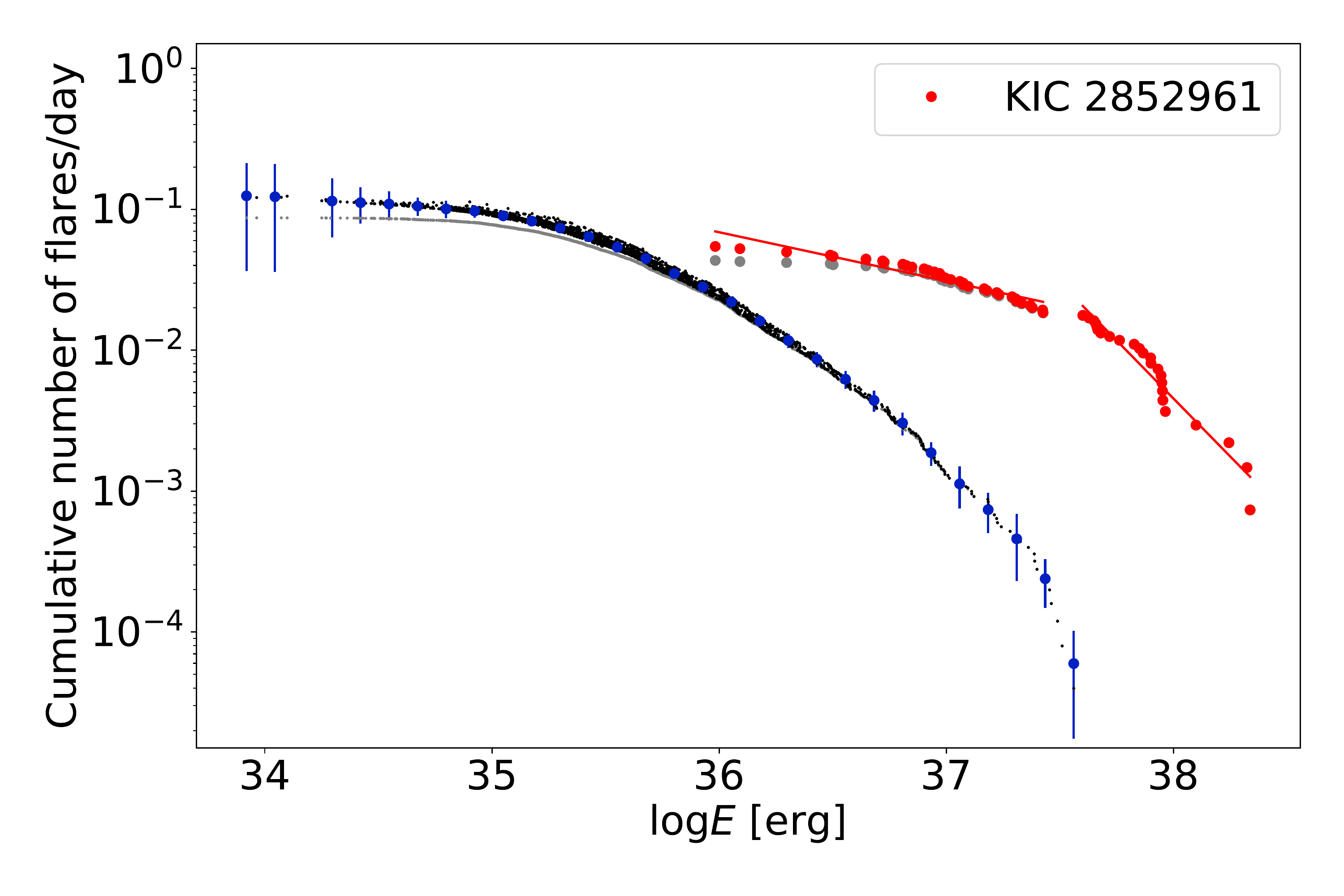} 
      \caption{Cumulative FFD for the most flaring 19 giant stars co-adding their 2188 flares(grey and black dots show the original and bias-corrected values, respectively), with an estimate of the range of values in energy bins (red dots and bars). The FFD of KIC~2852961 and the two-component fit is added separately (grey and green dots and lines for original and corrected values). Note the difference in the high energy range. Flare energies are derived for the \emph{Kepler} bandpass.}
         \label{common:ffd}
   \end{figure}

Cumulative FFD for the most flaring 19 stars co-adding 2188 flares is shown in Fig.~\ref{common:ffd} with grey and black dots for the original and bias corrected values. Red dots and bars show means and errorbars for 30 bins throughout the data. The errorbars were calculated from the $\sqrt{N}$ Poisson noise in each bin. Towards the low energy parts the individual FFDs bend at different energies, even after the correction for the recovery bias, therefore when summing up all flares a curvature on the overall FFD appears ending around $\log~E_{Kp} = 36$\,[erg]. Towards higher energies the flare frequency is nearly linearly decreasing in all FFDs, consequently also in the common FFD, except the highest energy part with a few flares only, where the decrease is faster.  Note the similarity of Fig.~\ref{common:ffd} with \citet[][their Fig. 15]{2019ApJS..241...29Y}. The FFD of KIC~2852961 is plotted separately into the cumulative FFD of the 19 giants (grey for the original data, and blue for bias corrected data and the fit). The high energy part of the FFD of KIC~2852961 starts at about the same energy ($\log E_{Kp} = 37.60$\,[erg]) where the highest energy flare of the 19 giants ($\log E_{Kp} = 37.56$\,[erg]) appears, meaning that in the high-energy range all flares of KIC~2852961 are more powerful than any flare from the 19 frequently flaring giant stars.

The recent works of \citet{2018ApJ...858...55P}, \citet{2013ApJS..209....5S} and the present paper allow an interesting comparison between the FFDs of set of ultracool dwarfs, G dwarfs and  flaring giants. The observations of ten ultracool dwarfs with spectral types between M6-L0 were made during the {\it Kepler} K2 mission, the flare recovery and the individual FFDs of their targets were calculated and presented in the same way as for the giants in this paper. For comparison see \citet[][their Figs. 3-4]{2018ApJ...858...55P} and Fig.~\ref{all:ffd} and \ref{common:ffd:fits} in this paper. The energy ranges of the flares on the ultracool dwarfs are between $29.0\leq\log~E_{Kp}\leq33.5$\,[erg] whereas the whole energy range of the giants is outside this interval at higher energies, between $33.9\leq\log E_{Kp}\leq38.7$\,[erg]. The average power law index of the individual fits to the FFDs of the ten ultracool dwarfs \citep{2018ApJ...858...55P} and 20 flaring giants (present paper) is the same: $\alpha \approx2.0$. Between the ultracool dwarfs and flaring giants are the flare energies of the (super)flaring G-dwarf stars with bolometric $33.0\leq\log E\leq36.0$\,[erg] energy studied by \citet{2013ApJS..209....5S} also from \emph{Kepler} data, who found power law indices $\alpha\approx $2.0-2.2 for slowly rotating G stars, and for all G-dwarfs together. The flare energy interval the \emph{Kepler} passband for the G-dwarfs is about $32.2\leq\log E_{Kp}\leq35.2$\,[erg] using the conversion from \citet{2015ApJ...809...79O}. \citet{2019ApJS..241...29Y} calculated FFDs for different spectral types of flaring stars (their Fig. 3.), gave $\alpha = 1.90\pm 0.10$ power law index especially for the giants and $\approx 2$ as an average for all spectral types except the A-stars.

These results present a consistent view of magnetically active stars with flares from the smallest L0-stars through G- and M-dwarfs to normal giants in a wide range of stellar mass, structure and age. In this pool all stars show flares of magnetic origin with similar shapes but of different timescales spanning a huge energy range between  $29.0\leq\log E_{Kp}\leq38.7$\,[erg]. The flares were observed with the same instrument (\emph{Kepler}) and treated with the same or very similar methods. The FFDs of the different groups of stars run in the same way with a common power law index of $\alpha\approx2.0$. The different flare energies from an L0 to a giant star result from only a scaling effect depending on the details of the magnetic field, on the size of the active regions from where the flares originate, and on the local environments like density and gravity. We discussed the details of the scaling effect in Paper~I (see Sect.~9.5 in that paper).

A recent paper of \citet[][their Fig. 14]{2020arXiv200914412M} compares the observed flare durations in the function of flare energies of the Sun, the flaring M-dwarf YZ\,CMi, G-dwarfs \citep{2013ApJS..209....5S} and KIC~2852961 from Paper~I. The slope of the relation is about 1/3 assuming constant magnetic filed strengths, and in this relation KIC~2852961 is on the same line as the Sun  with $B$=60\,G \citep{2020arXiv200914412M}. The selected most flaring 19 stars of the present paper have observed flare durations between about 1-30 hours (59-1766 minutes), and the energies are between $33.9\leq\log E_{Kp}\leq37.6$\,[erg] occupying a region in the upper right part of Fig.~14. in \citet{2020arXiv200914412M}. The slope of the flare energy-duration power-law relation of KIC~2852961 from Paper~I is $0.325\pm0.026$, while that from the present paper is higher, it is $0.50\pm0.04$ as seen from Fig.~\ref{duration:19stars} (upper left corner, red), still, the two results are within 3-$\sigma$ from each other. On this basis we may suspect that the slopes of the relations of the other 19 individual stars, plotted in Fig.~\ref{duration:19stars}, are probably also higher than real, due to the underestimation of the flare duration mostly for the shorter, lower energy flares, see Section~\ref{sect_methods} for explanation.  Nevertheless, all the slopes reveal positive correlation, they are different, but concerning the errors the difference is not significant.

For the question we ask in the Introduction about the correct definition of the term "superflare" we cannot give a definite answer. The flare energy ranges of stars embrace $\approx$10$^{10}$\,erg from the smallest flare on an ultracool dwarf to the most energetic one on a giant star. A specific, high flare energy limit to reach for being a "superflare" in the individual stars could be estimated through the maximum available magnetic energies as we did in Paper~I (see Sect.~9.4) for KIC~2852961. But this is a complicated task with many uncertainties like the sizes of the active regions (through spot modeling), the magnetic field strengths (from measurements) and flare loop sizes (from models). Perhaps the term "superflare" would better to reserve only for the Sun where these parameters can directly be measured.


\section{Summary and conclusions}\label{sect_summary}

We confirmed 61 flaring giant stars in the \emph{Kepler} field. This number can raise to 69 with those eight stars of unknown evolutionary status (Table~\ref{tab:long}), and could further increase by a few giants from the uncertain cases (Table~\ref{table:uncertain}) once any of those proves to be the origin of the observed flares.

Throughout this paper we used the recent study of KIC~2852961 (Paper~I) for verifying the new methods and results by re-analysing this star using the more automatized approaches introduced in this paper. 

We showed that the rotational periods of the flaring giants correlated only weakly with the number of flares, hence with this measure of the activity strength. No correlation was found between $S\ind{ph}$ and rotational periods, but the index was found to be significantly higher for all stars than for their inactive counterparts \citep{Gaulme_2020}. We found only six giants which show both flares and oscillations, and on one pulsating star (a $\gamma$\,Dor-$\delta$\,Sct hybrid) we identified a few significant flares.

Flare energies were calculated and energy distributions in histograms were drawn for all flaring giants and contaminating dwarfs from Table~\ref{tab:long} with peak energies at $\log E_{Kp}=35.6$\,[erg] and $\log E_{Kp}=34.2$\,[erg], respectively. Most of the histograms have nearly symmetric shapes with longer tails towards higher energies.

We constructed FFDs for a subsample 19 most flaring stars from Table~\ref{tab:long} plus a new FFD for KIC~2852961 from Paper~I. for comparison purposes. The individual FFDs were fitted first by a single power law. For two stars from this subsample broken power law approaches turned out to be more appropriate. These two stars have flares in the same energy range but the breakpoints of their fits are at different energies suggesting a difference between the local environments from where their flares originate. We also find that even the weakest flare from the high energy part of the KIC~2852961 FFD is stronger than any flare erupted in the subsample of the 19 most flaring stars.

\begin{figure}
   \centering
   \includegraphics[width=\columnwidth]{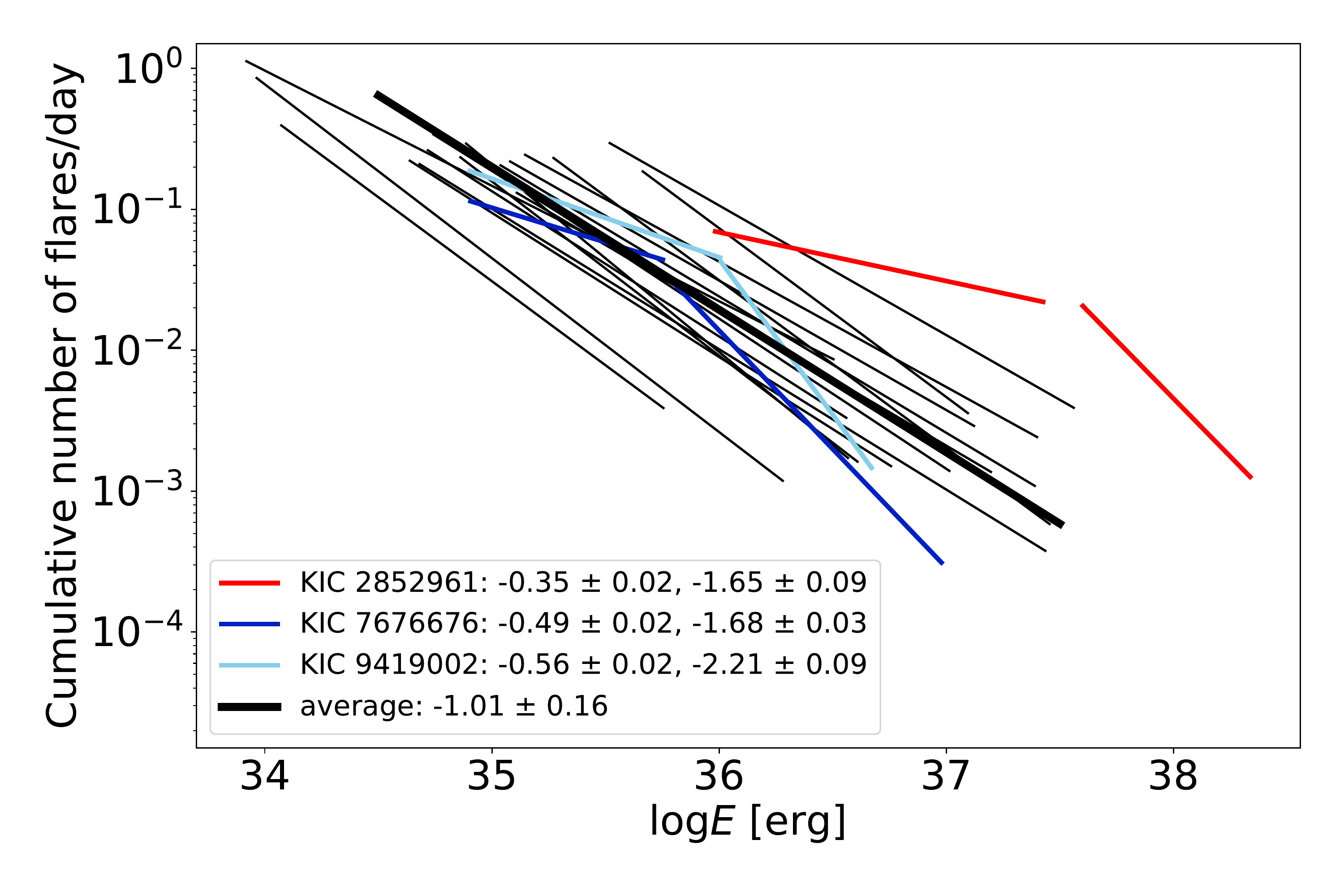} 
      \caption{Fits and slopes of the individual FFDs for the 19 most flaring giant stars from the \emph{Kepler} database, together with their mean (thick black line). For three stars broken power-law fits are also plotted. Flare energies are derived for the \emph{Kepler} bandpass.}
         \label{common:ffd:fits}
   \end{figure}

We find that the slopes of fits to the individual FFDs of the 19 most flaring stars are quite similar to each other, the average being $\alpha=2.01\pm 0.16$, though the difference between the energy ranges of the individual stars is close to two orders, see Fig.~\ref{common:ffd:fits}.

Positive flare energy-duration correlations were found for the subsample of the 19 most flaring giants with slopes somewhat higher than suspected due to the low sampling  of the data (30\,min cadence), since the observed flare durations were underestimated.

From the results summarized above we conclude that:
\begin {itemize}

\item Only about 0.3\% of the \emph{Kepler} giant stars with $\log g < 3.5$ show flare activity, that is, about 10\% of the result in D17, but agreeing well with the result of YL19.

\item No strong correlation was found between the stellar properties (rotation period, velocity, binarity) and the flaring characteristics. The wide scale of the flaring specialities points towards the heterogeneity of flaring giants. 

\item On flaring giant stars, similar to flaring dwarfs, observed flare durations correlate with flare energies, suggesting similar background physics and a scaling effect behind (cf. Paper~I).

\item The flaring nature of KIC~2852961 with higher energy flares in excess is markedly different from the rest of the studied giants. The reason of the dissimilarity, probably specific of the stellar structure, needs further investigation.

\item A large fraction of the flaring red giants are likely to belong to close binary systems, given the levels of photometric variability $S\ind{ph}$, the general absence of oscillations, and the fact that 11 out of the 14 giants that have spectroscopic data are binaries. 
\end{itemize}


\begin{acknowledgements}
 We thank an anonymous referee for useful comments which helped to express our results more clearly.
 This work was supported by the Hungarian National Research, Development and Innovation Office grant OTKA K131508 and by the Lend\"ulet Program  of the Hungarian Academy of Sciences, project No. LP2018-7/2019. Authors from Konkoly Observatory acknowledge the financial support  of the Austrian-Hungarian  Action  Foundation (95 \"ou3, 98\"ou5, 101\"ou13). B.S. was supported by the \'UNKP-19-3 New National Excellence Program of the Ministry for Innovation and Technology. M.N.G. acknowledges support from MIT’s Kavli Institute as a Torres postdoctoral fellow. P. Gaulme acknowledges funding from the German Aerospace Center (Deutsches Zentrum f\"ur Luft- und Raumfahrt) under PLATO Data Center grant 50OO1501.
This research has made use of the NASA Exoplanet Archive, which is operated by the California Institute of Technology, under contract with the National Aeronautics and Space Administration under the Exoplanet Exploration Program. This work presents results from the European Space Agency (ESA) space mission \emph{Gaia}. \emph{Gaia} DR2 data are being processed by the Gaia Data Processing and Analysis Consortium (DPAC). Funding for the DPAC is provided by national institutions, in particular the institutions participating in the \emph{Gaia} MultiLateral Agreement (MLA). The \emph{Gaia} mission website is \texttt{https://www.cosmos.esa.int/gaia}. The \emph{Gaia} archive website is \texttt{https://archives.esac.esa.int/gaia}. Guoshoujing Telescope (the Large Sky Area Multi-Object Fiber Spectroscopic Telescope LAMOST) is a National Major Scientific Project built by the Chinese Academy of Sciences. Funding for the project has been provided by the National Development and Reform Commission. LAMOST is operated and managed by the National Astronomical Observatories, Chinese Academy of Sciences.
\end{acknowledgements}

\bibliographystyle{aa}
\bibliography{bib.bib}

\begin{thebibliography}{50}
\expandafter\ifx\csname natexlab\endcsname\relax\def\natexlab#1{#1}\fi

\bibitem[{{Auri{\`e}re} {et~al.}(2015){Auri{\`e}re}, {Konstantinova-Antova},
  {Charbonnel}, {Wade}, {Tsvetkova}, {Petit}, {Dintrans}, {Drake}, {Decressin},
  {Lagarde}, {Donati}, {Roudier}, {Ligni{\`e}res}, {Schr{\"o}der},
  {Landstreet}, {L{\`e}bre}, {Weiss}, \& {Zahn}}]{2015A&A...574A..90A}
{Auri{\`e}re}, M., {Konstantinova-Antova}, R., {Charbonnel}, C., {et~al.} 2015,
  \aap, 574, A90

\bibitem[{{Beck} {et~al.}(2018){Beck}, {Mathis}, {Gallet}, {Charbonnel},
  {Benbakoura}, {Garc{\'{\i}}a}, \& {do Nascimento}}]{Beck_2018}
{Beck}, P.~G., {Mathis}, S., {Gallet}, F., {et~al.} 2018, \mnras, 479, L123

\bibitem[{{Berger} {et~al.}(2018){Berger}, {Huber}, {Gaidos}, \& {van
  Saders}}]{2018ApJ...866...99B}
{Berger}, T.~A., {Huber}, D., {Gaidos}, E., \& {van Saders}, J.~L. 2018, \apj,
  866, 99

\bibitem[{{Brun} \& {Browning}(2017)}]{2017LRSP...14....4B}
{Brun}, A.~S. \& {Browning}, M.~K. 2017, Living Reviews in Solar Physics, 14, 4

\bibitem[{{Cantiello} {et~al.}(2016){Cantiello}, {Fuller}, \&
  {Bildsten}}]{2016ApJ...824...14C}
{Cantiello}, M., {Fuller}, J., \& {Bildsten}, L. 2016, \apj, 824, 14

\bibitem[{{Catalano} \& {Frasca}(1994)}]{1994A&A...287..575C}
{Catalano}, S. \& {Frasca}, A. 1994, \aap, 287, 575

\bibitem[{{Charbonnel} {et~al.}(2017){Charbonnel}, {Decressin}, {Lagarde},
  {Gallet}, {Palacios}, {Auri{\`e}re}, {Konstantinova-Antova}, {Mathis},
  {Anderson}, \& {Dintrans}}]{2017A&A...605A.102C}
{Charbonnel}, C., {Decressin}, T., {Lagarde}, N., {et~al.} 2017, \aap, 605,
  A102

\bibitem[{{Che} {et~al.}(2011){Che}, {Monnier}, {Zhao}, {Pedretti}, {Thureau},
  {M{\'e}rand}, {ten Brummelaar}, {McAlister}, {Ridgway}, {Turner}, {Sturmann},
  \& {Sturmann}}]{2011ApJ...732...68C}
{Che}, X., {Monnier}, J.~D., {Zhao}, M., {et~al.} 2011, \apj, 732, 68

\bibitem[{{Cutispoto} {et~al.}(1992){Cutispoto}, {Pagano}, \&
  {Rodono}}]{1992A&A...263L...3C}
{Cutispoto}, G., {Pagano}, I., \& {Rodono}, M. 1992, \aap, 263, L3

\bibitem[{{Davenport}(2016)}]{2016ApJ...829...23D}
{Davenport}, J. R.~A. 2016, \apj, 829, 23

\bibitem[{{Davenport} {et~al.}(2019){Davenport}, {Covey}, {Clarke}, {Boeck},
  {Cornet}, \& {Hawley}}]{2019ApJ...871..241D}
{Davenport}, J. R.~A., {Covey}, K.~R., {Clarke}, R.~W., {et~al.} 2019, \apj,
  871, 241

\bibitem[{{Davenport} {et~al.}(2014){Davenport}, {Hawley}, {Hebb},
  {Wisniewski}, {Kowalski}, {Johnson}, {Malatesta}, {Peraza}, {Keil},
  {Silverberg}, {Jansen}, {Scheffler}, {Berdis}, {Larsen}, \&
  {Hilton}}]{2014ApJ...797..122D}
{Davenport}, J. R.~A., {Hawley}, S.~L., {Hebb}, L., {et~al.} 2014, \apj, 797,
  122

\bibitem[{{Frasca} {et~al.}(2016){Frasca}, {Molenda-{\.Z}akowicz}, {De Cat},
  {Catanzaro}, {Fu}, {Ren}, {Luo}, {Shi}, {Wu}, \&
  {Zhang}}]{2016A&A...594A..39F}
{Frasca}, A., {Molenda-{\.Z}akowicz}, J., {De Cat}, P., {et~al.} 2016, \aap,
  594, A39

\bibitem[{{Gaulme} {et~al.}(2014){Gaulme}, {Jackiewicz}, {Appourchaux}, \&
  {Mosser}}]{2014ApJ...785....5G}
{Gaulme}, P., {Jackiewicz}, J., {Appourchaux}, T., \& {Mosser}, B. 2014, \apj,
  785, 5

\bibitem[{{Gaulme} {et~al.}(2020){Gaulme}, {Jackiewicz}, {Spada}, {Chojnowski},
  {Mosser}, {McKeever}, {Hedlund}, {Vrard}, {Benbakoura}, \&
  {Damiani}}]{Gaulme_2020}
{Gaulme}, P., {Jackiewicz}, J., {Spada}, F., {et~al.} 2020, \aap, 639, A63

\bibitem[{{Gaulme} {et~al.}(2016){Gaulme}, {McKeever}, {Jackiewicz}, {Rawls},
  {Corsaro}, {Mosser}, {Southworth}, {Mahadevan}, {Bender}, \&
  {Deshpande}}]{2016ApJ...832..121G}
{Gaulme}, P., {McKeever}, J., {Jackiewicz}, J., {et~al.} 2016, \apj, 832, 121

\bibitem[{{G{\"u}nther} \& {Daylan}(2019)}]{allesfitter-code}
{G{\"u}nther}, M.~N. \& {Daylan}, T. 2019, {allesfitter: Flexible star and
  exoplanet inference from photometry and radial velocity}

\bibitem[{{G{\"u}nther} \& {Daylan}(2020)}]{allesfitter-paper}
{G{\"u}nther}, M.~N. \& {Daylan}, T. 2020, arXiv e-prints, arXiv:2003.14371

\bibitem[{{G{\"u}nther} {et~al.}(2020){G{\"u}nther}, {Zhan}, {Seager},
  {Rimmer}, {Ranjan}, {Stassun}, {Oelkers}, {Daylan}, {Newton}, {Kristiansen},
  {Olah}, {Gillen}, {Rappaport}, {Ricker}, {Vanderspek}, {Latham}, {Winn},
  {Jenkins}, {Glidden}, {Fausnaugh}, {Levine}, {Dittmann}, {Quinn},
  {Krishnamurthy}, \& {Ting}}]{2020AJ....159...60G}
{G{\"u}nther}, M.~N., {Zhan}, Z., {Seager}, S., {et~al.} 2020, \aj, 159, 60

\bibitem[{{Holzwarth} \& {Sch{\"u}ssler}(2001)}]{2001A&A...377..251H}
{Holzwarth}, V. \& {Sch{\"u}ssler}, M. 2001, \aap, 377, 251

\bibitem[{{Hon} {et~al.}(2019){Hon}, {Stello}, {Garc{\'\i}a}, {Mathur},
  {Sharma}, {Colman}, \& {Bugnet}}]{2019MNRAS.485.5616H}
{Hon}, M., {Stello}, D., {Garc{\'\i}a}, R.~A., {et~al.} 2019, \mnras, 485, 5616

\bibitem[{{Kallinger} {et~al.}(2014){Kallinger}, {De Ridder}, {Hekker},
  {Mathur}, {Mosser}, {Gruberbauer}, {Garc{\'{\i}}a}, {Karoff}, \&
  {Ballot}}]{Kallinger_2014}
{Kallinger}, T., {De Ridder}, J., {Hekker}, S., {et~al.} 2014, \aap, 570, A41

\bibitem[{{K{\H{o}}v{\'a}ri} \& {Ol{\'a}h}(2014)}]{2014SSRv..186..457K}
{K{\H{o}}v{\'a}ri}, Z. \& {Ol{\'a}h}, K. 2014, \ssr, 186, 457

\bibitem[{{K{\H{o}}v{\'a}ri} {et~al.}(2017){K{\H{o}}v{\'a}ri}, {Ol{\'a}h},
  {Kriskovics}, {Vida}, {Forg{\'a}cs-Dajka}, \&
  {Strassmeier}}]{2017AN....338..903K}
{K{\H{o}}v{\'a}ri}, Z., {Ol{\'a}h}, K., {Kriskovics}, L., {et~al.} 2017,
  Astronomische Nachrichten, 338, 903

\bibitem[{{K{\H{o}}v{\'a}ri} {et~al.}(2020){K{\H{o}}v{\'a}ri}, {Ol{\'a}h},
  {G{\"u}nther}, {Vida}, {Kriskovics}, {Seli}, {Bakos}, {Hartman}, {Csubry}, \&
  {Bhatti}}]{2020arXiv200505397K}
{K{\H{o}}v{\'a}ri}, {\mbox Zs}., {Ol{\'a}h}, K., {G{\"u}nther}, M.~N., {et~al.}
  2020, \aap, 641, A83

\bibitem[{{Maehara} {et~al.}(2020){Maehara}, {Notsu}, {Namekata}, {Honda},
  {Kowalski}, {Katoh}, {Ohshima}, {Iida}, {Oeda}, {Murata}, {Yamanaka},
  {Takagi}, {Sasada}, {Akitaya}, {Ikuta}, {Okamoto}, {Nogami}, \&
  {Shibata}}]{2020arXiv200914412M}
{Maehara}, H., {Notsu}, Y., {Namekata}, K., {et~al.} 2020, arXiv e-prints,
  arXiv:2009.14412

\bibitem[{{Mathur} {et~al.}(2017){Mathur}, {Huber}, {Batalha}, {Ciardi},
  {Bastien}, {Bieryla}, {Buchhave}, {Cochran}, {Endl}, {Esquerdo}, {Furlan},
  {Howard}, {Howell}, {Isaacson}, {Latham}, {MacQueen}, \&
  {Silva}}]{2017ApJS..229...30M}
{Mathur}, S., {Huber}, D., {Batalha}, N.~M., {et~al.} 2017, \apjs, 229, 30

\bibitem[{{Mathur} {et~al.}(2014){Mathur}, {Salabert}, {Garc{\'\i}a}, \&
  {Ceillier}}]{2014JSWSC...4A..15M}
{Mathur}, S., {Salabert}, D., {Garc{\'\i}a}, R.~A., \& {Ceillier}, T. 2014,
  Journal of Space Weather and Space Climate, 4, A15

\bibitem[{{McQuillan} {et~al.}(2014){McQuillan}, {Mazeh}, \&
  {Aigrain}}]{2014ApJS..211...24M}
{McQuillan}, A., {Mazeh}, T., \& {Aigrain}, S. 2014, \apjs, 211, 24

\bibitem[{{Mosser} {et~al.}(2014){Mosser}, {Benomar}, {Belkacem}, {Goupil},
  {Lagarde}, {Michel}, {Lebreton}, {Stello}, {Vrard}, {Barban}, {Bedding},
  {Deheuvels}, {Chaplin}, {De Ridder}, {Elsworth}, {Montalban}, {Noels},
  {Ouazzani}, {Samadi}, {White}, \& {Kjeldsen}}]{2014A&A...572L...5M}
{Mosser}, B., {Benomar}, O., {Belkacem}, K., {et~al.} 2014, \aap, 572, L5

\bibitem[{{Mullan} \& {Paudel}(2018)}]{2018ApJ...854...14M}
{Mullan}, D.~J. \& {Paudel}, R.~R. 2018, \apj, 854, 14

\bibitem[{{Ol\'ah} {et~al.}(1991){Ol\'ah}, {Hall}, \&
  {Henry}}]{1991A&A...251..531O}
{Ol\'ah}, K., {Hall}, D.~S., \& {Henry}, G.~W. 1991, \aap, 251, 531

\bibitem[{{Ol{\'a}h} {et~al.}(2009){Ol{\'a}h}, {Koll{\'a}th}, {Granzer},
  {Strassmeier}, {Lanza}, {J{\"a}rvinen}, {Korhonen}, {Baliunas}, {Soon},
  {Messina}, \& {Cutispoto}}]{2009A&A...501..703O}
{Ol{\'a}h}, K., {Koll{\'a}th}, Z., {Granzer}, T., {et~al.} 2009, \aap, 501, 703

\bibitem[{{Osten} \& {Wolk}(2015)}]{2015ApJ...809...79O}
{Osten}, R.~A. \& {Wolk}, S.~J. 2015, \apj, 809, 79

\bibitem[{{Paudel} {et~al.}(2018){Paudel}, {Gizis}, {Mullan}, {Schmidt},
  {Burgasser}, {Williams}, \& {Berger}}]{2018ApJ...858...55P}
{Paudel}, R.~R., {Gizis}, J.~E., {Mullan}, D.~J., {et~al.} 2018, \apj, 858, 55

\bibitem[{{Reiners}(2012)}]{2012LRSP....9....1R}
{Reiners}, A. 2012, Living Reviews in Solar Physics, 9, 1

\bibitem[{{Roettenbacher} {et~al.}(2016){Roettenbacher}, {Monnier}, {Korhonen},
  {Aarnio}, {Baron}, {Che}, {Harmon}, {K{\H{o}}v{\'a}ri}, {Kraus}, {Schaefer},
  {Torres}, {Zhao}, {Ten Brummelaar}, {Sturmann}, \&
  {Sturmann}}]{2016Natur.533..217R}
{Roettenbacher}, R.~M., {Monnier}, J.~D., {Korhonen}, H., {et~al.} 2016, \nat,
  533, 217

\bibitem[{{Shibata} \& {Magara}(2011)}]{2011LRSP....8....6S}
{Shibata}, K. \& {Magara}, T. 2011, Living Reviews in Solar Physics, 8, 6

\bibitem[{{Shibayama} {et~al.}(2013){Shibayama}, {Maehara}, {Notsu}, {Notsu},
  {Nagao}, {Honda}, {Ishii}, {Nogami}, \& {Shibata}}]{2013ApJS..209....5S}
{Shibayama}, T., {Maehara}, H., {Notsu}, S., {et~al.} 2013, \apjs, 209, 5

\bibitem[{{Stassun} {et~al.}(2019){Stassun}, {Oelkers}, {Paegert}, {Torres},
  {Pepper}, {De Lee}, {Collins}, {Latham}, {Muirhead}, {Chittidi},
  {Rojas-Ayala}, {Fleming}, {Rose}, {Tenenbaum}, {Ting}, {Kane}, {Barclay},
  {Bean}, {Brassuer}, {Charbonneau}, {Ge}, {Lissauer}, {Mann}, {McLean},
  {Mullally}, {Narita}, {Plavchan}, {Ricker}, {Sasselov}, {Seager}, {Sharma},
  {Shiao}, {Sozzetti}, {Stello}, {Vanderspek}, {Wallace}, \&
  {Winn}}]{2019AJ....158..138S}
{Stassun}, K.~G., {Oelkers}, R.~J., {Paegert}, M., {et~al.} 2019, \aj, 158, 138

\bibitem[{{Strassmeier} {et~al.}(2008){Strassmeier}, {Briguglio}, {Granzer},
  {Tosti}, {Divarano}, {Savanov}, {Bagaglia}, {Castellini}, {Mancini},
  {Nucciarelli}, {Straniero}, {Distefano}, {Messina}, \&
  {Cutispoto}}]{2008A&A...490..287S}
{Strassmeier}, K.~G., {Briguglio}, R., {Granzer}, T., {et~al.} 2008, \aap, 490,
  287

\bibitem[{{Strassmeier} {et~al.}(1990){Strassmeier}, {Fekel}, {Bopp},
  {Dempsey}, \& {Henry}}]{1990ApJS...72..191S}
{Strassmeier}, K.~G., {Fekel}, F.~C., {Bopp}, B.~W., {Dempsey}, R.~C., \&
  {Henry}, G.~W. 1990, \apjs, 72, 191

\bibitem[{{Strassmeier} {et~al.}(1994){Strassmeier}, {Handler}, {Paunzen}, \&
  {Rauth}}]{1994A&A...281..855S}
{Strassmeier}, K.~G., {Handler}, G., {Paunzen}, E., \& {Rauth}, M. 1994, \aap,
  281, 855

\bibitem[{{Uytterhoeven} {et~al.}(2011){Uytterhoeven}, {Moya},
  {Grigahc{\`e}ne}, {Guzik}, {Guti{\'e}rrez-Soto}, {Smalley}, {Hand ler},
  {Balona}, {Niemczura}, {Fox Machado}, {Benatti}, {Chapellier}, {Tkachenko},
  {Szab{\'o}}, {Su{\'a}rez}, {Ripepi}, {Pascual}, {Mathias},
  {Mart{\'\i}n-Ru{\'\i}z}, {Lehmann}, {Jackiewicz}, {Hekker}, {Gruberbauer},
  {Garc{\'\i}a}, {Dumusque}, {D{\'\i}az-Fraile}, {Bradley}, {Antoci}, {Roth},
  {Leroy}, {Murphy}, {De Cat}, {Cuypers}, {Kjeldsen}, {Christensen-Dalsgaard},
  {Breger}, {Pigulski}, {Kiss}, {Still}, {Thompson}, \& {van
  Cleve}}]{2011A&A...534A.125U}
{Uytterhoeven}, K., {Moya}, A., {Grigahc{\`e}ne}, A., {et~al.} 2011, \aap, 534,
  A125

\bibitem[{{Van Doorsselaere} {et~al.}(2017){Van Doorsselaere}, {Shariati}, \&
  {Debosscher}}]{2017ApJS..232...26V}
{Van Doorsselaere}, T., {Shariati}, H., \& {Debosscher}, J. 2017, \apjs, 232,
  26

\bibitem[{{Verbunt} \& {Phinney}(1995)}]{Verbunt_Phinney_1995}
{Verbunt}, F. \& {Phinney}, E.~S. 1995, \aap, 296, 709

\bibitem[{{Vida} \& {Roettenbacher}(2018)}]{2018A&A...616A.163V}
{Vida}, K. \& {Roettenbacher}, R.~M. 2018, \aap, 616, A163

\bibitem[{{Vrard} {et~al.}(2018){Vrard}, {Kallinger}, {Mosser}, {Barban},
  {Baudin}, {Belkacem}, \& {Cunha}}]{2018A&A...616A..94V}
{Vrard}, M., {Kallinger}, T., {Mosser}, B., {et~al.} 2018, \aap, 616, A94

\bibitem[{{Yang} \& {Liu}(2019)}]{2019ApJS..241...29Y}
{Yang}, H. \& {Liu}, J. 2019, \apjs, 241, 29

\bibitem[{{Zwintz} {et~al.}(2020){Zwintz}, {Neiner}, {Kochukhov}, {Rybchikova},
  {Pigulski}, {Muellner}, {Steindl}, {Kuschnig}, {Handler}, {Moffat}, {Pablo},
  {Popowicz}, \& {Wade}}]{2020arXiv200904784Z}
{Zwintz}, K., {Neiner}, C., {Kochukhov}, O., {et~al.} 2020, arXiv e-prints,
  arXiv:2009.04784

\end{thebibliography}

\begin{appendix} 
\onecolumn
\section{Comparison of temperature, radius and distance values from different data sources}\label{A1}

\begin{figure*}[h!!!]
   \centering
    \includegraphics[width=8.8cm]{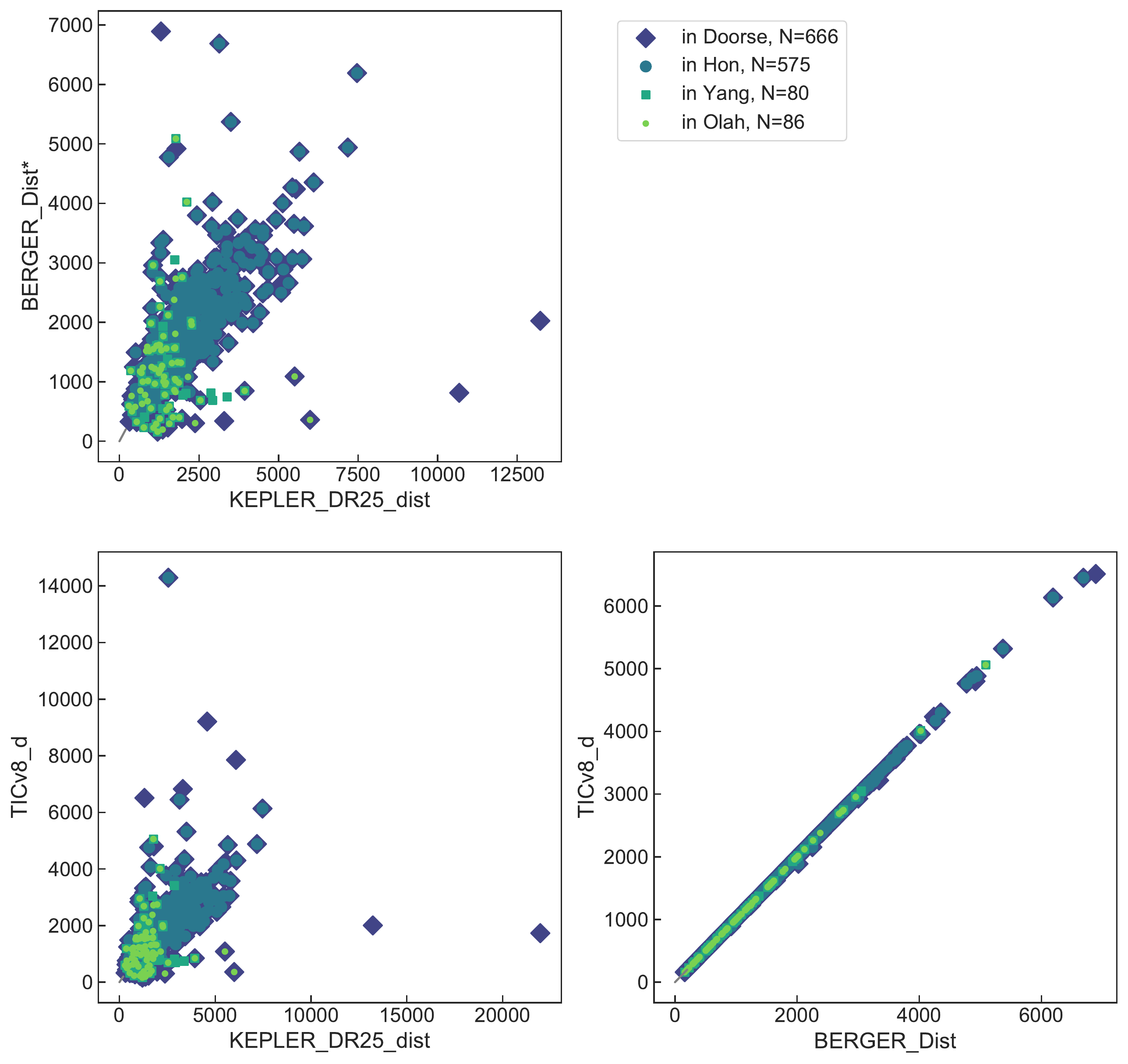}\hspace*{0.6cm}\includegraphics[width=8.8cm]{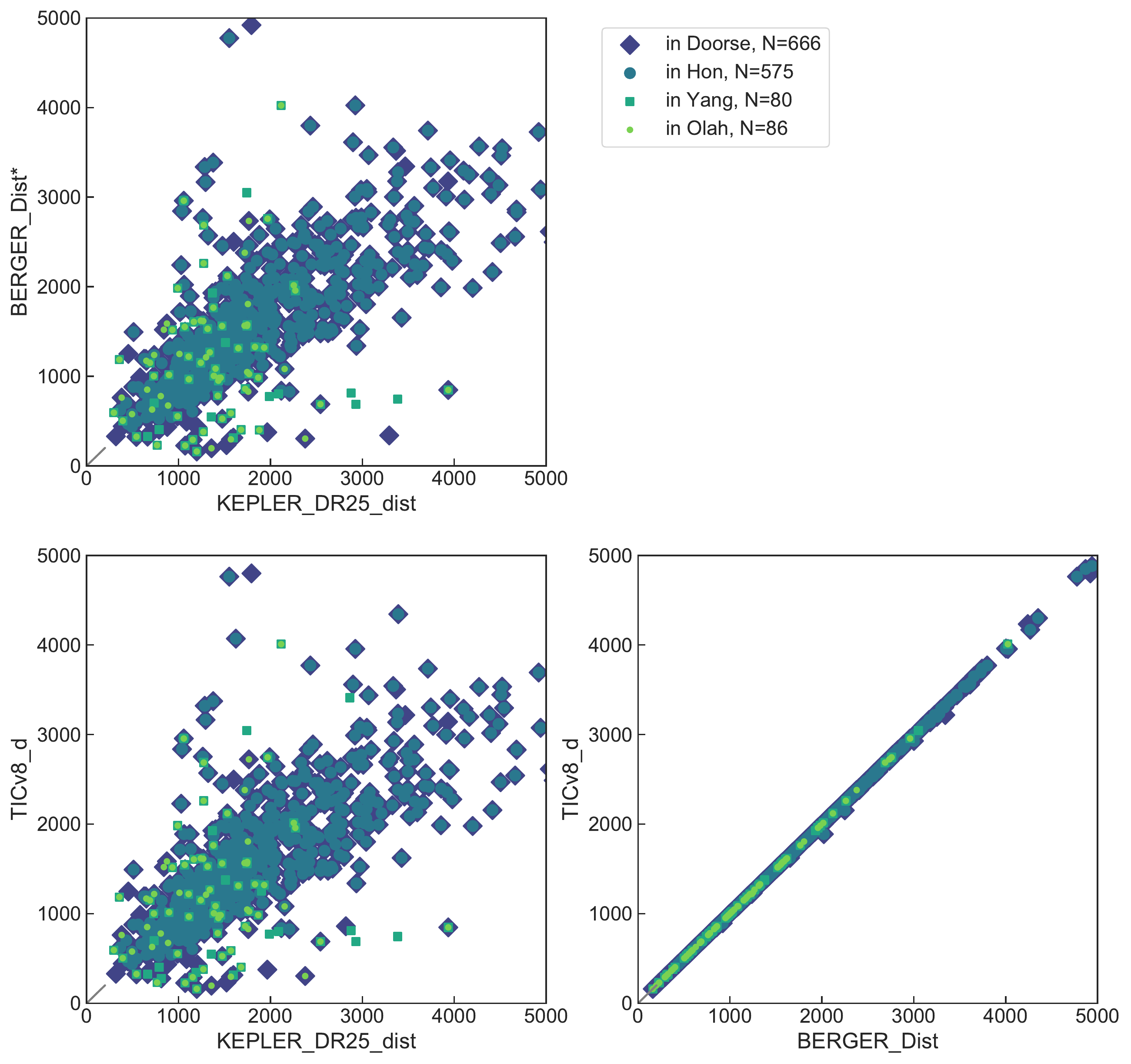}
    \caption{Left: Comparing distances of the flaring giant stars from D17 and YL19, and oscillating giant stars from \citet{2019MNRAS.485.5616H}, using distances from different catalogs marked in top right. The positions of the flaring stars verified in the present paper are also marked. Right: Zoom-in to Fig.~\ref{dist}.}
    \label{dist}
\end{figure*}

\begin{figure*}[h!!!]
\centering
    \includegraphics[width=8.8cm]{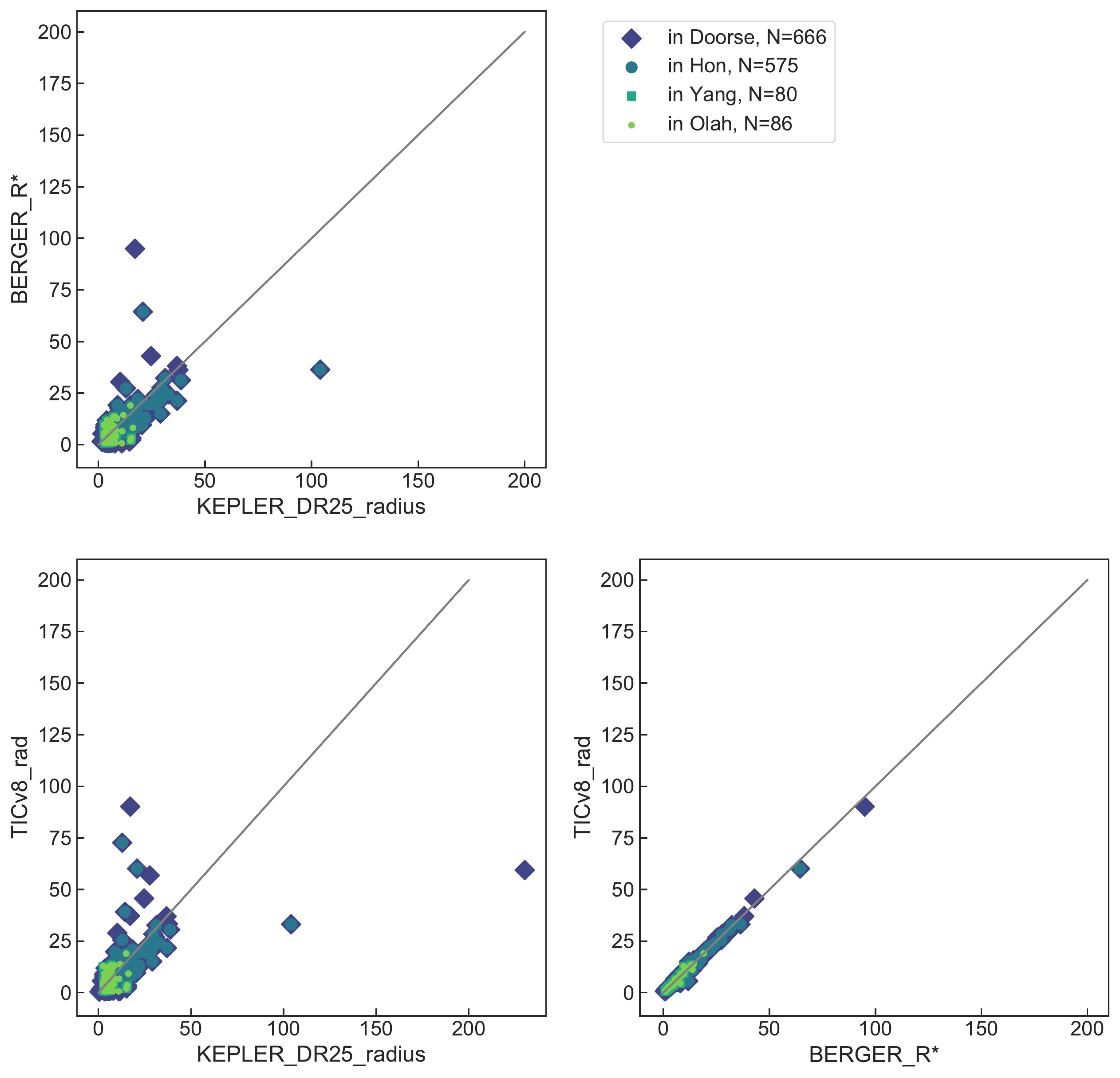}\hspace*{0.6cm}\includegraphics[width=8.8cm]{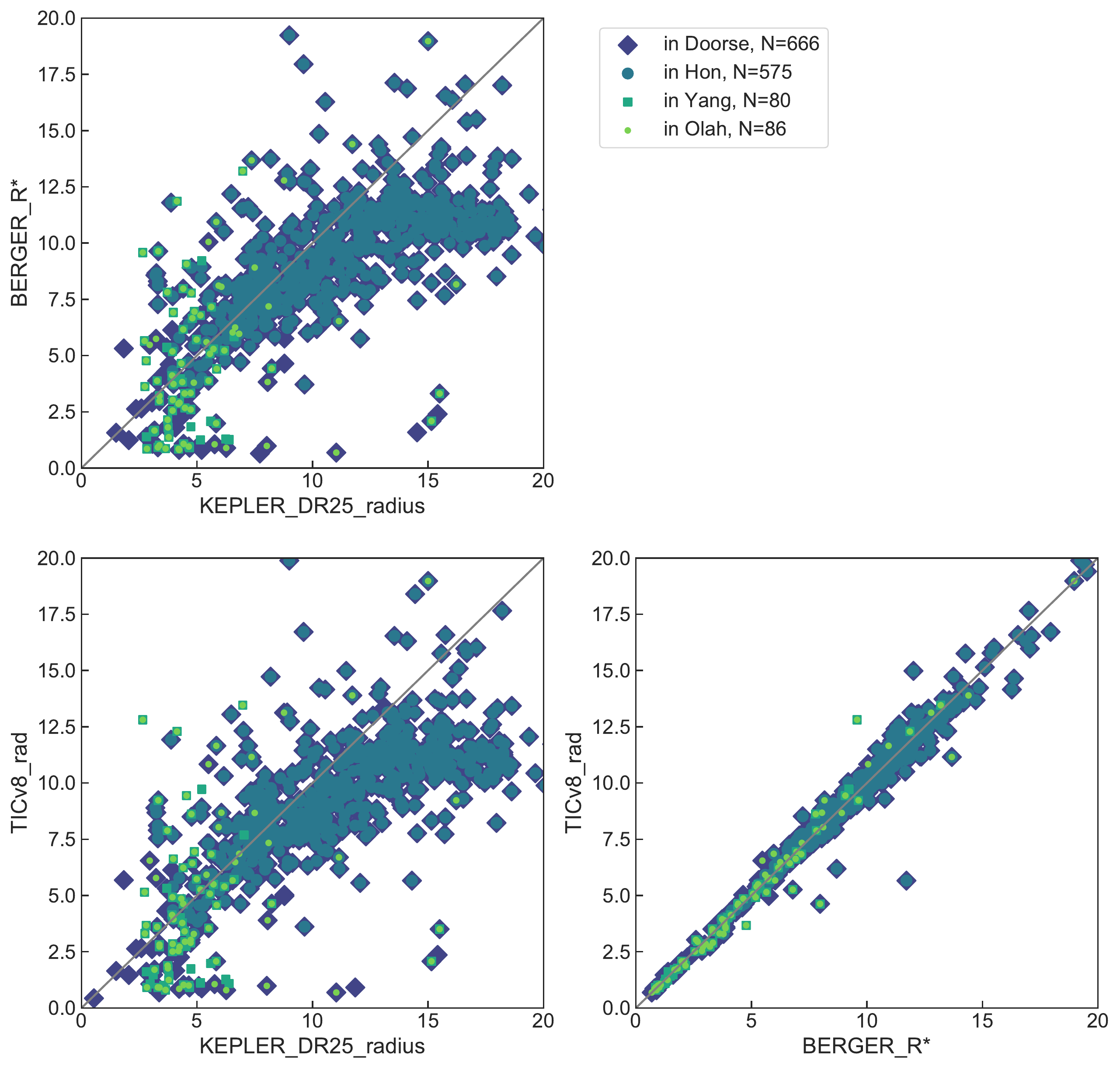}
    \caption{Left: Comparing radii of the flaring giant stars from D17 and YL19, and oscillating giant stars from \citet{2019MNRAS.485.5616H}, using radii from different catalogs marked in top right. The positions of the flaring stars verified in the present paper are also marked. Right: Zoom-in to Fig.~\ref{radius}.}
    \label{radius}
\end{figure*}

\begin{figure*}[h!!!]
    \centering
    \includegraphics[width=8.8cm]{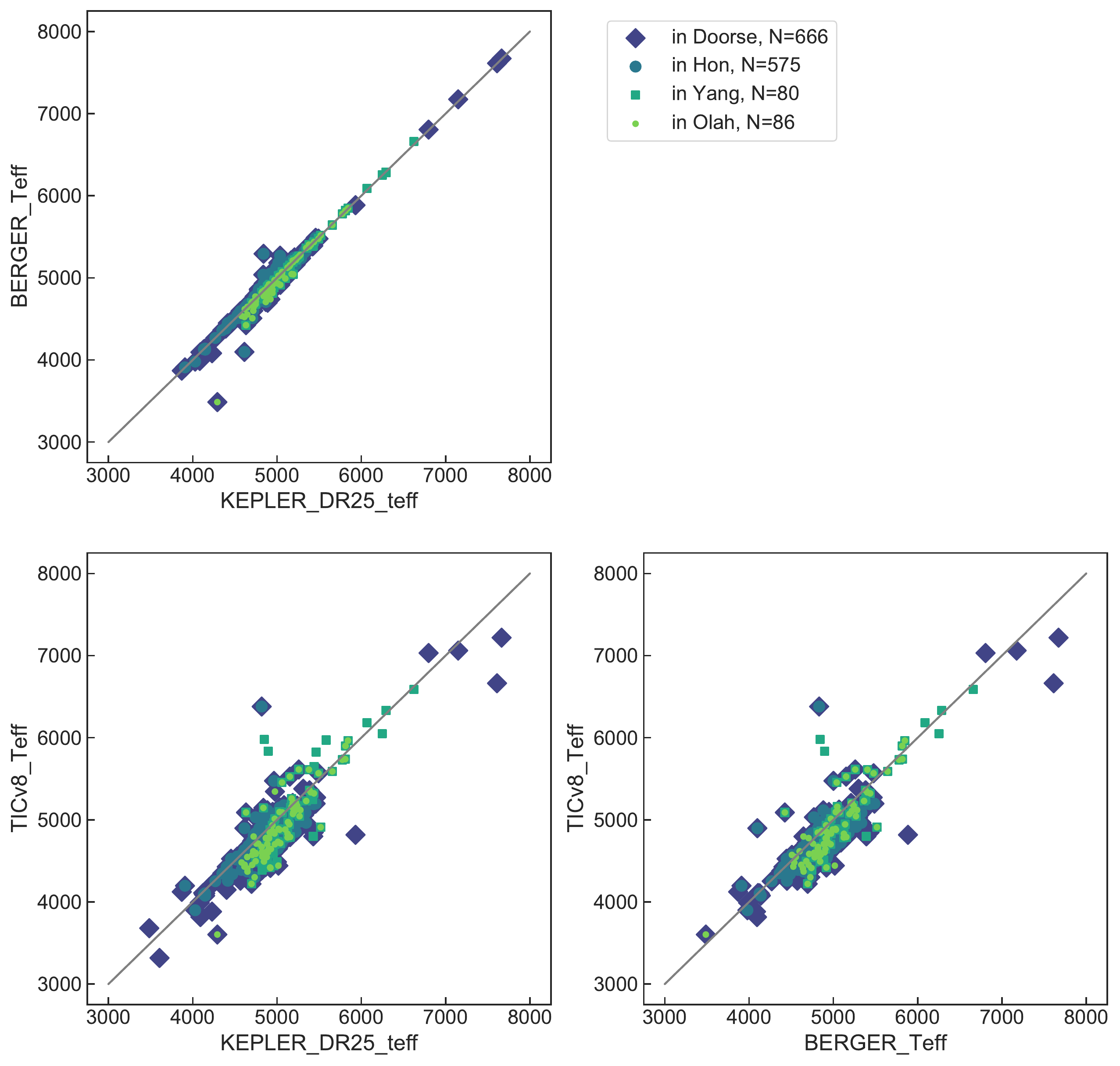}\hspace*{0.6cm}\includegraphics[width=8.8cm]{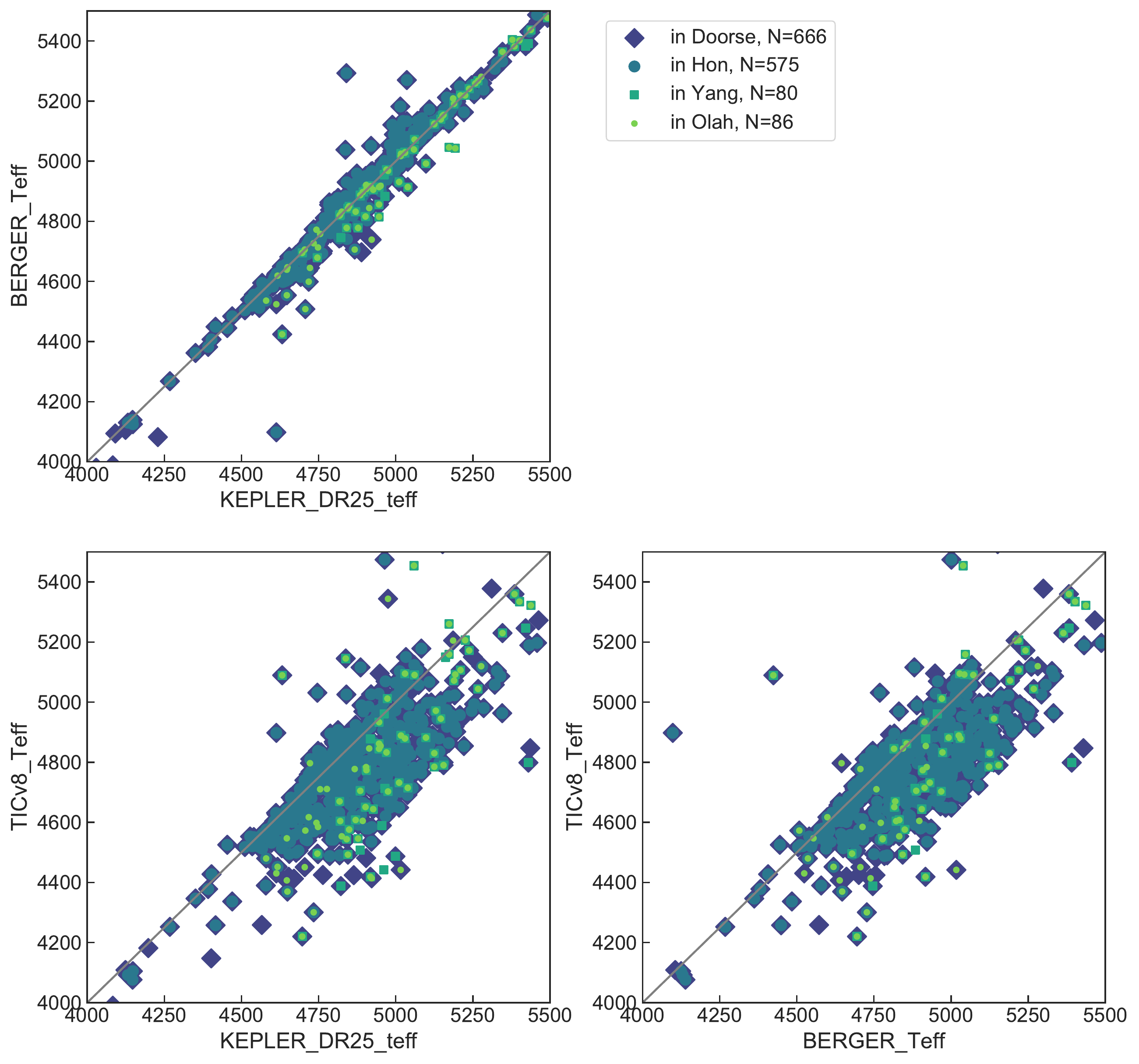}
    \caption{Left: Comparing temperatures of the flaring giant stars from D17 and YL19, and oscillating giant stars from \citet{2019MNRAS.485.5616H}, using temperatures from different catalogs marked in top right. The positions of the flaring stars verified in the present paper are also marked. Right: Zoom-in to Fig.~\ref{temp}.}
    \label{temp}
\end{figure*}

Figures \ref{dist}--\ref{temp} compare the temperature and radius values, and the distances of all \emph{Kepler} stars used throughout this paper originating from D17, YL19 and \citet{2019MNRAS.485.5616H} using temperature, radii and distances from KIC~DR25 and TIC catalogs, and from \citet{2018ApJ...866...99B}.

\bigskip
\bigskip

\bigskip
\bigskip

\clearpage

\newpage

\clearpage

\newpage

\section{Flaring giants and contaminating dwarfs in the \emph{Kepler} field}\label{B}

\begin{longtable}{r| r| l| l| c| r| r| c}
\caption{Flaring giants and contaminating dwarfs in the \emph{Kepler} field. Parameters are from TICv8, if not available, than from \emph{Gaia}~DR2 (G) or KIC (Kp). Notes of binarity are from \citet{Gaulme_2020}: ``SB'' stands for spectroscopic binary, ``SB1'', ``SB2'', and ``SB3'' for single, double, or triple-lined SBs. The flag ``RV-stable'' indicates that there are no significant radial-velocity variations and that the star is likely to be single. The flag ``unclear'' indicates that the red giant rotation is too large ($\simeq$100\,kms$^{-1}$) to give a reliable radial-velocity study \citep{Gaulme_2020}.} \label{tab:long} \\
\hline \hline
\multicolumn{1}{c|}{KIC No.} & \multicolumn{1}{c|}{Kpmag} & \multicolumn{1}{c|}{$T_{\rm eff}$ [K] } & \multicolumn{1}{c|}{$R/R_\odot$} & \multicolumn{1}{c|}{lum. class} & \multicolumn{1}{c|}{$P_{\rm rot}$ [d]} & \multicolumn{1}{c|}{no. of flares} & \multicolumn{1}{c}{binarity}\\
\hline
\endfirsthead
\caption{continued.}\\
\hline\hline
\multicolumn{1}{c|}{KIC No.} & \multicolumn{1}{c|}{Kpmag} & \multicolumn{1}{c|}{$T_{\rm eff}$ [K]} & \multicolumn{1}{c|}{$R/R_\odot$ } & \multicolumn{1}{c|}{lum. class} & \multicolumn{1}{c|}{$P_{\rm rot}$ [d]} & \multicolumn{1}{c|}{no. of flares} & \multicolumn{1}{c}{binarity}\\ 
\hline
\endhead
\hline
\endfoot
1573138    & 12.341  & 4838 Kp  &   2.871 Kp  &         &  4.687   &  113   &   \\         
1872340    & 12.770  & 5454    &   3.674    & GIANT\tablefootmark{a}  &  10.884   &  90    &   \\         
2142183    & 13.642  & 4738 G   &   1.206 G   &         &  3.637   &  113   &   \\         
2441154    & 10.326  & 4480    &   8.669    & GIANT   &  32.653  &  16    &   SB \\      
2585397    & 10.974  & 4975 Kp  &   8.554 Kp  &         &  49.048  &  14    &   \\         
2852961    & 10.146  & 4797    &   10.836   & GIANT   &  35.715  &  59    &   SB1 \\     
2968811    & 13.469  & 4220    &   3.504    & GIANT\tablefootmark{a}   &  14.814  &  118   &   \\        
3122450    & 13.891  & 4945    &   12.288   & GIANT   &  44.901   &  23    &   \\         
3324644    & 12.812  & 4451    &   3.897    & GIANT   &  2.875   &  42    &   \\         
3560427    & 13.775  & 4544    &   6.433    & GIANT   &  32.054  &  14    &   \\         
3561372    & 13.191  & 4414    &   5.671    & GIANT   &  15.115  &  33    &   \\         
4068539    & 13.569  & 4862    &   2.818    & GIANT\tablefootmark{a}   &  13.226  &  92    &   \\        
4157933    & 12.362  & 5567    &   3.547    & GIANT\tablefootmark{a}   &  6.914   &  139   &   \\        
4180534    & 13.018  & 4496    &   0.999    & DWARF   &  3.944   &  154   &   \\         
4263801    & 13.795  & 5206    &   1.883    & DWARF   &  4.304, 4.187  & 169   &  \\    
4273689    & 11.266  & 5159    &   3.309    & GIANT\tablefootmark{a}   &  29.301   &  134   &   \\         
4562996    & 13.013  & 5739    &   1.761    & DWARF   &  15.474   &  29    &   \\          
4680688    & 13.476  & 4846    &   8.037    & GIANT   &  9.190   &  19    &   \\          
4750889    & 13.448  & 5092    &   5.148    & GIANT   &  17.567   &  27    &   \\          
4920178    & 13.860  & 4452    &   5.922    & GIANT   &  55.359  &  18    &  \\
4937206    & 16.710  & 4605    &   0.851    & DWARF   &  4.030   &  26    &   \\         
5080290    & 9.507   & 5091    &   9.439    & GIANT   &        &  6     &   \\         
5181824    & 13.345  & 4732    &   9.231    & GIANT\tablefootmark{a}   &  6.861   &  79    &   \\        
5281818    & 12.309  & 5900    &   4.922    & GIANT   &  4.287   &  41    &   \\         
5286780    & 13.963  & 4919 G   &   0.687 G   &         &  8.625   &  12    &   \\         
5296446    & 12.631  & 5611    &   4.574    & GIANT   &  14.288   &  3     &   \\         
5428626    & 13.980  & 4664 G   &   0.880 G   &         &  2.607   &  128   &   \\         
5480528    & 13.061  & 4882    &   3.603    & GIANT\tablefootmark{a}   &  5.506   &  116   &   \\        
5482181    & 14.047  & 4889    &   3.779    & GIANT   &  8.799   &  15    &   \\         
5808398    & 11.455  & 4710    &   6.484    & GIANT   &  145.81   &  3     &   \\         
5821762    & 12.298  & 5205    &   13.125   & GIANT   &  9.252   &  11    &   unclear \\ 
6192231    & 12.796  & 4777 Kp  &   2.716 Kp  &         &  33.296  &  76    &   \\         
6206885    & 13.875  & 4644    &   4.133    & GIANT   &  15.925  &  40    &   \\         
6219880    & 13.060  & 5965    &   1.246    & DWARF   &  15.692   &  22    &   \\         
6233558    & 12.596  & 4617    &   6.854    & GIANT   &  29.247  &  39    & \\
6445442    & 13.194  & 4785    &   13.458   & GIANT   &  45.032   &  1     &  \\        
6619942    & 10.859  & 4598    &   18.979   & GIANT   &  10.627  &  3     &  \\        
6707805    & 12.447  & 5613    &   5.261    & GIANT\tablefootmark{a}   &  22.216  &  103   &  SB2 \\   
6861498    & 13.419  & 4670    &   5.498    & GIANT   &  89.276  &  16    & \\
7363468    & 12.386  & 5230    &   6.836    & GIANT   &  61.445  &  35    &  \\        
7433177    & 12.993  & 4442    &   6.550    & GIANT   &  5.935   &  25    &  \\        
7676676    & 13.853  & 4774    &   2.539    & GIANT\tablefootmark{a}   &  15.979  &  86    &  \\       
7696356    & 13.472  & 5044    &   5.396    & GIANT   &  40.657   &  64    &  \\        
7740188    & 12.581  & 4419    &   3.028    & GIANT\tablefootmark{a}   &  9.094   &  130   &  \\       
7838958    & 12.560  & 5360    &   5.727    & GIANT\tablefootmark{a}   &  27.369  &  119   &  \\        
7848068    & 13.501  & 5072    &   2.509    & GIANT   &  41.152  &  49    &  \\        
7869590    & 10.876  & 4370    &   7.337    & GIANT   &  40.568  &  47    &  SB1 \\    
7969754    & 12.269  & 4554    &   6.696    & GIANT   &  69.106  &  17    &  SB \\      
8022670    & 12.471  & 4573    &   3.290    & GIANT   &  4.126  &  55    &  \\         
8231401    & 13.824  & 4833    &   5.073    & GIANT   &  46.540   &  11    &  \\
8259835    & 13.112  & 4651    &   0.915    & DWARF   &  4.427   &  124   &  \\        
8515227    & 11.176  & 4778    &   11.653   & GIANT   &  28.157  &  23    &  SB3  \\   
8517303    & 11.430  & 4547    &   8.679    & GIANT   &  53.385  &  9     &  SB1 \\    
8713822    & 14.258  & 5095    &   1.702    & DWARF   &  3.223   &  41    &  \\        
8749284    & 12.190  & 5089    &   4.628    & GIANT   &  3.192   &  88    &  unclear \\   
8774912    & 12.995  & 4585    &   0.792    & DWARF   &  3.796   &  105   &  \\        
8776850    & 12.691  & 5172    &   1.027    & DWARF   &  5.080   &  174   &  \\        
8780458    & 13.589  & 4430    &   1.060    & DWARF   &  2.086   &  93    &  \\        
8894773    & 13.833  & 4845    &   0.910    & DWARF   &  1.892   &  106   &  \\        
8915957    & 10.918  & 4910    &   12.811   & GIANT   &  47.004   &  11    &  \\        
8951096    & 13.336  & 4879    &   6.628    & GIANT   &  9.074    &  9     &  \\        
9093349    & 13.642  & 5130 Kp  &   4.000 Kp  &         &  5.480   &  251   &  \\        
9116222    & 13.390  & 5526    &   2.715    & DWARF\tablefootmark{b}   &  7.443   &  260   &  \\        
9237305    & 12.644  & 5107    &   2.074    & GIANT\tablefootmark{a}   &  23.458, 29.136 &   95  & \\      
9419002    & 12.411  & 5322    &   4.857    & GIANT\tablefootmark{a}   &  25.640   &  104   &  \\       
9474208    & 15.941  & 3603    &   0.688    & DWARF   &  7.739   &  64    &  \\        
9716554    & 12.655  & 5120    &   5.779    & GIANT   &  68.318   &  1     &  \\        
9752982    & 12.224  & 4923 G   &   0.888 G   &         &  13.270   &  12    &  \\        
9770992    & 10.531  & 4784    &   13.892   & GIANT   &  17.817  &  25    &  SB2  \\   
9992402    & 15.208  & 4715    &   0.916    & DWARF   &  3.573   &  17    &  \\        
10603977   & 13.651  & 4933    &   2.728    & GIANT\tablefootmark{a}   &   17.962 &  112   &   \\        
10646009   & 12.697  & 4493    &   4.634    & GIANT\tablefootmark{a}   &   5.412  &  80    &   \\      
10666510   & 10.258  & 4790    &   2.074    & GIANT\tablefootmark{a}   &   20.845 &  165   &  \\       
10875937   & 14.092  & 4546    &   3.407    & GIANT   &   6.251   &  53    &  \\        
11087027   & 10.881  & 4854    &   11.157   & GIANT   &   30.427 &  10    &  RV-stable \\       
11135986   & 13.944  & 4830    &   2.899    & GIANT\tablefootmark{a}   &   8.692  &  173   &   \\       
11146520   & 12.422  & 4608    &   8.618    & GIANT   &   34.217  &  2     & SB1 \\     
11407895   & 11.870  & 5732    &   6.231    & GIANT   &   9.206   &  5     &  \\        
11515713   & 12.889  & 5589    &   2.847    & DWARF\tablefootmark{b}   &   19.849, 23.210 &  186  &  \\       
11551404   & 11.105  & 4703    &   2.944    & GIANT\tablefootmark{a}   &   11.120  &  155   &  SB2 \\    
11554998   & 11.779  & 4301    &   9.231    & GIANT   &   30.335 &  20    &  SB1 \\    
11568624   & 13.754  & 4407    &   0.983    & DWARF   &   1.646  &  49    &  \\        
11668891   & 13.722  & 4576    &   7.883    & GIANT   &   19.097  &  8     &  \\        
11753121   & 13.814  & 5335    &   0.802    & DWARF   &   22.410  &  11    &  \\        
11962994   & 13.984  & 4604    &   6.943    & GIANT   &   12.872  &  4     &  \\
11970692   & 13.516  & 4705    &   3.944    & GIANT\tablefootmark{a}   &   28.116 &  98    &  \\
\end{longtable}
\tablefoot{
\tablefoottext{a}{Beside the GIANT evolutionary status mark the 19 most flaring giant stars with constructed FFDs.}
\tablefoottext{b}{Beside the DWARF evolutionary status these stars are probably subgiants, see Section~\ref{energy_distribution}, Fig.~\ref{histo_all}}
}

\newpage

\begin{figure*}[h!!!]
\centering
    \includegraphics[width=16cm]{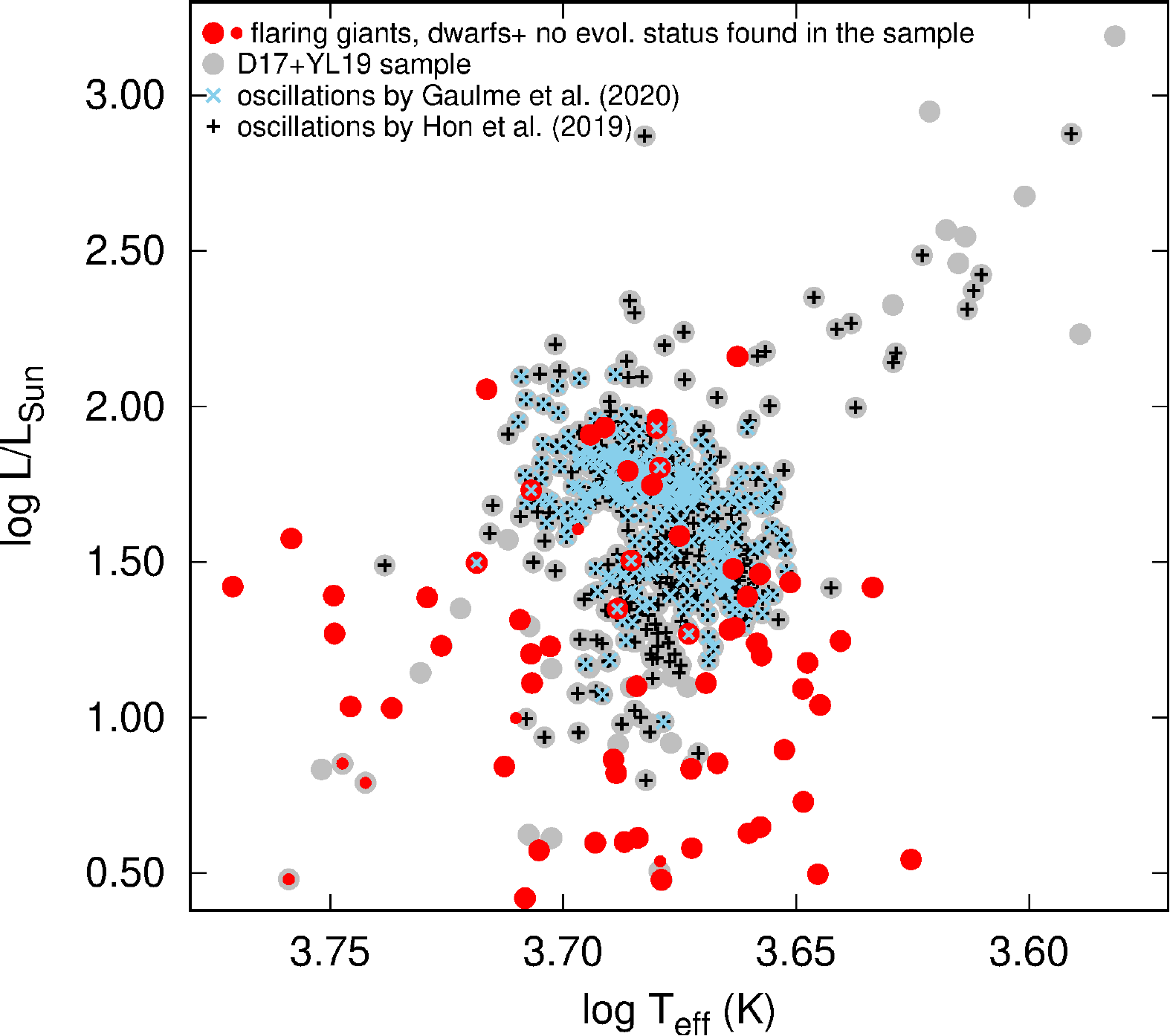}
    \caption{Enlarged centre of the HRD showing locations of the D17+YL19 giant sample (grey dots) and the derived flaring stars from this sample in the present paper (red dots). Light blue crosses mark the oscillating giant stars from \citet{Gaulme_2020} for stars brighter than Kpmag=12.5 and black plusses stand for the results of oscillating giants by \citet{2019MNRAS.485.5616H}, cross-matched with our D17+YL19 giant sample.}
    \label{HRD_enlarge}
\end{figure*}

\begin{table*}[ht]
 \caption{Uncertain cases from the D17+YL19 sample.}
 \label{table:uncertain}      
\centering 
\begin{tabular}{r r c r c r r l}
 \hline \hline
KIC No.     & Kpmag  & $T_{\rm eff}$ [K] &  $R/R_\odot$  & lum. class. & $P_{\rm rot}$ [d]  & no. of flares & remarks \\  
\hline
2019477  & 13.330 & 5647 &   2.733 & DWARF &  ---   &    2     & = KOI-6083 with two planets (1) \\  
3245458  & 13.795 & 5980 &   1.085 & DWARF & 12-16  &    9     & blend (2) \\  
3659584  & 13.045 & 4595 &   7.346 & GIANT & 68.41  &    1     & large ampl. rot. mod. (3) \\ 
4752302  & 12.687 & 4811 &   3.615 & GIANT &  ---   &    1     & oscillations (?) \citet{2019MNRAS.485.5616H} \\
5093168  & 13.421 &      &         & GIANT &  2.799 &   21     & two stars (4) \\
5961988  & 14.139 & 5837 &   1.115 & DWARF &  ---   &   12     & blend (5) \\ 
6190679  &  9.030 & 4570 &  15.147 & GIANT & 76.9   &    1     & oscillations (?) \citet{2019MNRAS.485.5616H} \\
6364525  & 10.993 & 5040 &   2.659 & GIANT &  ---   &    5     & (6) \\
6863731  & 11.611 & 5104 &   4.029 & GIANT & 17     &    2     & very small ampl., oscill. (?) \citet{2019MNRAS.485.5616H} \\
6869726  & 13.763 & 4855 &   1.074 & DWARF &  0.722 &    4     & too short period (7) \\ 
7272363  & 13.128 & 5058 &   3.833 & GIANT &  ---   &    2     & oscillations (?)  \citet{2019MNRAS.485.5616H}  \\
7292720  & 13.773 & 4822 &   4.539 & GIANT &  ---   &    1     & oscillations (?)  \citet{2019MNRAS.485.5616H}  \\
8106192  & 13.373 & 4763 &   6.366 & GIANT &  ---   &    1     & oscillations (?)  \citet{2019MNRAS.485.5616H}  \\
8112102  & 13.852 & 5150 &   7.691 & GIANT & 9.45   &    5     & blend? (8)  \\
9592627  & 16.955 & 5150 &   0.911 & DWARF & 3.598  &    1     & small amplitude regular variability (9) \\ 
9835672  &  8.694 & 4197 &  33.135 & GIANT &  ---   &    1     & oscillations (?)  \citet{2019MNRAS.485.5616H} \\
10421610 & 12.990 & 5826 &   1.001 & DWARF & 13.5   &  134     & looks as a single object, many flares (10) \\
\hline
\end{tabular}
\tablefoot{
(1) has a close companion of similar brightness (\emph{Gaia}~DR2 2051819028221333120, looks redder), both can flare
(2) measured together with KIC~3245449, an RGB star, 0.7\,mag brighter, with oscillations (e.g. \citet{2014A&A...572L...5M}), the periodograms of both stars shows peaks around 12-16 days. Both could flare.
(3) many faint stars around
(4) blended by the dwarf KIC~5093163 \citep{2014ApJS..211...24M}. The two rot. periods 2.799 (giant) and 0.402 (dwarf) are clearly seen in the Fourier spectrum, 2\,mag diff. in $J$ (dwarf is fainter), both can flare
(5) blended by the oscillating KIC~5961985, 2\,mag brighter \citep{2018A&A...616A..94V} 
(6) bright F-star around, not very near
(7) KIC 6869740 is close, 2\,mag brighter, FIII-IV, and pulsate (P=0.351441)
(8) a fainter star very close
(9) 2\,mag fainter star nearby, one uncertain flare
(10) Two distinct periods. First: at 3.18 days there is one sharp signal, second: around 13.5\,d a bunch of signals are seen, typical of pulsation and differential rotation, respectively.
}
\end{table*}

\newpage

\begin{table*}[h]
 \caption{Available LAMOST results}
 \label{table:LAMOST}     
\centering 
\begin{tabular}{r r | r r r r | r r | r r r r}
 \hline \hline
KIC No. & $T_{\rm eff}$ [K] &  $T_{\rm eff}$ [K] & $\log g$ & [Fe/H] & remark\tablefootmark{a} & KIC No. & $T_{\rm eff}$ [K] &  $T_{\rm eff}$ [K] & $\log g$ & [Fe/H] &  remark\tablefootmark{a} \\ 
    & \multicolumn{1}{c|}{TICv8}   & \multicolumn{4}{c|}{LAMOST} &  & \multicolumn{1}{c|}{TICv8}  & \multicolumn{4}{c}{LAMOST} \\
\hline

1573138  &  4838  &    5182 &   3.93 &   0.07 &  V? EM     &  8776850  &  5172  &   5140  &  4.53 &  0.12  &  ABS\\
1872340  &  5454  &    5454 &   3.80 &   0.34 &  ABS              &           &        &   5185  &  4.55 &  0.11  &  ABS\\
2585397  &  4975  &    5218 &   2.98 &   0.18 &  ABS              &           &        &   5179  &  4.58 &  0.09  &  ABS\\
2852961  &  4797  &    4747 &   2.22 &  $-$0.31 &  ABS              &  8780458  &  4430  &  4427 &   4.47 &   0.13  &  W ABS \\
4157933  &  5567  &   --    &   --   &  --    &  ABS              &  8894773  &  4845  &  4845 &   4.37 &  $-$0.23  &  filled in \\
4180534  &  4496  &   4496  &  4.32 &  0.20  &  filled in        &  8915957\tablefootmark{b}  &  4910  &  7001 &   4.25 &   0.39  &  H ABS \\
4263801  &  5206  &   5211  &  3.80  &  0.12  &  ABS              &           &        &  6972 &   4.28 &   0.42  &  H ABS \\
4750889  &  5092  &   --    &   --   &  --    &  ABS              &  8951096  &  4879  &  4827 &   2.71 &  $-$0.54  &  ABS \\
5281818  &  5900  &   5900  &  3.57  &  0.10  &  S ABS     &  9093349  &  5130  &  4971 &   3.87 &   0.17  &  filled in \\
5296446  &  5611  &   5616  &  3.50 &  0.10  &  ABS              &  9116222  &  5526  &  5526 &   3.69 &  $-$0.20  &  ABS \\
5808398  &  4710  &   4536  &  3.00  &  0.20  &  ABS              &  9237305  &  5107  &  5220 &   3.94 &   0.33  &  ABS \\
6192231  &  4777  &   5398  &  4.07  &  0.32  &  ABS              &  9419002  &  5322  &  5388 &   3.45 &   0.04  & S ABS \\
6219880  &  5965  &   5972  &  4.42 &  0.12  &  ABS              &  9716554  &  5120  &  5257 &   3.35 &  $-$0.36  & S ABS \\
7363468  &  5230  &   5237  &  3.09 &  0.05  &  ABS              &  9752982  &  4923  &  5191 &   4.73 &   0.09  &  ABS \\
         &        &   5225  &  3.09 &  0.07  &  ABS              &  10603977 &  4933  &  4933 &   3.80 &   0.09  &  filled in \\
7433177  &  4442  &   5031  &  3.37  &  0.09  &  W ABS         &  10646009 &  4493  &  4505 &   3.10 &  $-$0.12  &  EM \\
7676676  &  4774  &   4774  &  3.72  &  0.24  &  EM               &  10666510 &  4790  &  4784 &   3.68 &  $-$0.35  &  W ABS \\ 
         &        &   4920  &  3.82  &  0.35  &  ABS              &           &        &  4797 &   3.75 &  $-$0.33  &  filled in \\
7740188  &  4419  &   4741  &  3.57 &  0.24  &  filled in        &  11135986 &  4830  &  --   &    --  &   --    &  EM \\
         &        &   4812  &  3.61  &  $-$0.21 &  W ABS         &  11407895 &  5732  &  5796 &   3.43 &  $-$0.01  &  S ABS  \\
7848068  &  5072  &   5072  &  3.80  &  0.20  &  ABS              &  11515713 &  5589  &  5589 &   3.88 &   0.18  &  ABS \\
7869590  &  4370  &   4420  &  2.57  &  0.09  &  filled in?       &  11551404 &  4703  &  5046 &   3.77 &  $-$0.08  &  W ABS  \\
         &        &   4392  &  2.58  &  0.19  &  ABS              &           &        &  5181 &   3.95 &   0.01  &  filled in  \\
8022670  &  4573  &   4623  &  3.13 &  0.27  &  S EM       &  11554998 &  4301  &  4386 &   2.42 &  $-$0.25  &  W ABS  \\
         &        &   4525  &  3.04 &  0.35  &  S EM      &           &        &  4431 &   2.46 &  $-$0.22  &  W ABS  \\
         &        &   4366  &  2.96 &  0.41  &  S EM      &           &        &  4422 &   2.51 &  $-$0.25  &  W ABS  \\
8231401  &  4833  &   --    &   --   &  --    &  ABS              &  11753121 &  5335  &  5335 &   4.65 &   0.16  &  ABS  \\
8517303  &  4547  &   --    &   --   &  --    &  ABS              &           &        &  5437 &   4.66 &   0.17  &  ABS  \\
8749284  &  5089  &   5089  &  3.25  &  0.18  &  H EM?        &  11962994 &  4604  &  4604 &   2.69 &   0.01  &  ABS   \\
8774912  &  4585  &   --    &   --   &  --    &  EM               &           &        &       &        &         & \\
\hline                                                            
\end{tabular}                                                     
\tablefoot{                                 \tablefoottext{a}{ABS=absorption, EM=emission, V=variable, H=huge, S=strong, W=weak;} 
\tablefoottext{b}{This should be an identification problem because the LAMOST spectra indicate a main-sequence F-star.}
{For KIC~7433177, 7869590, 8022670, 8749284, 10666510 and 11551404 \citet{2016A&A...594A..39F} lists H$\alpha$ and Ca\,{\sc ii} infrared triplet equivalent widths.}
}
\end{table*}

\clearpage

\newpage
\section{Flare energy distributions and flare energy$-$observed flare duration correlations of the 19 most flaring stars and that of KIC~2852961.}\label{C}

\begin{figure*}[h!!!]
    \centering
    \includegraphics[width=0.88\columnwidth]{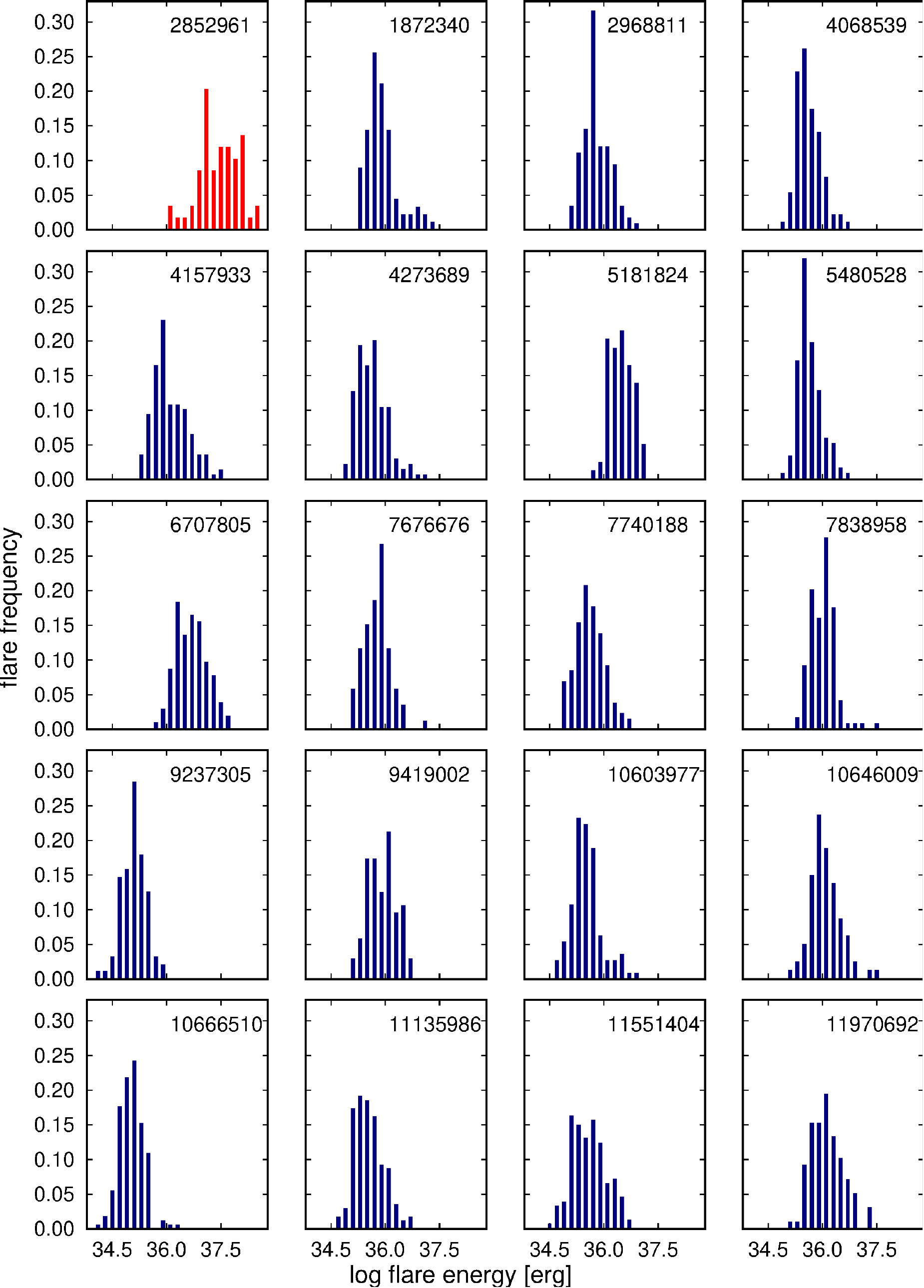}
    \caption{Flare energy distributions of the 19 most flaring stars (blue) plus KIC~2852961 (red).}
    \label{histo:19stars}
\end{figure*}

\begin{figure*}[h!!!]
    \centering
    \includegraphics[width=\columnwidth]{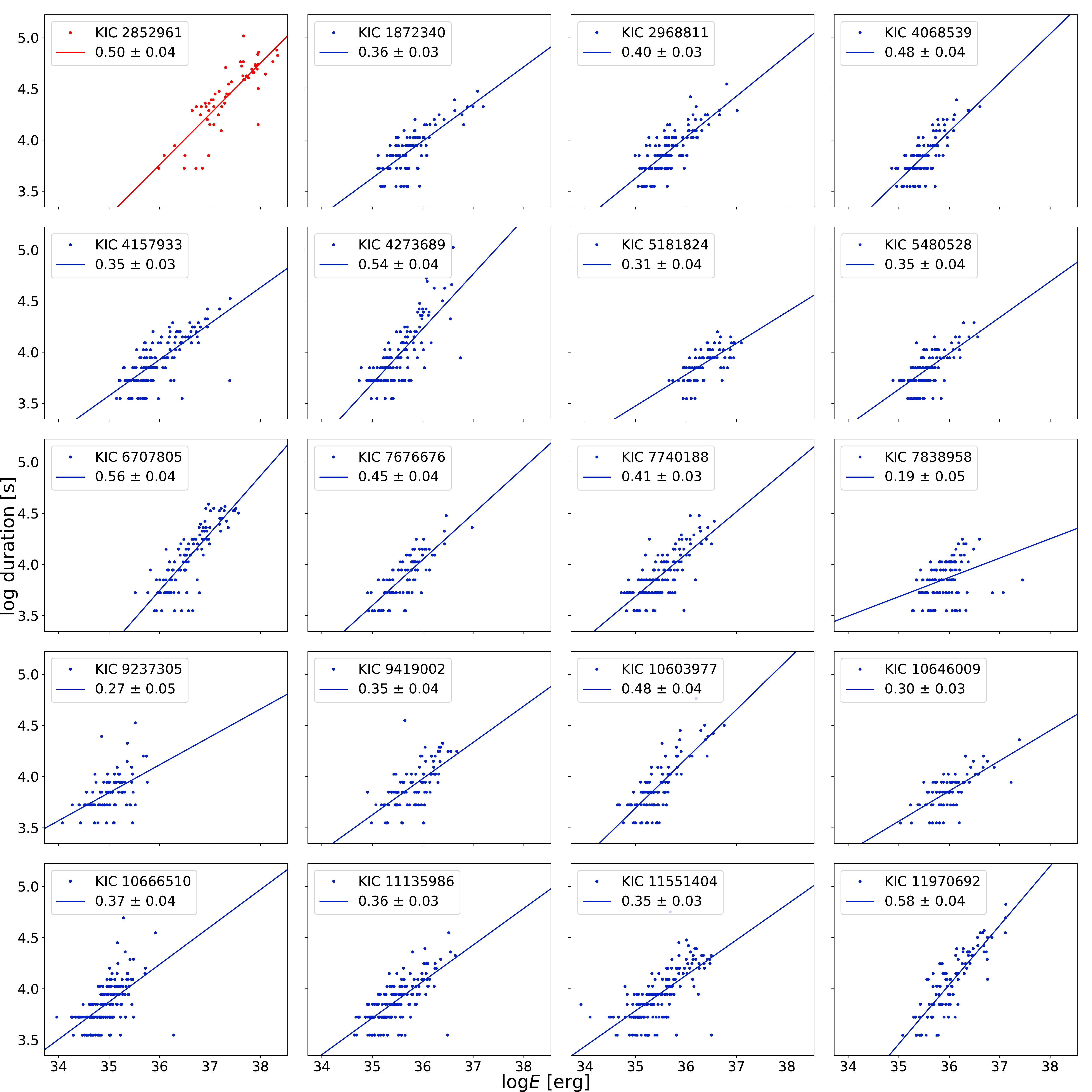}
    \caption{Flare energy$-$observed flare duration correlations of the 19 most flaring stars (blue) plus KIC~2852961 (red).}
    \label{duration:19stars}
\end{figure*}

\end{appendix}

\end{document}